\documentclass[final,3p,times,onecolumn]{elsarticle}

\usepackage{amssymb}
\usepackage{graphicx}
\usepackage{hyperref}
\usepackage{amsmath}
\usepackage{natbib}
\usepackage[utf8]{inputenc}		
\usepackage[T1]{fontenc}

\journal{Handbook of Astrochemistry}

\begin{document}

\begin{frontmatter}



\title{Observations of non complex organic molecules in the gas phase of the interstellar medium}


\author[inst1]{Charlotte VASTEL}

\affiliation[inst1]{organization={IRAP, Universit\'e de Toulouse, CNRS, UPS, CNES},
            city={Toulouse},
            postcode={31400}, 
            country={FRANCE}}

\author[inst2]{Francesco FONTANI}
\affiliation[inst2]{organization={INAF-Osservatorio Astrofisico di Arcetri},
            addressline={Largo E. Fermi 5}, 
            city={Florence},
            postcode={I-50125}, 
            country={ITALY}}

\begin{abstract}
The field of astrochemistry has seen major advances triggered by the completion of new powerful radio telescopes, with gains in sensitivity of receivers and in bandwidth. As of June 2026, about 347 molecular species have been detected, in interstellar clouds, circumstellar shells and even extragalactic sources. The first interstellar molecules were first discovered through their electronic transitions in the visual and near UV regions of the spectra in the 1930s. Then the discovery of (pure) rotational transitions of interstellar molecules dates back to the late 1960s. The improvement of detectors and the increase in telescope sizes really opened up the submillimeter sky. The radio and submillimeter ranges cover the lowest rotational lines of molecular species. The bigger the molecule, the more spectral lines at different frequencies it produces, with weaker line intensities. Over the past 30 years, we have discovered that we live in a molecular universe, where molecules are abundant and widespread, probing the structure and evolution of galaxies, as well as the temperature and density of the observed medium, opening a new field called astrochemistry. The progress has been dramatic, since the discovery of the first molecules about 100 years ago. We present in this review, the detection techniques that led to the discovery of the simple molecules in the gas phase and the methodology that lead to the abundances determinations and the comparison with chemical modelling.
\end{abstract}




\end{frontmatter}


\section{Detection techniques}
\label{sec:detection}

\subsection{Electromagnetic spectrum}
\label{subsec:spectrum}

The description of molecular structure is more complicated than that of isolated atoms.
In an atom, energy levels are determined only by the electronic state. Instead, a
molecule has more degrees of freedom, due to the presence of more than one nucleus.
The nuclei can oscillate around their relative equilibrium positions and
they can rotate about the mass center.
The problem is greatly simplified because the mass of the electrons is much smaller
than that of the nuclei, while the forces to which the electrons and the nuclei are submitted
are of comparable intensity. As a result, the motion of the nuclei is negligible
with respect to that of the electrons, so that, we can assume that the nuclei occupy
nearly fixed positions within the molecule.
One can demonstrate that under this approximation (Born-Oppenheimer approximation), the electronic,
vibrational and rotational eigenfunctions can be separated. 
Therefore the
total energy is given by the sum of the eigenvalues of the three contributions:

\begin{equation}
  E_{\rm tot}=E_{\rm el}+E_{\rm vib}+E_{\rm rot} \;,
  \label{eq:energy}
\end{equation}
where 
\begin{itemize}
    \item [$E_{\rm el}$] are the energies of the electronic states: typical energy jumps between
two adjacent electronic states are of a few eV. These lines are mostly in the ultraviolet (UV) and optical
region of the spectrum.
\item [$E_{\rm vib}$] are the energies of the vibrational states: typical energy jumps between
two adjacent vibrational states are of 0.1-0.01 eV. These lines are mostly in the
infrared (IR) region of the spectrum.
\item [$E_{\rm rot}$] are the energies of the rotational states: typical values energy jumps
between two adjacent rotational states are of 0.001 eV. These lines are in the millimeter
and centimeter (that is radio) region of the spectrum.
\end{itemize}

Given the energies shown above, and the typical physical conditions of molecular clouds characterised by kinetic temperatures of $\sim 10-100$~K and average H$_2$ volume densities of $\sim 10^3 - 10^7$~cm$^{-3}$, only rotational levels in the ground electronic and vibrational states are easily populated. For this reason, in the following we focus our attention on them.

The rotational hamiltonian for a generic molecule, parametrized as a rigid rotor, is:

\begin{equation}
    H = \frac{J_{\rm a}^2}{2 I_{\rm a}}+\frac{J_{\rm b}^2}{2 I_{\rm b}}+\frac{J_{\rm c}^2}{2 I_{\rm c}}\;,
\label{eq:hamiltonian}
\end{equation}
where $a$, $b$, and $c$ label the three coordinate principal axes of the molecule, $J_{\rm a}$, $J_{\rm b}$, and $J_{\rm c}$ are the projections of the total angular momentum along the corresponding principal axis and $I_{\rm a}$, $I_{\rm b}$, and $I_{\rm c}$ are the momenta of inertia of the molecule along the three axes.
It is possible to solve the Schr\"{o}dinger equation and find the energy levels based on the geometry of the molecule. Based on the values of the principal momenta of inertia, molecules can be divided in the following categories:
linear, symmetric top, asymmetric top and spherical rotors.

\begin{figure}
    \centering
    \includegraphics[width=1\textwidth]{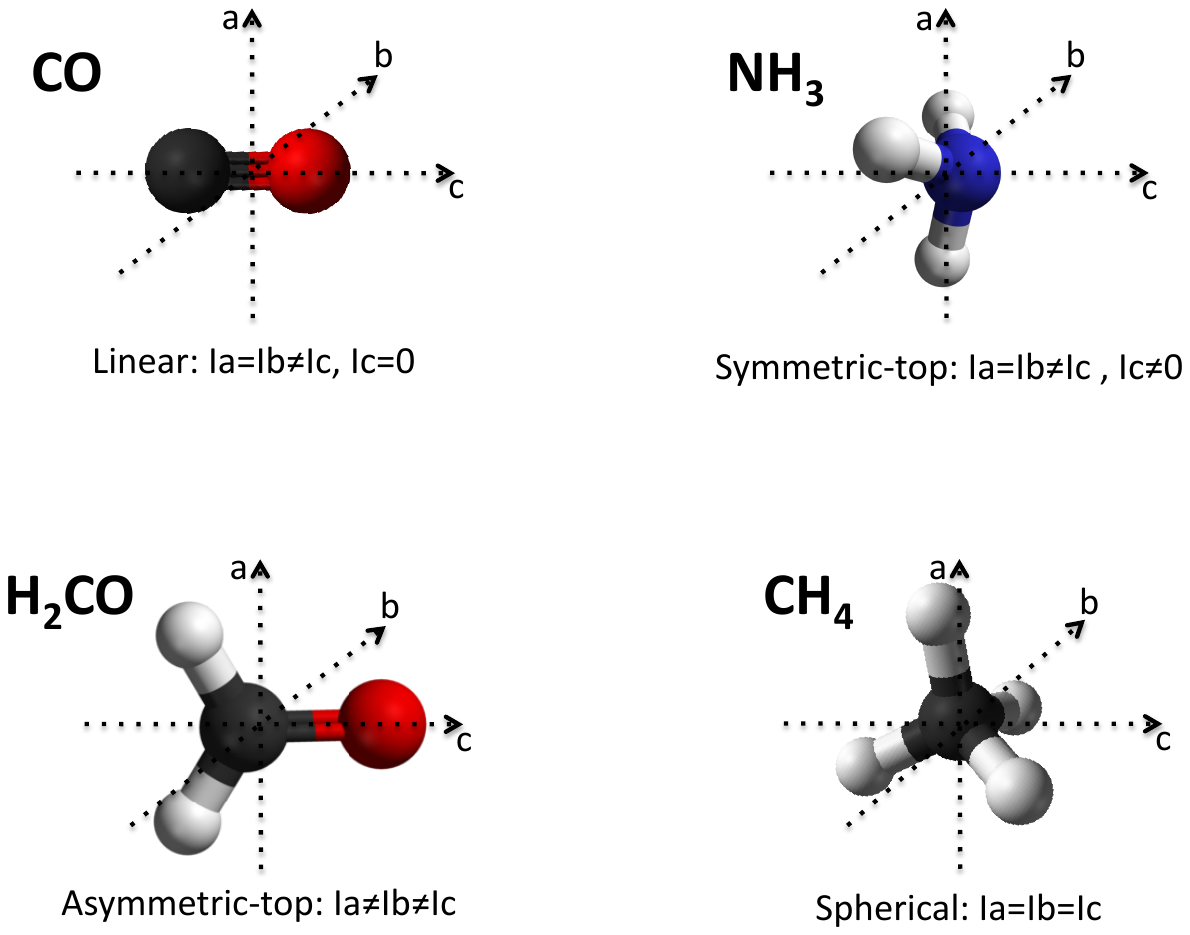}
    \caption{Classification of molecules according to their principal momenta of inertia.}
    \label{fig:molecule-type}
\end{figure}

Examples of molecules belonging to the four groups are illustrated in Fig.~\ref{fig:molecule-type}. In the following we give a brief description of the spectrum and the characteristic quantum numbers of the energy levels of each group. For a detailed description, we refer to \citet{Townes1955}.

\subsubsection{Linear rotors}

Linear rotors are molecules characterized by a linear geometry, so that one of the principal momentum of inertia, e.g. $I_{\rm c}$, is zero, and the others are equal ($I_{\rm a} = I_{\rm b} = I_{\perp}$). In this case $H$ can be written as:
\begin{equation}
    H=\frac{J_{\rm a}^2}{2 I_{\perp}}+\frac{J_{\rm b}^2}{2 I_{\perp}} = \frac{J^2}{2 I_{\perp}}\;,
\end{equation}
whose eigenvalues are:
\begin{equation}
    E_{\rm J}=\frac{\hbar}{2 I_{\perp}}J(J+1)=B_{\rm rot}hJ(J+1)\;,
\end{equation}
where $B_{\rm rot}\equiv h / 8\pi^2 I_{\perp}$ is the rotational constant of the molecule, and $h$ is the Planck constant. In the dipole approximation,
the selection rules allow transitions with $\Delta J = \pm 1$, and the energy of the photon emitted is thus:
\begin{equation}
    E_{\rm J \rightarrow J-1}=B_{\rm rot}hJ(J+1)-(J-1)J=B_{\rm rot}hJ \;.
\end{equation}
Due to this, the rotational spectrum of linear molecules has the peculiar feature that the distance
between two lines is constant in frequency, and proportional to $B_{\rm rot}$. 
This implies that molecules with higher momentum of inertia (i.e. lower $B_{\rm rot}$) have closer energy levels and emit transitions $J \rightarrow J-1$ at lower frequencies with respect to molecules with lower momentum (i.e. higher $B_{\rm rot}$).

Some abundant symmetric, and hence non-polar (i.e. with permanent molecular dipole moment $\mu=0$), molecules such as H$_2$, N$_2$, or CO$_2$ are linear rotors. However, because the Einstein spontaneous coefficient for dipole transitions is proportional to $\mu^2$, these species do not emit dipole transitions, unless they are deformed. 

Examples of polar, linear rotors broadly used in astrochemical studies are: carbon monoxide (CO, Fig.~\ref{fig:molecule-type}), formyl cation (HCO$^+$), hydrogen (iso-)cyanide (HCN and HNC), diazenylium (N$_2$H$^+$, often used as proxy of the non-polar N$_2$), silicon monoxyde (SiO), phosphorus mononitride (PN), etc.

\subsubsection{Symmetric-top rotors}

Symmetric-top molecules are characterised by three-fold (or higher) symmetry axis, which implies that $I_{\rm a} = I_{\rm b} = I_{\perp}$, like for linear rotors, but $I_{\rm c} = I_{\parallel} \neq 0$. In this case the Hamiltonian can be written
as:
\begin{equation}
    H=\frac{J_{\rm a}^2}{2 I_{\perp}}+\frac{J_{\rm b}^2}{2 I_{\perp}} + \frac{J_{\rm c}^2}{2 I_{\parallel}}=\frac{J^2}{2I_{\perp}}+J_{\rm c}^2\left( \frac{1}{2 I_{\parallel}} - \frac{1}{2 I_{\perp}}\right)\;,
\end{equation}
where $J^2 = J^2_{\rm a} +J^2_{\rm b} +J^2_{\rm c}$ 
is the total angular momentum, and $J_{\rm c}$ is the projection of it along the $c$ axis. 
The eigenvalues of the two quantum operators $J^2$ and
$J_{\rm c}$ are $\hbar^2 J(J+1)$ and $\hbar K$ ($-J\leq K \leq +J$), respectively, and thus the energy of the eigenvectors is:
\begin{equation}
E_{\rm J,K}=\frac{\hbar}{2I_{\perp}J(J+1)} + \hbar^2 \left(\frac{1}{2I_{\parallel}} - \frac{1}{2I_{\perp}}\right)K^2 \;.
\end{equation}
In analogy with linear rotors, we define $B_{\rm rot}=h/8\pi^2I_{\perp}$ and $A_{\rm rot}=h/8\pi^2I_{\parallel}$.
In dipole approximation the allowed transitions have
$\Delta J = \pm 1$ and $\Delta K = 0$.
Thus, the energy of the levels are:
\begin{equation}
E_{\rm J \rightarrow J-1}=B_{\rm rot}hJ(J+1)+(A_{\rm rot}-B_{\rm rot})hK^2 \;
\end{equation}
and that of the emitted photons are:
\begin{equation}
E_{J,K \rightarrow J-1, K}=2hB_{\rm rot}J\;.
\end{equation}

Examples of symmetric-top molecules are ammonia (NH$_3$, Fig.~\ref{fig:molecule-type}), methyl cyanide (CH$_3$CN), methyl acetylene (CH$_3$CCH), phosphine (PH$_3$), benzene (C$_6$H$_6$), etc. 

\subsubsection{Asymmetric-top rotors}

Asymmetric-top molecules, such as water (H$_2$O), formaldehyde (H$_2$CO, Fig.~\ref{fig:molecule-type}), methanol (CH$_3$OH), and many others present no threefold symmetry
and the momenta of inertia along the three axis are all different. Their rotational emission
spectra are very complex and a universal formula to describe the energy levels of
all these molecules does not exist. Generalizing from the $(J, K)$ notation for symmetric
tops, the rotational states are labeled with three quantum numbers $J$, $K_{-1}$, and $K_{+1}$.
The dipole selection rules allow $\Delta J = 0;\pm 1$, and $\Delta K = \pm 1$,$\pm 3$ transitions. 

\subsubsection{Spherical rotors}

For these molecules, such as methane (CH$_4$, Fig.~\ref{fig:molecule-type}), the momentum of inertia is the same along the three principal axes: $I_{\rm a} = I_{\rm b} = I_{\rm c} = I$. Thus, the rotational hamiltonian is:
\begin{equation}
    H=\frac{J^2}{2 I}\;,
\end{equation}
and the energy levels are given by:
\begin{equation}
    E_{\rm J}=\frac{\hbar^2}{2I}J(J+1)\;.
\end{equation}
However, these species always have $\mu=0$, and hence they do not emit dipole rotational transitions.

\subsection{Radio astronomy detection techniques}
\label{subsec:techniques}

\begin{figure*}
    \centering
    \includegraphics[width=15cm]{{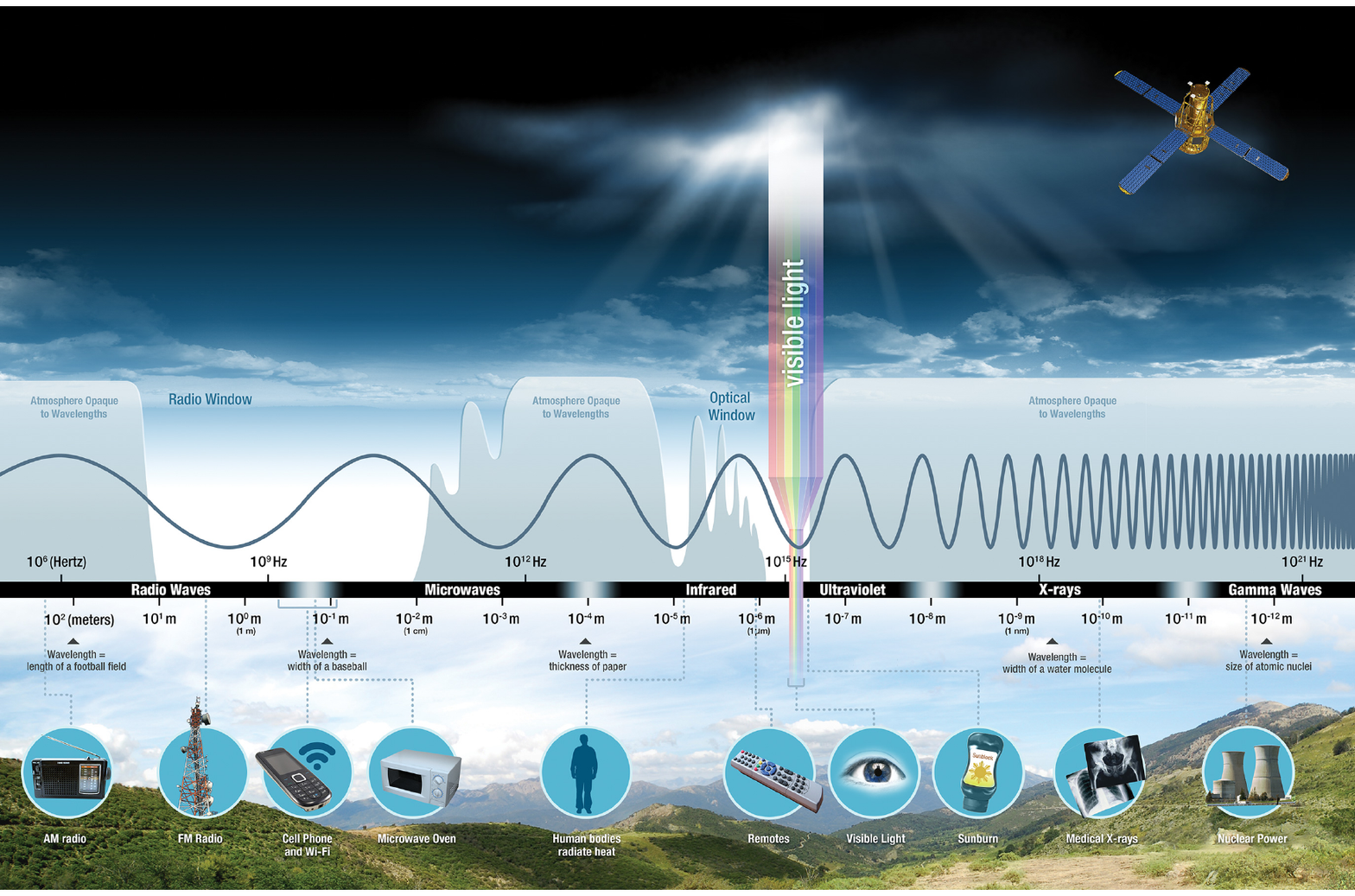}}
    \caption{The transmission of the Earth's atmosphere for electromagnetic radiation with examples of emitters. Credit: NASA.}
    \label{fig:earthatmosphere}
\end{figure*}

For a long time, visible light was the only available way for astronomers to observe
the night sky. With the invention of radio antennas and the discovery of extraterrestrial
radio sources, a new kind of astronomy began. This was allowed by the fact
that Earth's atmosphere is not homogeneous in absorbing and transmitting light; in
particular, together with the optical window, there is another window (see Fig.~\ref{fig:earthatmosphere}), with wavelength in the range $\sim 30$~m -- 0.2~mm (in frequency $\sim 10$ MHz -- 1.5~THz), in which the atmosphere is mostly transparent. This is called the radio window. 
This transparency depends on where you are on Earth and the conditions in the atmosphere such as the quantity of water vapour.

At wavelengths longer than $\sim 30$~m the free electrons in the ionosphere make the
atmosphere opaque to radiation. At wavelengths shorter than $\sim 0.2$~mm the radiation is absorbed by rotational lines of
molecules in the troposphere. The presence of these molecules creates absorption
features also at some frequencies inside the radio window, such as water vapor H$_2$O
transitions at $\sim 22.2$ GHz and $\sim 183$ GHz, or O$_2$ group of lines at $\sim 60$~GHz. In order to
minimize the atmospheric effects, observations should be performed in places with the lowest possible water vapor: this is the reason why observatories, especially working in the (sub-)mm regime, should be at high altitude and with a dry
climate.

In the following, we give a brief overview of the detection techniques used in radio astronomy, bearing in mind that we will not present
a systematic and complete exposition of these techniques, but only the basics to understand how molecular emission can be detected and mapped.

\subsubsection{Single dish telescopes}
\label{subsec:singledish}

A radio telescope is composed of three principal elements: the antenna, the receiver
and the backend.
The aim of the antenna, as for telescopes in the optical range, is to collect the radiation
coming from the target source, and to direct it toward the focus of the optical system,
where the receiver is placed. To properly reflect the light at a certain wavelength $\lambda$, the
inhomogeneities of the reflecting surface must be below $\lambda$/20. Hence, for radiation at millimeter wavelength, the antenna surface must be levigated with micrometer accuracy.
The receiver transforms the incoming radiation into a voltage as a function of time.
After the calibration, which transforms the receiver output in its unit into temperature
units, the signal is sent to the backend, which analyzes the signal in
frequency.

\begin{figure}
    \centering
    \includegraphics[width=0.5\textwidth]{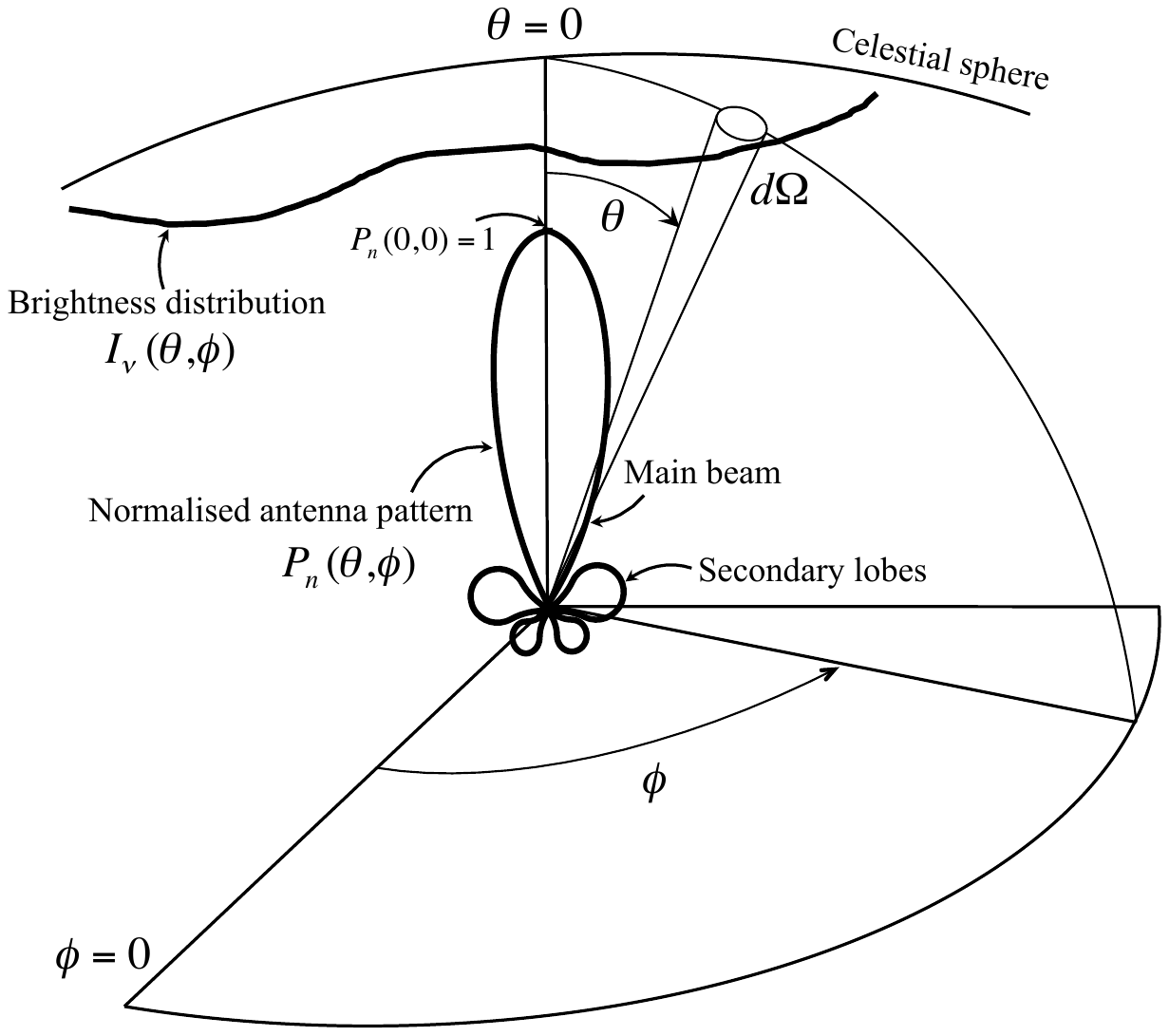}
    \caption{Normalized power pattern for an arbitrary antenna observing a source of specific intensity $I_{\nu}(\theta, \phi)$.}
    \label{fig:antennapattern}
\end{figure}

For any single-dish radio antenna, the power radiated in different directions is described by a
{\em power pattern} $P$, which, according to the reciprocity theorem (see e.g.~Wilson~\citeyear{Wilson2013}), represents also the 
power received by the antenna. Often, this is represented by the {\em normalised antenna pattern} (see Fig.~\ref{fig:antennapattern}):
\begin{equation}
    P_{\rm n}(\theta,\phi)=\frac{P(\theta,\phi)}{P_{\rm max}}\;,
    \label{eq:pattern}
\end{equation}
where $P(\theta,\phi)$ expresses the power received as a function of the angular distance from the optical axis of the telescope, and $P_{\rm max}$ is its maximum.
The normalized antenna pattern presents several lobes, in which the central one, called main beam, collects the majority of the received power.

The integral of the normalized power pattern over the
entire solid angle is called {\em beam solid angle}:
\begin{equation}
    \Omega_{\rm A}=\int_{4\pi}P_{\rm n}(\theta,\phi){\rm d}\Omega\;.
\end{equation}
The integral of the normalized power pattern over the main beam is defined as the {\em main beam solid angle}, or just {\em main beam}:
\begin{equation}
    \Omega_{\rm MB}=\int_{MB}P_{\rm n}(\theta,\phi){\rm d}\Omega\;,
\end{equation}
and the ratio between the main beam solid angle and the beam solid angle is the {\em beam efficiency}:
\begin{equation}
    B_{\rm eff}=\frac{\Omega_{\rm MB}}{\Omega_{\rm A}}\;.
\end{equation}
In radioastronomy, the signal emitted by a source and received by the telescope is often parametrized through a quantity called {\em brightness temperature}. This is related to another parameter, the {\em specific intensity} or {\em brightness}, defined as follows:
when considering an astrophysical source which emits electromagnetic radiation, the {\em specific intensity} or {\em brightness} $I_{\nu}$ is the electromagnetic energy ${\rm d}E$ with frequency in the range [$\nu$; $\nu$+d$\nu$] received by the surface ${\rm d}A$ in the time ${\rm d}t$ in the solid angle ${\rm d}\Omega$:

\begin{equation}
    I_{\nu}=\frac{{\rm d}E}{\cos{\theta} {\rm d}A {\rm d}t {\rm d}\Omega {\rm d}\nu}\,,
\end{equation}
where $\theta$ is the angle between the propagation direction of the radiation and the normal to the receiving surface (see Fig.~\ref{fig:antennapattern}).

It can be shown that if the electromagnetic radiation propagates in empty space, the
specific intensity remains constant along its path. Otherwise, if the radiation passes through
a medium, absorption and emission by this medium must be considered. The variation
of brightness along the line of sight due to emission and absorption by a slab of
material is described by the radiative transfer equation, which will be discussed in detail in Sect.~\ref{sec:RT}.

If the radiation field is in {\em Local Thermodynamic Equilibrium} (LTE) conditions at temperature $T$, the brightness corresponds to the brightness of a black-body:
\begin{equation}
    I_{\nu}=\frac{2h \nu^3}{c^2}\frac{1}{\exp({\frac{h\nu}{kT}})-1}\,.
    \label{eq:blackbody}
\end{equation}
At radio frequencies the Rayleigh-Jeans approximation is valid (e.g. $h\nu/kT \ll 1$), except in the cold phase of the interstellar medium, and Eq.(\ref{eq:blackbody}) can be written as:
\begin{equation}
    I_{\nu}= B_{\nu}(T) = \frac{2k\nu^2}{c^2}T\,.
    \label{eq:brightness}
\end{equation}
Even when the source is not in LTE, one can always define a {\em Brightness temperature}, $T_{\rm B}$, which is the temperature of the equivalent black-body, namely the black-body that would emit the same amount of brightness as the observed one at frequency $\nu$:

\begin{equation}
    I_{\nu}\equiv B_{\nu}(T_{\rm B})=\frac{2k\nu^2}{c^2}T_{\rm B}\,.
    \label{eq:TB}
\end{equation}

If the brightness of an astronomical source is $I_{\nu}(\theta,\phi)$, the brightness temperature will be $T_{\rm B}(\theta,\phi)$, and the output of a single-dish radio telescope will be the convolution of $T_{\rm B}(\theta,\phi)$ with the beam pattern of the telescope.
Usually this quantity is called {\em antenna temperature} $T_{\rm A}^*$, and it is defined as:
\begin{equation}
    T_{\rm A}^*(\theta,\phi)=\frac{\int_{\Omega_{\rm A}} T_{\rm B}(\theta,\phi)P_{\rm n}(\theta - \theta_0,\phi - \phi_0){\rm d}\Omega}{\int_{\Omega_{\rm A}} P_{\rm n}(\theta,\phi){\rm d}\Omega}\;,
    \label{eq:ta}
\end{equation}
where $\theta_0$ and $\phi_0$ are the angular coordinates of the telescope pointing direction.
Similarly, the {\em main beam brightness temperature} is defined as:
\begin{equation}
    T_{\rm MB}(\theta,\phi)=\frac{\int_{\Omega_{\rm MB}} T_{\rm B}(\theta,\phi)P_{\rm n}(\theta - \theta_0,\phi - \phi_0){\rm d}\Omega}{\int_{\Omega_{\rm MB}} P_{\rm n}(\theta,\phi){\rm d}\Omega}\;.
    \label{eq:tmb}
\end{equation}
$T_{\rm MB}$ can be seen as the mean of the brightness temperature over the main beam. Clearly, $T_{\rm MB}$ and
$T_{\rm A}^*$ are related, and indeed one can demonstrate that:
\begin{equation}
T_{\rm MB}=\frac{T_{\rm A}^*}{B_{\rm eff}}.
\end{equation}
The intensity of the observed emission measured by radio telescopes is usually in $T_{\rm MB}$ or $T_{\rm A}^*$ units.

In the case that both $T_{\rm B}(\theta,\phi)$ and $P_{\rm n}(\theta,\phi)$ have a Gaussian profile, Eq.(\ref{eq:tmb}) can be solved to give:
\begin{equation}
    T_{\rm MB}=T_{\rm B}\left(\frac{\theta_{\rm s}^2}{\theta_{\rm s}^2+\theta_{\rm MB}^2}\right)\;,
    \label{eq:dilution}
\end{equation}
where $\theta_{\rm MB}$ is the angular size subtended by the main beam solid angle at half maximum, and $\theta_{\rm s}$ that of the source.
From Eq.~(\ref{eq:dilution}), we can see that if $\theta_{\rm s}$ is much smaller than $\theta_{\rm MB}$, then $T_{\rm MB}\ll T_{\rm B}$.
This means that if the angular size of the emission of a molecular line is (much) more compact than the beam size, the observed line intensity will be (much) fainter than the intrinsic intensity. 
This effect, also called "beam dilution", can make the detection of lines intrinsically faint and arising from compact angular regions challenging with single-dish telescopes. For a detailed exposition, we refer to \citet{Wilson2013}.\\

The telescope sensitivity is a crucial parameter for the detection of molecular species (See Sect. 2). It depends on the antenna size, the accuracy of its reflecting surfaces and the atmosphere transparency. Due to absorption by atmospheric water vapour, sensitivity is degraded at shorter radio wavelengths. That is why telescopes operating below 2 mm (i.e. above 150 GHz) are built on high altitude locations. High frequency telescopes must have more accurate surfaces to keep a high aperture efficiency. Larger antenna that require a high surface accuracy, which are located at high altitude are much more difficult to build and need protection against the harsh weather conditions. This limits the diameter to about 30 m to operate efficiently at 1 mm or below. So, in order to increase the effective area of a telescope operating at wavelengths shorter than 2
mm, one chooses, to build an interferometer instead of increasing the diameter of the single dish.

\subsubsection{Interferometers}
\label{subsec:interferometers}

The need to solve the beam dilution effect, and to improve the angular resolution in images of sources that have a small angular size compared to the beam size of single-dish telescopes, has led to develop (radio-)interferometers. 
These are telescopes composed by several (radio-)antennas of diameter
$D$ separated by a certain distance $b$. 
Because diffraction laws predict that the angular resolution of a filled aperture telescope is given by $\theta_{\rm MB}\sim 1.22 \lambda / D$, where $\lambda$ is the observed wavelength, for large $\lambda$ one would need a telescope with large $D$ to provide an image with limited $\theta_{\rm MB}$. For example, to have $\theta_{\rm MB}=20^{\prime\prime}$ at $\lambda=3$~mm, a telescope with a diameter of 55~m is required, and at $\lambda=3$~cm, a telescope with a diameter of 550~m would be needed.
Therefore, because such huge dishes would be complicated to manage due to mechanical problems, the primary scope of an interferometer is to solve this problem by means of the {\em aperture synthesis technique}. This uses a combination of smaller telescopes to produce an image with angular resolution equivalent to that of a filled telescope with diameter equal to their separation, $b$, which can thus be as large as needed.

If the angular size of a (radio-)source is small enough, the electromagnetic waves arriving on Earth can be considered as parallel planes. In a {\em two-element interferometer}, the incoming signal from two separate telescopes is appropriately {\em cross-correlated} to obtain a resulting signal which depends only on the geometry of the telescope pair with respect to the source structure.

\begin{figure}
    \centering
    \includegraphics[width=0.8\textwidth]{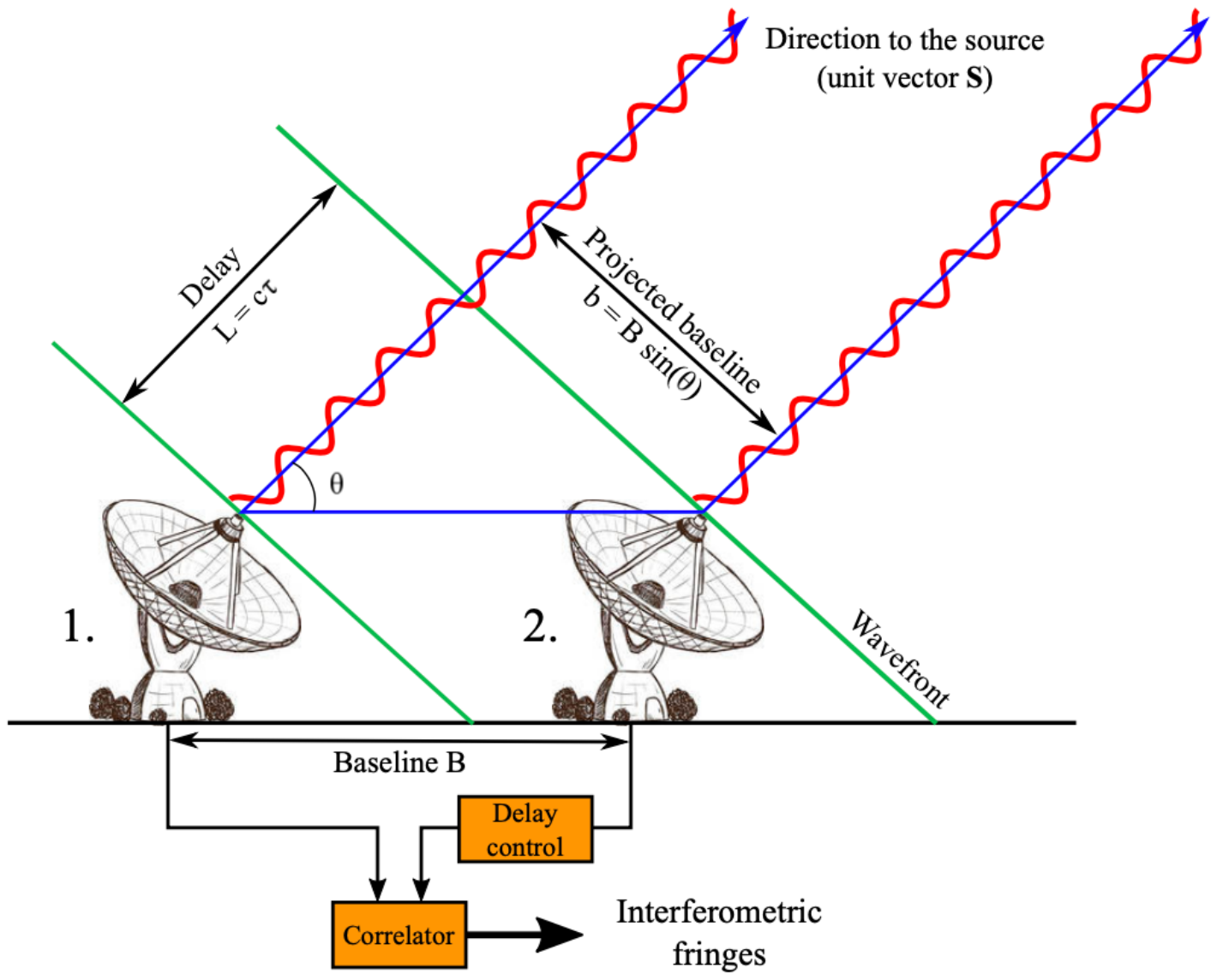}
    \caption{Sketch of a two-element interferometer. Adapted from \citet{Quenard2016}}
    \label{fig:interfero1}
\end{figure}

In summary, a two-element interferometer works like this: the signals on each telescope are collected by a feed at the centre of the focal plane. 
The output of an interferometer is obtained by multiplying and integrating over time the voltages received by the two telescopes. This process is done in the part of the interferometer called {\em correlator} which computes the {\em mutual spatial coherence function}, or {\em cross-correlation function}. 
Let us see first the simple case of a point-like source at field center and two antennas. We will assume that the distance between the two antennas is much smaller than their distance to the source (which is always the case for astronomical radio-sources), so that the wavefront is plane.
If $E_1(t)=E_0\exp{i2\pi \nu(t+\tau)}$ and $E_2(t)=E_0\exp{i2\pi \nu (t)}$ are the electromagnetic fields received by Antenna 1 and Antenna 2 (see Fig.~\ref{fig:interfero1}), where $E_0$ is the electromagnetic field amplitude, $t$ is the time, $\nu$ is the observing frequency, and $\tau$ is the delay between the two signals (that will be defined below), the output of the correlator is:
\begin{equation}
    R(\tau)= E_1 \otimes E_2 = \lim_{T\rightarrow \infty} \frac{1}{2T} \int_{-T}^{T} E_1(t) E^*_2(t) dt=E_0^2\exp{(i2\pi \nu \tau)}.
    \label{eq:Rtau}
\end{equation}
In theory the integration time $2T$ should be infinite. In practice, it is sufficient that $2T \gg \nu^{-1}$, which is usually the case.

\begin{figure}
    \centering
    \includegraphics[width=0.8\textwidth]{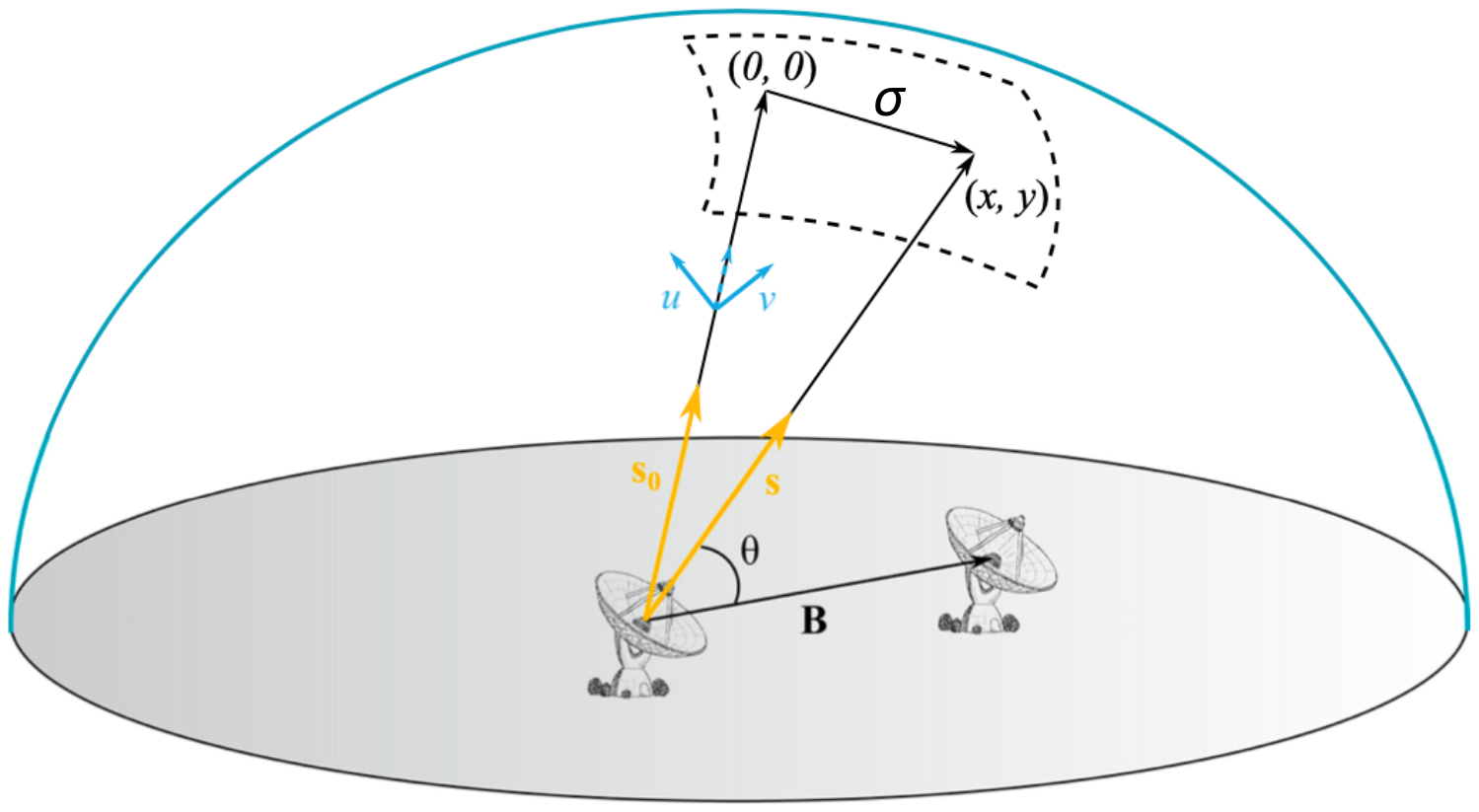}
    \caption{Sketch showing the relation between vectors $\vec{s}$, $\vec{s_0}$, $\vec{\sigma}$, and $\vec{B}$. Adapted from \citet{Quenard2016}.}
    \label{fig:aperturesynthesis}
\end{figure}

The Van Cittert-Zernike theorem (\citet{Vancittert1934}; \citet{Zernike1938}) states that $R(\tau)$ defined in Eq.(\ref{eq:Rtau}) is related to the brightness $I_{\nu}$ (or simply $I$) of the source in the sky. 
More precisely, one single $R(\tau)$, obtained when the interferometer observes a non-variable source over a short time, is one Fourier component of $I$, 
and it is a complex number called {\em complex visibility} or just {\em visibility}, $V$. 
Which Fourier component is measured is determined by the {\em projected baseline}, that is the physical distance between the two telescopes, projected on the plane perpendicular to a reference direction called {\em phase centre} (which is usually the pointing direction of the telescopes).
The coordinates on this plane are often labelled as $(u,v)$ (Fig.~\ref{fig:aperturesynthesis}), and thus the visibility is a function of these coordinates, $V(u,v)$.
Because Earth rotates, the position on the sky of the target changes with time, and so do the projected baselines.
For different projected baselines the interferometer will measure a different Fourier component of $I$. 
Therefore, following the source on the celestial sphere for a long time, an interferometer can measure several different visibilities without moving physically the distance among the telescopes.

The finite number of antennas and baselines implies that an interferometer cannot measure all visibilities but only a sample of the {\em Visibility function}.
Let us define approximately this function, bearing in mind that a rigorous exposition goes beyond the scope of this chapter.

First, let us now consider the general case in which the source is extended, and its brightness in direction $\vec{s}$ is given by $I_{\nu}(\vec{s})$. 
We can assume that the source is made by several point-like sources, each one subtending an infinitesimal solid angle ${\rm d}\Omega$ in direction $\vec{s}$, so that what discussed so far is valid for each of these infinitesimal source elements. 
The power received per bandwidth ${\rm d}\nu$ from the source element ${\rm d}\Omega$ is $A(\vec{s})I_{\nu}(\vec{s}){\rm d}\Omega{\rm d}\nu$, 
where $A(\vec{s})$ is the effective collecting area of each antenna in the direction $\vec{s}$ (assuming the same $A(\vec{s})$ for each telescope). Since the power received is also proportional to $E_0^2$, from Eq.\ref{eq:Rtau} and the definition of $I_\nu$
the output of the correlator for radiation from the direction $\vec{s}$ is hence: 

\begin{equation}
    R(\tau)=A(\vec{s})I_{\nu}(\vec{s})\exp{(i2\pi \nu \tau)}{\rm d}\Omega{\rm d}\nu .
\end{equation}
$\tau$, the delay between the two signals, is the difference between the geometrical and instrumental delays, $\tau_{\rm g}$ and $\tau_{\rm i}$.
If $\vec{B}$ is the baseline vector for the two antennas:
\begin{equation}
    \tau = \tau_{\rm g}-\tau_{\rm i} = \frac{1}{c}\vec{B}\cdot \vec{s}-\tau_{\rm i}\;,
\end{equation}
and the total response is obtained integrating over the source solid angle:
\begin{equation}
    R(\vec{B})=\int \int_{\Omega_{\rm s}} A(\vec{s})I_{\nu}(\vec{s})\exp{[i2\pi \nu (\frac{1}{c}\vec{B}\cdot \vec{s}-\tau_{\rm i})]}{\rm d}\Omega {\rm d}\nu\;.
    \label{eq:rb}
\end{equation}

Interferometers measure the response $R(\vec{B})$ for different $\vec{B}$ to find $I_{\nu}$ from Eq.~(\ref{eq:rb}). To do this, a convenient coordinate
system must be introduced for the two vectorial quantities $\vec{s}$ and $\vec{B}$. 
Fig.~\ref{fig:aperturesynthesis} illustrates the relation among the various planes and vectors.
For $\vec{s}$, one can define $\vec{s_0}$ as the vector identifying the pointing direction, so that the direction of radiation coming from a source element ${\rm d}\Omega$ off the pointing direction is parametrised as:
\begin{equation}
    \vec{s}=\vec{s_0}+\vec{\sigma} \;,
    \label{eq:s}
\end{equation}
where $\vec{\sigma}$ are the coordinates of the tangent plane in the sky centred on the phase centre (Fig.~\ref{fig:aperturesynthesis}). Substituting Eq.~(\ref{eq:s}) in Eq.~\ref{eq:rb}, and transforming the integral over the solid angle in the integral over the source surface S, $R(\vec{B})$ can be written as:
\begin{equation}
    R(\vec{B})=\exp{[i2\pi \nu (\frac{1}{c}\vec{B}\cdot \vec{s_0}-\tau_{\rm i})]} {\rm d}\nu \int \int_{\rm S} A(\vec{\sigma})I_{\nu}(\vec{\sigma})\exp{[i2\pi \nu (\frac{1}{c}\vec{B}\cdot \vec{\sigma})]}{\rm dS}\;.
    \label{eq:rb-2}
\end{equation}
The visibility function is the integral part of Eq.~(\ref{eq:rb-2}):
\begin{equation}
    V(\vec{B})=\int \int_{\rm S} A(\vec{\sigma})I_{\nu}(\vec{\sigma})\exp[i2\pi \nu (\frac{1}{c}\vec{B}\cdot \vec{\sigma})]{\rm dS}\;.
    \label{eq:visibility}
\end{equation}
Considering now the coordinates of $\vec{B}$ in the $(u,v)$ plane and the Cartesian coordinates $(x,y)$ in the tangent plane of the sky with origin in direction $\vec{s_0}$ (Fig.~\ref{fig:aperturesynthesis}), Eq.~(\ref{eq:visibility}) can be written as:
\begin{equation}
    V(u,v)=\int \int_{-\infty}^{+\infty} A(x,y)I_{\nu}(x,y)e^{[i2\pi (ux+vy)]}{\rm d}x{\rm d}y;.
    \label{eq:visibility-2}
\end{equation}
Performing the inverse Fourier transform on Eq.~(\ref{eq:visibility-2}), we obtain:
\begin{equation}
   \widetilde{I}(x,y)=A(x,y)I(x,y)=\int \int_{-\infty}^{+\infty} V(u,v)e^{-i2\pi(ux+vy)}{\rm d}u{\rm d}v \;.
    \label{eq:fourier}
\end{equation}

As stated above, the finite number of antennas and baselines implies that the Fourier transform in Eq.~(\ref{eq:fourier}) will always be a ”discrete” Fourier transform. 
Thus, to improve as much as possible the Fourier transform, one has to measure as many visibilities as possible, and interpolate the values that are missing in the $(u,v)$ plane.
We stress that the relations derived above between $V(u,v)$ and $I(x,y)$ are correct for sources that subtend small angles on the sky, so that we can assume that the source brightness lies on a plane. Because the linear size of radio-sources is typically much smaller  than their distance, this assumption is easily and almost always fulfilled.   

Finally, Eq.(\ref{eq:fourier} states that the brightness measured offset by coordinates $(x,y)$ from the pointing direction is the sky brightness multiplied by the effective collecting area $A(x,y)$. The latter follows the normalized antenna pattern of the individual antenna. Therefore,
the measured sky brightness $\widetilde{I}(x,y)$ is essentially the true sky brightness distribution $I(x,y)$ close to the pointing direction, and goes to zero at large $(x,y)$ following the normalized antenna pattern. 
This is why the main beam of the individual telescope, also called {\em primary beam}, defines the field-of-view of an interferometer. 

For a rigorous and detailed exposition, we refer to \citet{Thompson2017}.\\

\subsubsection{Output units}
\label{subsec:output}

As stated in Sect.~\ref{subsec:singledish}, the output of a single-dish telescope is in temperature units, usually either $T_{\rm MB}$ (Eq.\ref{eq:tmb}), or $T^*_{\rm A}$ (Eq.\ref{eq:ta}). On the other hand, the output of an interferometer is usually in {\em flux density} units, a physical parameter derived from the source specific intensity and defined as the integral of $I_{\nu}$ over the source solid angle $\Omega_{\rm s}$:
    \begin{equation}
     F_{\nu} = \int_{\Omega_{\rm s}}I_{\nu}\cos{\theta}{\rm d}\Omega \sim \int_{\Omega_{\rm s}}I_{\nu}{\rm d}\Omega\;.
     \label{eq:fluxdensity}
    \end{equation}
Other useful parameters derived from the specific intensity are:
\begin{itemize}    
     \item the {\it Bolometric flux}, or simply {\it Flux $F$}, is the integral of $F_{\nu}$ over the whole frequency range:
    \begin{equation}
     F = \int_{0}^{+\infty}F_{\nu}{\rm d}\nu\;;
    \end{equation}
    \item the {\it Monocromatic luminosity $L_{\nu}$} is the total power at frequency $\nu$ radiated by the source per unit frequency, namely the integral of $F_{\nu}$ over the surface
    \begin{equation}
     L_{\nu} = \int_{A}F_{\nu}{\rm d}A\;;
    \end{equation}
    \item the {\it Bolometric luminosity $L_{\nu}$}, or simply {\it Luminosity}, is the total power receiver, namely the integral of $L_{\nu}$ over the whole frequency range:
    \begin{equation}
     L = \int_{0}^{+\infty}L_{\nu}{\rm d}\nu\;.
    \end{equation}
\end{itemize}
    
From Eq.(\ref{eq:TB}) and (\ref{eq:fluxdensity}), it follows that in Rayleigh-Jeans approximation:
\begin{equation}
    F_{\nu}=\frac{2k \nu^2}{c^2}\int_{\Omega_{\rm s}} T_{\rm B}(\Omega){\rm d}\Omega\;.
    \label{eq:fluxRJ}
\end{equation}
From Eq.(\ref{eq:fluxRJ}), the flux density is related to the brightness temperature $T_{\rm B}(\Omega)$, and hence also to the main beam temperature, $T_{\rm MB}$, from its definition (Eq.(\ref{eq:tmb}). 

The relation between $F_{\nu}$ and $T_{\rm MB}$ depends on the brightness temperature distribution in the sky, $T_{\rm B}(\Omega)$, and on the beam shape too. 
In practical cases, for a single-dish antenna the beam is a two-dimensional Gaussian with full width at half maximum $\theta_{\rm MB}$, and that of an interferometer an elliptical two-dimensional Gaussian with major and minor angular sizes $\theta_{\rm maj}$ and $\theta_{\rm min}$, respectively.
For a single-dish, if the source is also Gaussian with full width at half maximum $\theta_{\rm s}$ and point-like, namely $\theta_{\rm s} \ll \theta_{\rm MB}$, one can demonstrate that:
 \begin{equation}
     F^{point}_{\nu}=\left(\frac{2k \nu^2}{c^2}\right)\left(\frac{\pi \theta^2_{\rm MB}}{4\ln{2}}\right)T_{\rm MB}\;,
     \label{eq:JytoK1}
 \end{equation}
and for an interferometer, if $\theta_{\rm s} \ll \theta_{\rm maj},\theta_{\rm min}$:
 \begin{equation}
     F^{point}_{\nu}=\left(\frac{2k \nu^2}{c^2}\right)\left(\frac{\pi \theta_{\rm maj}\theta_{\rm min}}{4\ln{2}}\right)T_{\rm MB}\;.
     \label{eq:JytoK2}
 \end{equation}
Inserting numerical values, Eq.(\ref{eq:JytoK2}) is equivalent to:
\begin{equation}
     F^{point}_{\nu}(Jy)\simeq \left(\frac{\nu^2(GHz) \theta_{\rm maj}(^{\prime\prime})\theta_{\rm min}(^{\prime\prime})}{1222}\right)T_{\rm MB}(K)\;,
     \label{eq:JytoK3}
 \end{equation} 
where 1 Jansky (Jy) corresponds to $10^{-23}$ erg s$^{-1}$cm$^{-2}$Hz$^{-1}$.

\subsubsection{Mapping techniques}
\label{subsec:mapping}

For sources extended with respect to the beam of a radio-telescope, astronomers need to resort to techniques that allow to map the emission in the sky.
For interferometers, as discussed in Sect.~\ref{subsec:interferometers}, the aperture synthesis technique allows to map an angular field which corresponds to the HPBW of the single antenna, usually much wider than the synthesised beam of the interferometer. For example, at an observing frequency of 100 GHz, the 15m antennas of the Northern Extended Millimeter Array (NOEMA) or the 12m antennas of the Atacama Large Millimeter Array (ALMA) allow to map an angular field of $\sim 50^{\prime \prime}$ and $\sim 63^{\prime \prime}$, respectively.
Interferometers are hence ideal instruments to map sources in which the structure of the emission can contain both compact end extended components, but only if the whole source angular size does not exceed the field of view. 
Interferometers have the additional limitation that only structures as extended as the so-called Maximum Recoverable Scale (MRS), associated with the smallest baseline (see Sect.~\ref{subsec:interferometers}), can be retrieved.
In case the most extended structures are larger than the interferometer field-of-view and/or the MRS (e.g. large clouds with angular sizes of several primes or degrees), mosaicing techniques are required.

To map with a single-dish antenna a source more extended than its HPBW one needs scanning techniques. In the simplest technique, called sometimes "step-and-integrate" or "point-and-shoot", the telescope is moved on a grid of discrete positions in the sky. The data are recorded in each position, and then a two-dimensional map of the emission is obtained through interpolation algorithms. 
A more accurate technique is the On-The-Fly imaging: the emission field to map is observed by slewing the telescope in a two-dimensional raster pattern over the source, while data and antenna position information are recorded continuously (e.g. \citet{Mangum2007}). 
This technique has a higher observing efficiency than the step-and-integrate one, because it allows to reduce the telescope "dead times", and the entire field is covered more rapidly thus minimising changes in the atmospheric and instrumental properties. 

\begin{figure}
    \centering
    {\includegraphics[width=0.98\textwidth]{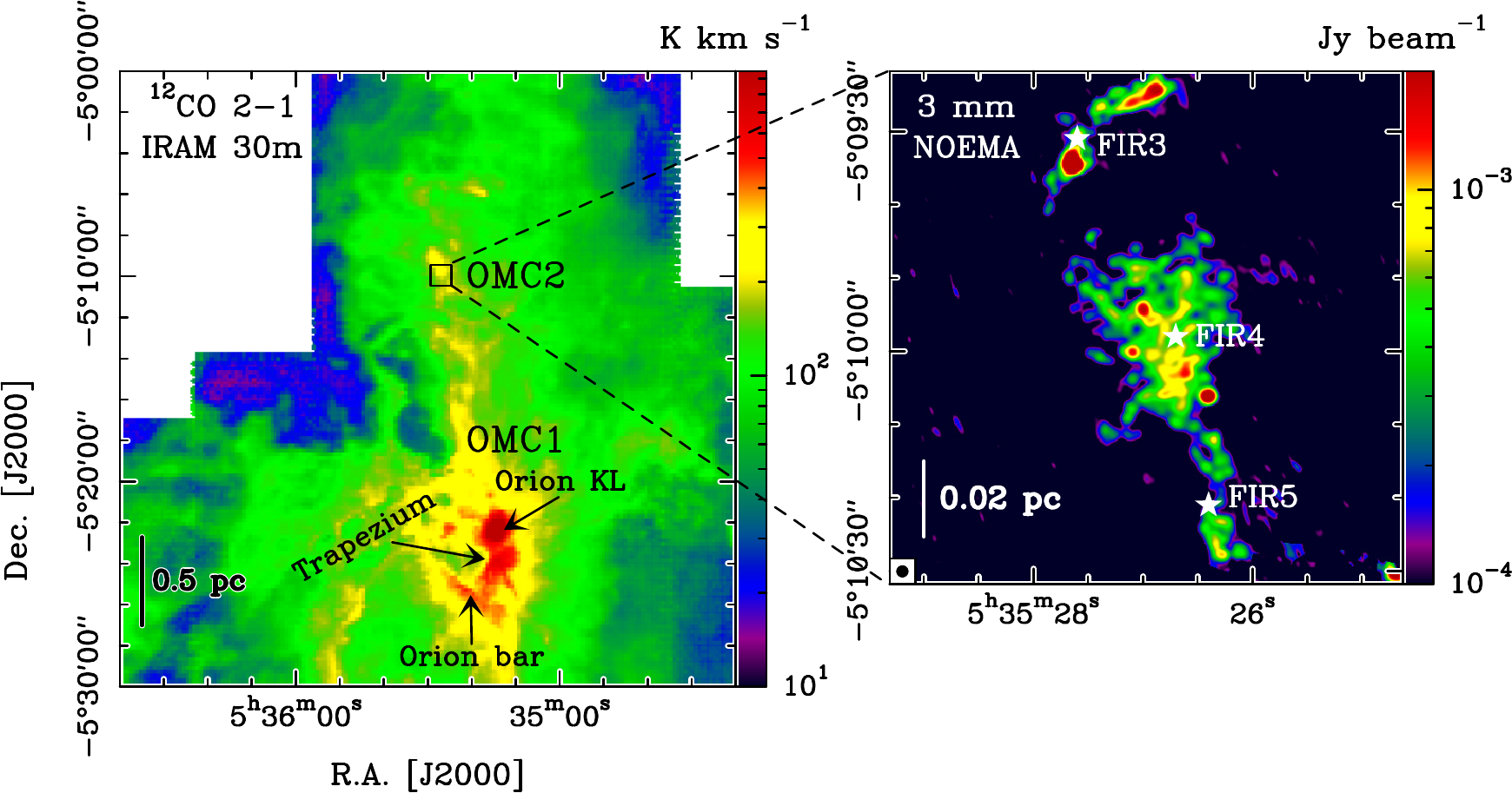}} 
    \caption{{\it Left:} Map of the Orion Molecular Cloud (OMC) complex obtained with the IRAM 30m telescope in CO 2--1. The data are taken from \citet{Berne2014}. We indicate the OMC-1 and OMC-2 clouds, and other distinctive objects within the nebula. The angular resolution of the map is $\sim 10.7^{\prime \prime}$ {\it Right:} zoom of 1$^{\prime}$ square in cloud OMC-2 as mapped with the NOrthern Extended Millimeter Array (NOEMA) in the 3~mm continuum emission (Neri et al., in prep.). The mapped region has a size of 70$^{\prime \prime}$, and an angular resolution of $\sim 1.3^{\prime \prime}$. The stars indicate the far-infrared sources FIR3, FIR4, and FIR5 located in the region, and the ellipse in the bottom-left corner is the synthesised beam.}
    \label{fig:Orion}
\end{figure}

The choice clearly depends on the expected source size and structure: for objects as compact as a few arcseconds, or objects extended up to 10-20$^{\prime\prime}$, for which we are interested in the source structure at (sub-)arcsecond scales, interferometer maps are the obvious best choice. 
For objects with size of several primes or even degrees, single-dish maps are preferred, provided that the angular resolution needed is not higher than $\sim 10-20^{\prime\prime}$. Single-dish and interferometer maps are often complementary, as the single-dish maps illustrate the large-scale morphology, in which interferometers can resolve small-scale structures.
As an example, we show maps of the Orion Molecular Cloud (OMC) complex, an extensively studied star-forming regions, in Fig.~\ref{fig:Orion} 

The choice of the molecular transition to use is also crucial. Usually, astronomers consider several properties, among which the most relevant are (i) the critical density and energy of the upper level of the transition, and (ii) the fractional abundance of the molecule. 
The critical density will be defined in a rigorous way in Sect.~\ref{sec:RT}, but roughly identifies the density of H$_2$ at which the two levels of a transition are populated according to the Boltzmann distribution at the gas kinetic temperature. 
At lower densities, the upper level is not sufficiently populated by collisions, thus reducing to zero the line emission.
Transitions from abundant species, such as CO, characterized by low critical densities ($\sim 10^{2-3}$ cm$^{-3}$, e.g. CO 1--0), are hence appropriate to trace diffuse and extended gas in which the average volume density of H$_2$ is higher than or comparable to their critical densities.
Instead, transitions of less abundant species, such as C$^{18}$O, CS, and N$_2$H$^+$, or characterized by high critical densities ($\geq 10^{4}$ cm$^{-3}$), such as CO 6--5, are appropriate to trace higher density structures.

\section{Census of the detected molecules (galactic and extragalactic)}
\label{sec:census}

Over the past 30 years, we have discovered that we live in a molecular universe, where molecules are abundant and widespread, probing the structure and evolution of galaxies, as well as the temperature and density of the observed medium, opening a new field called astrochemistry. The progress has been dramatic, since the discovery of the first molecules about 100 years ago. We present, in the following, a review of simple molecules, namely molecular species from two to five atoms. The so-called complex organic molecules (COM) have been defined, in space, as molecules with at least 6 atoms, with at least 1 carbon atom \citep{Herbst2009}. The species are indicated with their atomic mass unit appearing after their name in parenthesis and are listed from the low mass to the high mass for each group.

\subsection{Two atoms}

\subsubsection{\texorpdfstring{H$_2$ (2)}{}}

Molecular hydrogen is the most abundant molecule in the universe. It has been detected three decades after the first interstellar detection of CH, CH$^+$ and CN (see below). H$_2$ is symmetric homonuclear molecule and has no permanent dipole moment. Because of this it does not have rotation vibration spectrum. Electronic transition produces spectral lines in the optical, IR and UV. The transition between the vibrational states produces lines in the IR. Transitions between the rotational states of molecules produce least energetic lines, in the microwave and radio region. The first rotational line comes from the J=2 level, through quadrupole radiation, 512 K above from the ground state at 28 $\mu$m. The cold H$_2$ medium cannot therfore be traced. The first detection was made possible through Lyman absorption bands and was reported by \citet{Carruthers1970} in the FUV spectrum of the star $\xi$ Per using an Aerobee-150 rocket. In extragalactic sources, the first detection is reported by \citet{Thompson1978} towards NGC 1068 at $\sim 2.2\,\mu$m.

\subsubsection{\texorpdfstring{HeH$^+$ (5)}{}}

The helium hydride cation has been detected at 2.0102 THz (J=1--0) using the Stratospheric Observatory for Infrared Astronomy (SOFIA)  in emission towards NGC 7027 \citep{Gusten2019}.

\subsubsection{\texorpdfstring{CH (13)}{}}

CH (methylidyne) is the first detected species in the ISM. It was first detected at $\lambda=4300$ \AA\ by \citet{Dunham1937} with the Mount Wilson Observatory in diffuse gas, and confirmed in several additional transitions by \citet{McKellar1940}. It was suggested by \citet{Swings1937} as methylidyne. The first detection in the radio was obtained by \citet{Rydbeck1973} with the Onsala telescope in more than a dozen Galactic clouds, and the rotational spectrum was first directly measured by \citet{Brazier1983}. CH has also been detected in external galaxies, first by \citet{Whiteoak1980} towards three galaxies including the Large Magellanic Cloud.

\subsubsection{\texorpdfstring{CH$^+$ (13)}{}}

Methylidyne cation is one of the first molecules to be identified in space. During a conference at the Yerkes Observatory in 1941 , P. Swings called attention to three sharp interstellar lines in absorption at $\lambda = 4232.57, 3957.71$, and $3745.30$ \AA\ (detected with the Mount Wilson Observatory) and suggested that they belong to light-ionized molecules such as CH$^+$, CN$^+$, C$_2^+$, NH$^+$ or NO$^+$. E. Teller and G. Herzberg then suggested that CH$^+$ is the most likely species. Later on, \citet{Douglas1941} attributed these bands to CH$^+$. In the radio, \citet{Cernicharo1997} reported the first detection of a rotational transition with the infrared space observatory (ISO). \citet{Magain1987} reported the first extragalactic detection, at 4232 \AA\ towards supernova 1987A, inside the Large Magellanic Cloud, using the European Southern Observatory (ESO) 1.4m telescope.

\subsubsection{\texorpdfstring{NH (15)}{}}

The imidogen radical was first detected in the ISM by \citet{Meyer1991} in absorption towards $\zeta$ Per and HD 27778, with the the Kitt Peak National Observatory (KPNO) 4m telescope. In an external galaxy, NH was first detected by \citet{Gonzalez-Alfonso2004} towards Arp220 using the ISO satellite.

\subsubsection{\texorpdfstring{OH (17)}{}}

The first detection of OH (hydroxyl radical) in absorption was reported by \citet{Weinreb1963} towards the supernova remnant Cas A at $\sim 18$~cm. Since then, several lines of OH were seen in emission, including maser emission (\citet{Weaver1965}, \citet{Elitzur1992}), in Galactic star-forming regions. \citet{Weliachew1971} reported the first detection in two external galaxies, NGC 253 and M 82 with the Owens Valley Radio Observatory (OVRO).

\subsubsection{\texorpdfstring{OH$^+$ (17)}{}}

The N=1--0, J=0--1, fine structure component of oxidaniumylidene (or hydroxylium), near 909 GHz was first detected in absorption with the APEX 12m towards SgrB2 \citep{Wyrowski2010}. It was later detected in the external galaxy Mrk 231 using the Herschel/SPIRE instrument with three rotational transitions \citep{vanderWerf2010}. 

\subsubsection{\texorpdfstring{HF (20)}{}}

HF (hydrogen fluoride) was first detected by \citet{Neufeld1997} with the ISO satellite through the J=2--1 transition towards SgrB2. In external galaxies, it was detected through Herschel/SPIRE observations towards Mrk 231 \citep{vanderWerf2010}, Arp 2220 \citep{Rangwala2011}, and in the Cloverleaf quasar at the Caltech Submillimeter Observatory (CSO) \citep{Monje2011}.

\subsubsection{\texorpdfstring{C$_2$ (24)}{}}

\citet{Souza1977} detected for the first time dicarbon towards Cygnus OB2 with the Smithsonian Institution's Mount Hopkins Observatory. In an external galaxy (the Small Magellanic Cloud), the first detection was reported by \citet{Welty2013} with the ESO Very Large Telescope (VLT).

\subsubsection{\texorpdfstring{CN (26)}{}}

CN (cyano radical) is the second molecular species detected in the ISM. It was first detected in the optical around $\lambda = 3875$ \AA\ by \citet{McKellar1940} and \citet{Adams1941} using the Mount Wilson observatory. In the radio/millimeter, it was detected by \citet{Jefferts1970} with the National Radio Astronomy Observatory (NRAO) 11m (12m after 1981) telescope towards the Orion nebula and W51. In extragalactic sources, CN was detected first by \citet{Henkel1988} towards NGC 253, IC 342, and M 82 in the N=2--1 and 1--0 lines with the IRAM-30m telescope.

\subsubsection{\texorpdfstring{CN$^-$ (26)}{}}

The cyanide ion has been detected with the IRAM 30m towards the famous late-type star IRC+10216 in 3 rotational transitions from J=1--0 to 3--2 between 110 and 340 GHz \citep{Agundez2010}. 

\subsubsection{\texorpdfstring{CO (28)}{}}

Carbon monoxide was first detected by \citet{Wilson1970} in the J=1--0 transition at 115 GHz with the NRAO 11m telescope towards the Orion nebula. The line is very bright was used to perform the first maps in the ISM \citep{Penzias1971a}. The first extragalactic detection was reported by \citet{Rickard1975} towards M 82 and NGC 253.

\subsubsection{\texorpdfstring{CO$^+$ (28)}{}}

Carbon monoxide cation was first detected by \citet{Latter1993} in the ISM (M17SW) and a planetary nebula (NGC 7027) at $\sim 236$~GHz based on observations of three millimeter and sub-millimeter transitions performed with the NRAO 12m telescope. \citet{Fuente2006} reported its detection in the nucleus of the external galaxy M82.

\subsubsection{\texorpdfstring{N$_2$ (28)}{}}

The nitrogen molecule is the most abundant molecule in the Earth's atmosphere. It was first detected with the FUSE telescope in the diffuse medium towards  HD 124314 via transitions at 958.6 \AA\ and 960.3 \AA\ \citep{Knauth2004}.

\subsubsection{\texorpdfstring{NO (30)}{}}

The first detection of nitric oxide was reported by \citet{Liszt1978} at 150.2 and 150.5 GHz towards SgrB2 using the NRAO 11m telescope. The first extragalactic detection was performed using the IRAM 30m towards NGC 253 \citep{Martin2003} at 150.2, 250.4 and 250.8 GHz.

\subsubsection{\texorpdfstring{CF$^+$ (31)}{}}

Fluoromethylidynium was first detected towards the Orion Bar PDR by \citet{Neufeld2006} in its J=1--0 (102.6 GHz) and 2--1 (205.2 GHz) transitions at the IRAM 30m as well as the 3--2 (307.7 GHz) transition with the APEX 12m. The 2--1 transition was also observed in absorption in the z = 0.89 foreground galaxy towards the quasar PKS 1830-211 \citep{Muller2016}. 

\subsubsection{\texorpdfstring{O$_2$ (32)}{}}

Molecular oxygen was first tentatively detected  with the SWAS satellite through its N$_J$=3$_3$--1$_2$ line of molecular oxygen at 487249.270 MHz with the SWAS satellite \citep{Goldsmith2002}. However, the weak 1$_1$--1$_0$ line at 118.750 GHz was tentatively reported towards the same source by \citet{Larsson2007} with an even lower column density challenging the previous tentative detection. Later on, Herschel/HIFI was used to study lines at 487.2 GHz and at 773.8 GHz \citep{Liseau2012} securing the former identification by SWAS. Three lines, 3$_3$--1$_2$, 5$_4$--3$_4$, and 7$_6$--5$_6$, at 487.2, 773.8 and 112.1 GHz were securely confirmed by \citet{Goldsmith2011} towards Orion.

\subsubsection{\texorpdfstring{SH (33)}{}}

The first detection of the Sulfanyl radical has been reported in absorption in the diffuse medium along the sight-line to the submillimeter continuum source W49N by \citep{Neufeld2012}, using the SOFIA/GREAT instrument near 1382.91 and 1383.24 GHz (the J=5/2--3/2 transition of the lower energy fine structure component 2$\Pi_{3/2}$). 

\subsubsection{\texorpdfstring{SH$^+$ (33)}{}}

Sulfaniumylidene (or sulfanylium) was first detected in its N=1--0, J=1--1, fine structure component near 683 GHz has been detected in absorption with  APEX 12m towards SgrB2 \citep{Menten2011}. The N$_J$=1$_2$--0$_1$ fine structure transitions of SH$^+$ were observed in detection with the Atacama Large Millimeter Array (ALMA) interferometer, redshifted near 526 GHz towards PKS 1830-211 \citep{Muller2017}.

\subsubsection{\texorpdfstring{HCl (36)}{}}

Hydrogen chloride was detected first by \citet{Blake1985} in Orion with the Kuiper Airborne Observatory (KAO) in its J=1--0 transition at 625.9 GHz. 

\subsubsection{\texorpdfstring{HCl$^+$ (36)}{}}

Detection of Chloroniumyl was made possible using the Herschel/HIFI high spectral resolution instrument, with the J= 5/2--3/2 transition of the lower energy fine structure component 2$\Pi_{3/2}$ in absorption towards the HII regions W31C and W49N \citep{DeLuca2012}.

\subsubsection{\texorpdfstring{ArH$^+$ (37)}{}}

The J=1--0 (617.5 GHz) and 2--1 (1234.6 GHz) transitions of argonium were detected in emission with the Herschel/SPIRE instrument towards the Crab Nebula supernova remnant (M1) \citep{Barlow2013}. It was later detected in absorption in the z = 0.89 foreground galaxy towards the quasar PKS 1830-211 using ALMA, with redshifted transitions from 617 GHz to 325.2 GHz \citep{Muller2015}.

\subsubsection{\texorpdfstring{SiC (40)}{}}

Silicon carbide has only been detected towards IRC+10216 near 80 GHz (J=2--1), 160 GHz (J=4--3), and 236 GHz (J=6--5) using the IRAM 30m telescope \citep{Cernicharo1989}.

\subsubsection{\texorpdfstring{SiN (42)}{}}

Using the NRAO 12m telescope, the N=2--1 line at $\sim$ 87 GHz and N=6--5 line at 262 GHz of silicon nitride were detected towards IRC+10216 \citep{Turner1992a}. 

\subsubsection{\texorpdfstring{CP (43)}{}}

Carbon phosphide (phosphaethynyl radical) was detected for the first time towards IRC+10216 with the IRAM 30m telescope by \citet{Guelin1990}.

\subsubsection{\texorpdfstring{AlO (43)}{}}

The aluminum monoxide radical has first been detected towards the oxygen-rich supergiant star VY Canis Majoris (VY CMa) \citep{Tenenbaum2009} using the Arizona Radio Observatory (ARO). The N=7--6 and 6--5 rotational transitions at 267.937 and 229.670 GHz were observed using the ARO Submillimeter Telescope (SMT) and the N=4--3 transition at 153.124 GHz was detected using the ARO 12m telescope.

\subsubsection{\texorpdfstring{CS (44)}{}}

Carbon monosulphide, the first sulphur-bearing molecule found in the ISM, was detected first by \citet{Penzias1971b} (J=3--2) with the NRAO 11m telescope towards the Orion nebula, W51, IRC+10216, and DR2. \citet{Henkel1985} reported the first extragalactic detection in M 82 and IC 342 of the J=2--1 rotational line with the 7m telescope at AT\&T Bell Laboratories.

\subsubsection{\texorpdfstring{SiO (44)}{}}

The first detection of silicon monoxide was reported by \citet{Wilson1971} towards SgrB2 with the NRAO 11m telescope at $\sim 130.246$~GHz (J=3--2), and in an external galaxy (NGC 253) by \citet{Mauersberger1991} (J=2--1 and J=3--2). This species is the first silicon-bearing molecule detected in the ISM.

\subsubsection{\texorpdfstring{PN (45)}{}}

A feature at 234.936 GHz in the line survey performed by \citet{Sutton1985} towards Orion with OVRO was suggested to be the J=5--4 rotational line of phosphorus nitride. Two years later, \citet{Turner1987} and \citet{Ziurys1987} simultaneously confirmed this detection and detected multiple lines of PN towards several star-forming regions with the NRAO 12m telescope and the 14m Five College Radio Astronomical Observatory (FCRAO), respectively. PN is the first phosphorus-bearing molecules detected in the ISM. The 2--1 and 3--2 transitions near 93.980 and 140.968 GHz, respectively, were observed with ALMA towards the central molecular zone (CMZ) of the nearby starburst galaxy NGC 253 \citep{Haasler2022}.

\subsubsection{\texorpdfstring{NS (46)}{}}

The first detection of nitrogen sulfide was reported simultaneously and independently by \citet{Gottlieb1975} and \citet{Kuiper1975}, both towards SgrB2 with the 16 ft antenna at the University of Texas Millimeter Wave Observatory, and the NRAO 11m telescope, respectively. The J=5/2--3/2 $^2\Pi_{1/2}$ transitions at 115.16 GHz (c-state) and 115.6 GHz (d-state) were reported. \citet{Martin2003} detected it for the first time in an external galaxy, the nucleus of the starburst galaxy NGC 253. 

\subsubsection{\texorpdfstring{NS$^+$ (46)}{}}

The thionitrosylium molecule has first been detected through its J=2--1 (100.2 GHz), 3--2 (150.3 GHz), and 5--4 (250.5 GHz) transitions at the IRAM 30m towards the B1b dark cloud \citep{Cernicharo2018}.

\subsubsection{\texorpdfstring{AlF (46)}{}}

The first tentative detection of Aluminium fluoride was reported by \citet{Cernicharo1987b} towards IRC+10216 with the IRAM 30m telescope, and confirmed by \citet{Ziurys1994b} with CSO observations.

\subsubsection{\texorpdfstring{PO (47)}{}}

The phosphorus monoxide radical has been detected with the ARO 10m SMT in four lines of two rotational transitions (J= 5.5--4.5 and 6.5--5.5) near 240 and 284 GHz \citep{Tenenbaum2007}. It is the first new species to be identified in an oxygen-rich, as opposed to a carbon-rich, circumstellar envelope.

\subsubsection{\texorpdfstring{PO$^+$ (47)}{}}

Phosphorus monoxide ion was detected by \citet{Rivilla2022} in its J=1--0 and 2--1 transitions near 47.0 and 94.0 GHz, using the IRAM 30m and Yebes 40m telescopes towards the Galactic center cold molecular cloud G+0.693-0.027.

\subsubsection{\texorpdfstring{SO (48)}{}}

\citet{Gottlieb1973} detected for the first time sulphur monoxide in two rotational lines (J=3$_2$--2$_1$ and J=4$_3$--3$_2$) towards seven galactic sources with the NRAO 11m telescope. \citet{Johansson1991} reported the first detection in the Magellanic Clouds of the J=3$_2$--2$_1$ transition with the SEST telescope. 

\subsubsection{\texorpdfstring{SO$^+$ (48)}{}}

The sulfur monoxide cation was first identified by \citet{Turner1992b} towards IC 443G with the NRAO 12m telescope. \citet{Muller2011} reported the first detection in the external galaxy PKS 1830-211. 

\subsubsection{\texorpdfstring{NaS (55)}{}}

Sodium sulfide was detected by \citet{ReyMontejo2024} towards the Galactic Center molecular cloud G+0.693--0.027 employing the Yebes 40m and IRAM 30m radio telescopes with 5 of the 9 transitions 4.5 $\le$ J $\le$ 11.5 unblended.

\subsubsection{\texorpdfstring{MgS (56)}{}}

Magnesium sulfide was detected by \citet{ReyMontejo2024} towards the Galactic Center molecular cloud G+0.693--0.027 employing the Yebes 40m and IRAM 30m radio telescopes with 12 unblended transitions with 9 $\le$ J $\le$ 15 and one J = 2--1 blended transition.

\subsubsection{\texorpdfstring{NaCl (58)}{}}

Sodium chloride was detected for the first time in IRC+10216 by \citet{Cernicharo1987b} in several rotational lines with the IRAM 30m telescope. \citet{Ginsburg2019} reported the first detection obtained with ALMA in the ISM not associated with the ejecta of evolved stars (towards the Orioni SrcI star-forming region).

\subsubsection{\texorpdfstring{SiP (59)}{}}

SiP was detected in the circumstellar shell of IRC+10216, using the ARO 12m at 2 mm through the J = 13.5 -- 12.5 at 215.05 GHz and 16.5 -- 15.5 at 262.81 GHz transitions, clean of contamination \citep{Koelemay2022}.

\subsubsection{\texorpdfstring{SiS (60)}{}}

Silicon monosulfide is, so far, the simplest and unique molecule containing silicon and sulphur. It was detected for the first time by \citet{Morris1975} using the 11m NRAO telescope towards the carbon star IRC+10216 through the J = 6--5 and 5--4 transitions near 108.9 and 90.8 GHz, respectively.

\subsubsection{\texorpdfstring{AlCl (62) \& KCl (74)}{}}

Similarly to NaCl, \citet{Cernicharo1987b} detected millimeter lines of both AlCl (aluminium chloride) and KCl (potassium chloride) towards IRC+10216.

\subsubsection{\texorpdfstring{TiO (64)}{}}

\citet{Hyland1969} first suggested an association with TiO band heads in some of their unidentified IR emission lines towards the  O-rich late-type star, VY Canis Majoris. Additional lines were assigned to TiO by \citet{Wallerstein1971}. 

\subsubsection{\texorpdfstring{FeC (68)}{}}

\citet{Koelemay2023} detected iron carbide using the ARO towards IRC+10216 on the basis of three successive rotational transitions measured in the 2 and 1.3 mm bands.

\subsection{Three atoms}

\subsubsection{\texorpdfstring{H$_3^+$ (3)}{}}

The presence of H$_3^+$ in the ISM was first suggested by \citet{Martin1961}, and its IR spectrum was measured in the laboratory by \citet{Oka1980}. Attempts to detect this molecule has proven unsuccessful for a long time before the detection, 35 years later, by \citet{Geballe1996} towards the massive star-forming regions GL2136 and W33A. They used the UKIRT telescope and detected the IR transitions of the $\nu_2$ fundamental band around 3.7 $\mu$m. H$_3^+$ has been detected for the first time by \citet{Geballe2006} in an extragalactic object, the highly obscured ultraluminous galaxy IRAS 08572+3915 NW.

\subsubsection{\texorpdfstring{CH$_2$ (14)}{}}

Methylene has first been detected in the Orion nebula by \citet{Hollis1989}, although with a low s/n. It was later confirmed by \citet{Hollis1995} in the hot core of the Orion-KL nebula and the molecular cloud in proximity to the continuum source W51 M with the NRAO 12m telescope. 

\subsubsection{\texorpdfstring{NH$_2$ (16)}{}}

Amidogen was first detected in absorption in the SgrB2 line of sight by \citet{vanDishoeck1993} using the CSO. Three transitions have been detected through the J=3/2--3/2, 1/2--1/2 and 3/2--1/2 at 462.4, 469.4 and 461.4 GHz respectively. It has also been detected at 462.4 and 461.4 GHz with ALMA within cycle 0 observations in the unnamed foreground galaxy at z=0.89 toward the blazar PKS 1830-211 by \citet{Muller2014}.

\subsubsection{\texorpdfstring{H$_2$O (18)}{}}

The observation of water is often hampered because of its presence in high abundance in Earth's atmosphere. It was detected for the first time in the ISM by \citet{Cheung1969} towards SgrB2, the Orion Nebula and the W49 HII region. The emission is associated with the maser emission of the 6$_{1,6}$--5$_{2,3}$ rotational transition at 22 GHz using the Hat Creek Observatory. This detection was unexpected due to the presence of water in the atmosphere and non-maser emission was only made possible by the Herschel telescope with its HIFI and PACS instruments and the Spitzer Space Telescope. The 22 GHz water vapor maser has also been detected for the first time in an external galaxy (M33) by \citet{Churchwell1977} towards M33.

\subsubsection{\texorpdfstring{H$_2$O$^+$ (18)}{}}

Oxidaniumyl has been detected for the first time using the Herschel/HIFI instrument at 1115 GHz along the line of sight towards the star-forming regions DR21, Sagittarius B2(M), NGC 6334 and G10.6-0.4 \citep{Ossenkopf2010,Gerin2010}. The same transition has been detected towards an extragalactic source, M82 \citep{Weiss2010}. 

\subsubsection{\texorpdfstring{CCH (25)}{}}

The ethynyl radical has been detected for the first time by \citet{Tucker1974} near 87.3 GHz (N=1--0) in the Orion nebula and DR21 using the NRAO 11m. The same transition has latter been detected towards M82 using the IRAM-30m.

\subsubsection{\texorpdfstring{HCN (27)}{}}

Hydrogen Cyanide is abundant in all kinds of environments, from dark clouds to star-forming regions and circumstellar envelopes. The first detection was reported by \citet{Snyder1971} at 88.6 and 86.3 GHz using NRAO 11m towards a sample of star forming regions: W3 (OH), SgrA, W49, W51, and DR 21 (OH). Its high abundance made it one of the first to be detected in extragalactic sources such as NGC 253 and M82 \citet{Rickard1977}.

\subsubsection{\texorpdfstring{HNC (27)}{}}

Hydrogen isocyanide was among the early molecules detected in space and it was one of the molecules detected before laboratory spectroscopic information was available for its identification. It has been detected in 1972 using the NRAO 11m in emission  at 90.665 GHz (J=1--0) towards SgrB2, W51, DR21(OH) and NGC 2264 \citep{Snyder1972,Zuckerman1972}. The same transition was later detected in an extragalactic source towards IC 342 \citet{Henkel1988} using the IRAM-30m.

\subsubsection{\texorpdfstring{HCO$^+$ (29)}{}}

The unidentified interstellar line discovered by \citet{Buhl1970} with estimated rest frequency of 89.190 GHz was suggested by \citet{Klemperer1970} to the first rotational transition of HCO$^+$. The attribution was was later confirmed by \citet{Woods1975}. The same transition was later detected towards an extragalactic source, M82 by \citet{Stark1979}. 

\subsubsection{\texorpdfstring{HOC$^+$ (29)}{}}

Hydroxymethyliumylidene was first detected in 1983 with the J=1--0 transition at 89.487 GHz towards SgrB2 \citep{Woods1983} using the FCRAO 14m telescope. It was surprisingly later detected (8.5$\sigma$ on its integrated intensity) in the circumnuclear disk of the active galactic nuclei NGC 1068 \citep{Usero2004}.

\subsubsection{\texorpdfstring{N$_2$H$^+$ (29)}{}}

In 1974 an unidentified new interstellar triplet of microwave lines has been discovered at 93.174 GHz with the NRAO 11m towards a sample of star-forming regions \citep{Turner1974}. In the same journal, \citet{Green1974} suggest protonated nitrogen as an identification. The molecule is widespread in the Galaxy and has been detected and even mapped in nearby galaxies (NGC 253, Maffei 2, IC 342, M 82, and NGC 6946) by \citet{Mauersberger1991}.

\subsubsection{\texorpdfstring{HCO (29)}{}}

Formyl radical has been detected in the direction of W3, NGC 2024, W51, and K3-50 (although with a weak intensity) through its N$_{K^-,K^+}$=1$_{0,1}$--0$_{0,0}$, J=3/2--1/2, F=2--1 transition at 86670.65 GHz using the NRAO 11m telescope, based on the spectroscopic parameters  reported by \citet{Saito1972}. It has been mapped for the first time at high angular resolution ($\sim$1$^{\prime\prime}$, $\sim$ 140 au), towards the Solar-type protostellar binary IRAS 16293-2422 using ALMA \citep{Rivilla2019}. \citep{Sage1995} pointed a possible detection towards the 2 nearby galaxies NGC253 and M82 although highly affected by blending with SiO (2--1) and H$^{13}$CO$^+$ (1--0). A more definitive detection has been made by \citet{Garcia2002} using high-resolution ($\sim$5$^{\prime\prime}$) image at the IRAM-PdB interferometer of the nucleus of M82 showing the presence of widespread emission of HCO. 

\subsubsection{\texorpdfstring{HNO (31)}{}}

The nitroxyl radical was first detected by \citet{Ulich1977} using the NRAO 11m towards the galactic sources SgrB2 and NGC 2024 with the  J$_{K^{-},K^{+}}$=1$_{0,1}$--0$_{0,0}$ transition at 81.477 GHz.

\subsubsection{\texorpdfstring{HO$_2$ (33)}{}}

The hydroperoxyl radical has been detected by \citet{Parise2012} using the IRAM 30m and the APEX telescopes towards the SM1 core of the $\rho$ Oph A cloud. The J$_{K_a,K_c}$=2$_{0,2}$--1$_{0,1}$ and 4$_{0,4}$--3$_{0,3}$ transitions around 130.35 and 260.67 GHz, respectively, were detected, displaying fine structure splitting of about 200 MHz. In addition, H-hyperfine splitting was resolved for the lower quantum number transition. 

\subsubsection{\texorpdfstring{H$_2$S (34)}{}}

Hydrogen sulfide was first detected by \citet{Thaddeus1972} using the NRAO 11m through the 1$_{1,0}$--1$_{0,1}$ rotational transition at 168.7 GHz in 7 galactic sources with high abundances. The same transition was later detected outside the galaxy by \citet{Heikkila1999} with the 15m Swedish-ESO Submillimetre Telescope (SEST) in the Large Magellanic Cloud.

\subsubsection{\texorpdfstring{C$_3$ (36)}{}}

Propadienediylidene was first detected by \citet{Hinkle1988} observed towards the famous carbon star IRC+10216 using the KPNO 4m telescope through the vibration-rotation lines 2040 cm$^{-1}$. \citet{Welty2013} later detected, for the first time outside the Galaxy, the J = 0--12 transition, in absorption, towards the Small Magellanic Cloud using the ESO/VLT telescope. 

\subsubsection{\texorpdfstring{C$_2$O (36)}{}}

Dicarbon monoxide has been detected first toward the TMC-1 dark cloud by \citet{Ohishi1991}. The N$_J$=1$_2$--0$_1$ and 2$_3$--1$_2$ transitions were detected with the 43m telescope at Green Bank and the 45m Nobeyama telecope, respectively, at 22.258 and 45.827 GHz.

\subsubsection{\texorpdfstring{H$_2$Cl$^+$ (37)}{}}

Chloronium has been detected in absorption with Herschel/HIFI towards the star-forming regions NGC 6334I and Sagittarius B2(S) by \citet{Lis2010} through the ortho 2$_{1,2}$--1$_{0,1}$ transition at 781.6 GHz and the para 1$_{1,1}$--0$_{0,0}$ transition at 485.4 GHz.  

\subsubsection{\texorpdfstring{CCN (38)}{}}

The ARO 12m and the ARO SMT were used to detect cyanomethylidyne with its two $\Lambda$ doubling components and each of three rotational transitions in the 2$\Pi_{1/2}$ lower energy spin ladder: J=9.5-–8.5, 6.5--5.5, and 4.5--3.5 transitions at 224.45, 153.63, and 106.36 GHz in the circumstellar envelope of CW Leonis \citep{Anderson2014}.

\subsubsection{\texorpdfstring{NCO (42)}{}}

The first detection of the isocyanate radical was made at the IRAM 30m towards the dense core L483 \citep{Marcelino2018} with the  $^{2}\Pi_{3/2}$ state (the strongest components of the $J$=7/2--5/2 and $J$=9/2--7/2 at 81.4 and 104.7 GHz respectively).

\subsubsection{\texorpdfstring{CO$_2$ (44)}{}}

Unfortunately, carbon dioxyde cannot be traced in the (sub)millimeter regime because it lacks a permanent dipole moment. Therefore it can only be sought toward sources with a bright IR continuum. CO$_2$ was first detected by \citet{DHendecourt1989} in the solid state, before being detected in the gas phase, in the direction of compact HII regions and star-forming regions. It was detected with the IRAS/LRS instrument in the $\nu_2$ bending mode at 15.2 $\mu$m with an abundance roughly equal to that of solid CO. Later, \citet{vanDishoeck1996} searched for gas-phase CO$_2$ features in the ISO/SWS IR spectra of four deeply embedded massive young stars, which all show strong solid CO$_2$  absorption. They computed an abundance of gas-phase CO$_2$, less than 5$\%$ of that in the solid phase. 

\subsubsection{\texorpdfstring{N$_2$O (44)}{}}

Nitrous oxide (also called {\it laughing gas}) was first detected towards SgrB2 using the NRAO 12m \citep{Ziurys1994} through the J=3--2, 4--3, 5--4, and 6--5 rotational transitions at 75, 100, 125, and 150 GHz, respectively.

\subsubsection{\texorpdfstring{HCP (44)}{}}

Phosphaethyne has been detected in the J=2--1 (79.9 GHz), 4--3 (159.8 GHz), 5--4 (199.7 GHz), 6--5 (239.6 GHz), and 7--6 (279.6 GHz) rotational transitions toward IRC+10216 \citep{Agundez2007} with the IRAM 30m. It is the third phosphorus-bearing molecule identified in the ISM. 

\subsubsection{\texorpdfstring{AlOH (44)}{}}

Aluminum hydroxide has first been detected by \citet{Tenenbaum2010} towards the oxygen-rich supergiant star VY Canis Majoris (VY CMa). The N=7--6 and 6--5 rotational transitions at 268 and 230 GHz were observed using the ARO SMT and the N=4--3 line was detected using the ARO 12m telescope. 

\subsubsection{\texorpdfstring{HCS$^+$ (45)}{}}

Protonated carbon monosulfide (or thioformyl ion) has been detected by \citet{Thaddeus1981} prior to the laboratory measurements towards the hot-core sources Orion KL, Sgr B2(OH), and DR 21 as well as toward the dense cold cloud TMC-1, using the NRAO 11m and the Bell 7m telescopes. The identified transitions are the J=2--1, 3--2, 5--4, 6--5 transitions at 85.348, 128.021, 213.361 and 256.028 GHz respectively. An extragalactic detection was made in the foreground galaxy in direction of the quasar PKS 1830-212 using ATCA through the HCS$^+$ 2--1 near 45.3 GHz (near 85.3 GHz) but blended with c-C$_3$H$_2$ \citep{Muller2013}.

\subsubsection{H\texorpdfstring{CS and HSC (45)}{}}

The 2$_{0,2}$--1$_{0,1}$ transition of thioformyl (HCS) and its metastable isomer (HSC) was only  detected near 82 GHz towards the the young stellar object IRAS 18148-0440 in the L483 dense core \citep{Agundez2018} using the IRAM 30m.

\subsubsection{\texorpdfstring{MgC$_{2}$ (48)}{}}

The centimeter wavelength transitions of magnesium dicarbide has been published and  $^{24}$MgC$_2$, $^{25}$MgC$_2$, and $^{26}$MgC$_2$ have been detected through 14 previously unidentified lines towards IRC+10216 \citep{Changala2022}. 

\subsubsection{\texorpdfstring{NaCN (49)}{}}

The detection of sodium cyanide was reported by \citet{Turner1994} using the NRAO 12m towards IRC+10216. Four transitions were observed: 5$_{0,5}$--4$_{0,4}$, 6$_{0,6}$--5$_{0,5}$, 7$_{0,7}$--6$_{0,6}$ and 9$_{0,9}$--8$_{0,8}$ at 77.8, 93.2, 108.5 and 138.7 GHz respectively.

\subsubsection{\texorpdfstring{HSO (49)}{}}

The 1$_{0,1}$--0$_{0,0}$ HSO transition has been detected towards several cold dark clouds using the Yebes 40 m and IRAM 30 m telescopes \citep{Marcelino2023}.

\subsubsection{\texorpdfstring{MgNC (50)}{}}

Magnesium isocyanide was observed first as an unidentified species in the circumstellar envelope of CW Leo (IRC+10216) using the IRAM 30m \citep{Guelin1986}. The molecule was identified few years later as MgNC via measurement of its rotational spectrum by \citet{Kawaguchi1993} with the N=7--6, 8--7 and 9--8 transitions (83.538, 95.4693 and 107.3998 GHz respectively).

\subsubsection{\texorpdfstring{MgCN (50)}{}}

The magnesium cyanide isomer was detected in three transitions 8--7, 9--8, and 10--9 by \citet{Ziurys1995} towards the late-type star IRC+10216, using the NRAO 12m and IRAM 30m telescopes. 

\subsubsection{\texorpdfstring{c-SiC$_2$ (52)}{}}

Silacyclopropynylidene is a ring molecule and has been detected towards IRC+10216 by \citet{Thaddeus1984} NRAO 11m and the Bell 7m with nine transitions between 93 and 171 GHz (based on previously unidentified lines).

\subsubsection{\texorpdfstring{AlNC (53)}{}}

Aluminum isocyanide was observed in five transitions (J=10--9, 11--10, 12-11, 17--16 and 20--19 at 131.6, 143.6, 155.6, 215.4 and 251.2 GHz) towards CW Leo by \citet{Ziurys2002} using the IRAM 30m. 

\subsubsection{\texorpdfstring{SiCN (54)}{}}

Cyanosilylidyne (or silicon monocyanide radical) was identified towards IRC+10216 by \citet{Guelin2000} with the IRAM 30m through the J=7.5--6.5, 8.5--7.5 and 9.5--8.5, at 83.0, 94.0, and 105.1 GHz respectively.

\subsubsection{\texorpdfstring{CCP (55)}{}}

The phosphapropynylidyne radical has been detected in five lines using ARO 12m towards IRC+10216 \citep{Halfen2008}: J = 9.5 -- 10.5, F =	10 -- 11, F = 9 -- 10 (e and f parity) at 133.6 GHz, and J = 20.5 -- 21.5 at 273.5 GHz.

\subsubsection{\texorpdfstring{CCS (56)}{}}

\citet{Kaifu1987} carried out a spectral survey towards TMC-1 with the 45m Nobeyama telescope covering the 8--50 GHz range. Thioxoethenylidene was later identified in the spectral survey from laboratory spectroscopic measurements by \citet{Saito1987}: N$_J$=1$_2$--2$_1$, 3$_3$--2$_2$, 4$_3$--3$_2$, and 3$_4$--2$_3$ transitions at 22.344, 38.866, 43.981, and 45.379 GHz respectively. It  was identified by its four N$_J$=10$_{11}$--9$_{10}$ to 11$_{12}$--10$_{14}$ transitions with rest frequencies between 131.5 and 144.3 GHz towards the starburst Galaxy NGC 253 \citep{Martin2006} using the IRAM 30m.

\subsubsection{\texorpdfstring{NCS (58)}{}}

Thiocyanogen has been detected towards the TMC-1 cloud by \citet{Cernicharo2021} using the Yebes 40m. Three $^{14}$N hyperfine components of the $^2\Pi_{3/2}$ J=5/2--3/2 transition were detected near 42.7 GHz

\subsubsection{\texorpdfstring{SO$_2$ (64)}{}}

Sulfur dioxide was first detected by \citet{Snyder1975} towards Orion and SgrB2 using the NRAO 11m through its 8$_{1,7}$--8$_{0,8}$ transition at 83.688 GHz. It was also detected towards an extragalactic source, the nucleus of the starburst galaxy NGC 253 by \citet{Martin2003} with five transitions:  8$_{2,6}$--8$_{1,7}$ at 134.0 GHz, 5$_{1,5}$--4$_{0,4}$ at 135.7 GHz, 6$_{2,4}$--6$_{1,5}$ at 140.3 GHz, 4$_{2,2}$--4$_{1,3}$ at 146.6 GHz and 2$_{2,0}$--2$_{1,1}$ at 151.4 GHz.

\subsubsection{\texorpdfstring{CaC$_2$ (64)}{}}

Calciumcyclopropynylidene has been detected in the expanding molecular envelope of the evolved carbon star IRC+10216 with the GBT 100m, Yebes 40m, and IRAM 30m dishes \citep{Gupta2024}. With these observations, 14 a-type lines with 0 $\le$ J $\le$ 8 and K$_{a}$ $\le$ 4 between 14 and 115 GHz were identified.

\subsubsection{\texorpdfstring{S$_2$H (65)}{}}

Thiosulfeno radical was detected at the IRAM 30m, with its 6$_{0,6}$--5$_{0,5}$ and 7$_{0,7}$--6$_{0,6}$ transitions near 94.6 and 110.4 GHz by \citet{Fuente2017} towards the Horsehead photodissociation region. 

\subsubsection{\texorpdfstring{KCN (65)}{}}

Potassium cyanide was detected by \citet{Pulliam2010} towards IRC+10216 using the ARO 12m, the IRAM 30m, and the ARO SMT. Ten transitions have been identified including the K$_a$ = 1 and 2 asymmetry components of the J=11--10 and 10--9 transitions in the frequency range of 83--250 GHz.

\subsubsection{\texorpdfstring{CaNC (66)}{}}

Calcium isocyanide was detected towards IRC+10216 through nine rotational transitions with N=9--8 to 21--20, observed in emission between 72.8 and 169.9 GHz using the IRAM 30m telecope, although with a low S/N \citep{Cernicharo2019}.

\subsubsection{\texorpdfstring{SiCSi (68)}{}}

Disilylidynemethylene (also known as disilicon carbide) has been detected in more than hundred lines between 82 and 351 GHz towards IRC+10216 with the IRAM 30m telescope \citep{Cernicharo2015}.

\subsubsection{\texorpdfstring{TiO$_2$ (78)}{}}

Titanium dioxide has been detected towards IRC+10216 between 279 and 355 GHz  and around 222.5 GHz using SMA and PdB \citep{Kaminski2013} with 27 non-blended transitions with J up to 42 and K$_a$ up to 9, and upper state energies between 25 and about 730 K.

\subsubsection{\texorpdfstring{FeCN (82)}{}}

Iron cyanide was detected with the ARO 12m towards CW Leo by \citet{Zack2011} with eight successive rotational transitions in the lowest spin ladder, $\Omega$=7/2 between 75 and 150 GHz.

\subsection{Four atoms}

\subsubsection{\texorpdfstring{CH$_3$ (15)}{}}

Methyl radical was detected in the line-of-sight towards SgrA using ISO/SWS at 16 and 16.5 $\mu$m \citep{Feuchtgruber2000}.

\subsubsection{\texorpdfstring{CH$_3^{+}$ (15)}{}}

Methyl cation was detected in the d203-506 protoplanetary disk in the Orion star forming region using the James Webb Space Telescope (JWST) \citep{Berne2023}.

\subsubsection{\texorpdfstring{NH$_3$ (17)}{}}

Ammonia was first detected in its lowest J=K=1 inversion transition at 1.25 cm towards SgrB2 \citep{Cheung1968} with the 6m Hat Creek Observatory. The same transition was later detected towards the galaxies IC 342 and NGC 253 \citep{Martin1979} 100-m telescope of the MPIfR.

\subsubsection{\texorpdfstring{H$_3$O$^+$ (19)}{}}

Two groups reported almost simultaneously the detection of hydronium (or oxydanium) with its P(2,1) transition at 307192.41 MHz \citep{Hollis1986,Wootten1986} using the NRAO 12m towards Orion KL, OMC-1 and SgrB2. The 364 GHz transition was later detected towards M82 and Arp 220 using the JCMT \citep{Vandertak2008}.

\subsubsection{\texorpdfstring{C$_2$H$_2$ (26)}{}}

Acetylene was detected in absorption at the Mayall 4m telescope at Kitt Peak towards IRC+10216 at 4091.0 cm$^{-1}$ \citep{Ridgway1976} and later reported in the Magellanic clouds where broad absorptions at 3 and 3.3 $\mu$m have been presented \cite{Vanloon1999} however largely blended with HCN.

\subsubsection{\texorpdfstring{H$_2$CN (28)}{}}

Methyleneamidogen (methaniminyl) was first very weakly detected near 73.35 GHz with two noisy features at the NRAO 12m  \citep{Ohishi1994} towards TMC-1. More recent and more sensitive observations at the IRAM 30m led to the detection of six hyperfine transitions around 73.4 GHz towards L1544 \citep{Vastel2019}. It was detected in a foreground galaxy at z=0.89 in absorption toward the quasar PKS 1830-211 using Yebes in the 7mm range through the 1$_{0,1}$--0$_{0,0}$ transition \citep{Tercero2020}.

\subsubsection{\texorpdfstring{H$_2$NC (28)}{}}

The aminomethylidyne isomer was detected towards the L483 and B1-b galactic protostellar objects and the z=0.89 galaxy in front of the quasar PKS 1830-211 \citep{Cabezas2021}. The observations were made at 3 mm using the IRAM 30m. 

\subsubsection{\texorpdfstring{HCNH$^+$ (28)}{}}

J=1--0, 2--1, 3--2 rotational transitions of iminomethylium have been detected at 74, 148, and 222 GHz towards SgrB2 using using the NRAO 12m and the Texas MWO 4.9m \citep{Ziurys1986}.

\subsubsection{\texorpdfstring{H$_2$CO (30)}{}}

The 1$_{1,0}$--1$_{1,1}$ transition near 4.83 GHz was detected in absorption with the 43m NRAO telescope toward numerous galactic and extragalactic continuum sources \citep{Snyder1969}. Formaldehyde was the first organic polyatomic molecule ever detected in the ISM and its widespread distribution and brightness of its transitions indicated at the time that processes of interstellar chemical evolution may be much more complex than previously assumed. Formaldehyde was only the third poly-atomic molecule detected in space, the fourth molecule by radio astronomy, and the seventh molecule in space.

\subsubsection{\texorpdfstring{PH$_3$ (34)}{}}

Phosphine was detected by two groups almost simultaneously through the J$_K$=1$_0$--0$_0$ transition at 267 GHz. The first used the ARO SMT towards IRC+10216 and CRL 2688 \citep{Tenenbaum2008} and the second used the IRAM 30m towards IRC+10216 \citep{Agundez2008}. 

\subsubsection{\texorpdfstring{c-C$_3$H (37)}{}}

Cyclopropanediylidenyl was detected first through tow fine structure components of the 2$_{1,2}$--1$_{1,1}$ transition in the dense cold molecular cloud TMC-1 using the Nobeyama 45m \citep{Yamamoto1987}. It was then detected in the NGC 253 galaxy \citep{Martin2006}.

\subsubsection{\texorpdfstring{l-C$_3$H (37)}{}}

Propynylidyne was detected first in the dense cold molecular cloud TMC-1 and towards IRC+10216 \citep{Thaddeus1985a} with four HFS components of the J=3/2--1/2 transition near 32.6 GHz using the Onsala 20m telescope. It was also detected in absorption towards the quasar PKS 1830-211 at 76.2 GHz \citep{Muller2011}.

\subsubsection{\texorpdfstring{C$_3$H$^+$ (37)}{}}

Propynylidynium was detected with the IRAM 30m telescope in nine transitions (J$_{\rm up}$=3 to 11) towards the Horsehead Nebula in Orion, covering the 3, 2, and 1.3 mm regions \citep{Pety2012}. It was also detected in a foreground galaxy at z=0.89 in absorption toward the quasar PKS 1830-211 using the Yebes telescope in the 7 mm region \citep{Tercero2020}.

\subsubsection{\texorpdfstring{HNCN (41)}{}}

Cyanoamidogen was detected towards the Galactic center cold molecular cloud G+0.693-0.027 carried out with the Yebes 40m and IRAM 30m \citep{Rivilla2021} through the N=6--5 doublet at 132 GHz and the N=4--3 transition at 88 GHz (unblended transitions).

\subsubsection{\texorpdfstring{HCCO (41)}{}}

The ketenyl radical was first detected with four $\Delta$F=$\Delta$J=$\Delta$N hyperfine components with N=4--3 near 86.65 GHz toward the dense core Lupus-1A using the IRAM 30m \citep{Agundez2015a}

\subsubsection{\texorpdfstring{HNCO (43)}{}}

Isocyanic acid was among the very early molecules to be detected in space. The 4$_{0,4}$--3$_{0,3}$ transition at 3.4 mm was the first one to be detected in the Galactic center source SgrB2 \citep{Snyder1972}. It was later detected in an external galaxy (NGC 253) through its 4$_{0,4}$--3$_{0,3}$ and 6$_{0,6}$--5$_{0,5}$ transitions \citep{Nguyen1991}.

\subsubsection{\texorpdfstring{HCNO (43)}{}}

Fulminic acid was first detected through the J=4--3 and 5--4 transitions towards the B1 and L1527 protostars \citep{Marcelino2009}.

\subsubsection{\texorpdfstring{HOCN (43)}{}}

Cyanic acid was first detected in the laboratory and tentatively detected through archival data towards an astronomical source (SgrB2) by \citet{Bruenken2009}.

\subsubsection{\texorpdfstring{HOCO$^+$ (45)}{}}

The first tentative detection of protonated carbon dioxide in a star-forming regions (SgrB2) was reported by \citet{Thaddeus1981} in several rotational transitions (J$_{\rm up}$=4, 5, and 6) with the Bell 7m telescope. All transitions were confirmed by laboratory measurements by \citet{Bogey1984}. The species was also identified in an external galaxy (NGC 253) by \citet{Martin2006} and \citet{Aladro2015}.

\subsubsection{\texorpdfstring{H$_2$CS (46)}{}}

Through observations obtained with the Parkes 64m antenna of the J=$2_{1,1}-2_{1,2}$ transition, \citet{Sinclair1973} reported the first detection of Thioformaldehyde towards the star-forming region SgrB2. \citet{Martin2006} identified three transitions (J=$4_{1,4}-3_{1,3}$, J=$4_{1,3}-3_{1,2}$, and J=$5_{1,5}-4_{1,4}$) of H$_2$CS for the first time in the external galaxy NGC 253 with the IRAM 30m telescope.

\subsubsection{\texorpdfstring{HONO (47)}{}}

Several lines of the nitrous acid ($E_{\rm u}\sim$ 32 to 367~K) molecule were first detected in the ISM by \citet{Coutens2019} towards the protostellar binary IRAS 16293-2422 through ALMA observations.

\subsubsection{\texorpdfstring{MgC$_2$H (49)}{}}

\citet{Agundez2014} reported the first tentative detection of magnesium monoacetylide in the ISM with the IRAM 30m telescope towards the evolved star IRC+10216 (transitions N=9--8 and 10--9), confirmed by \citet{Cernicharo2019a}. 

\subsubsection{\texorpdfstring{C$_3$N (50)}{}}

Cyanoethynyl radical was first tentatively detected in the ISM towards IRC+20126 by \citet{Guelin1977} with the NRAO 11m telescope. A firmer detection was then obtained towards the dark clouds TMC1 and TMC2 by \citet{Friberg1980} with the Onsala 20m telescope.

\subsubsection{\texorpdfstring{C$_3$N$^-$ (50)}{}}

The cyanoethynyl anion was first detected by \citet{Thaddeus2008} with the IRAM 30m telescope in several transitions (J$_{\rm up}$ = 10 to 15) towards IRC+10216, based on their laboratory measurements.

\subsubsection{\texorpdfstring{HMgNC (51)}{}}

Hydromagnesium isocyanide was identified both in the laboratory and in the ISM (towards IRC+20126) with the IRAM 30m telescope by \citet{Cabezas2013} in several rotational lines (J$_{\rm up}$=8 to 13).

\subsubsection{\texorpdfstring{CNCN (52)}{}}

Interstellar isocyanogen was first detected by \citet{Agundez2018a} towards the dark cloud L483, and tentatively towards TMC-1, with the IRAM 30m telescope in several rotational transitions (J=8--7, 9--8, and 10--9).

\subsubsection{\texorpdfstring{C$_3$O (52)}{}}

\citet{Matthews1984} identified the J=2--1 transition of tricarbon monoxide with the NRAO 43m telescope at Green Bank towards the dark cloud TMC-1. 

\subsubsection{\texorpdfstring{HCCN (52)}{}}

The cyanomethylene radical was first detected by \citet{Guelin1991} with the IRAM 30m telescope towards IRC+10216 in several transitions (N$_{\rm up}$ = 4 to 10).

\subsubsection{\texorpdfstring{HCCS (57)}{}}

\citet{Cernicharo2021} detected for the first time in space the ethenthionyl radical towards the dark cloud TMC-1 with the IRAM 30m telescope in the strongest hyperfine components of the J=7/2--5/2 transition. 

\subsubsection{\texorpdfstring{HCCS$^+$ (57)}{}}

Thioketenylium was detected by \citet{Cabezas2022} in twenty-six hyperfine components from twelve rotational transitions (N$_{\rm up}$=2 to 8), observed with the Yebes 40m and IRAM 30m radio telescopes.

\subsubsection{\texorpdfstring{HNCS (59)}{}}

Isothiocyanic acid was detected in the interstellar medium for the first time by \citet{Frerking1979} in the J=8--7, 9--8, and 11--10 with the Bell 7m and NRAO 36 ft telescopes towards SgrB2.

\subsubsection{\texorpdfstring{HSCN (59)}{}}

\citet{Halfen2009} reported the first identification of thiocyanic acid in the ISM (SgrB2) though observations of several transitions (J$_{\rm up}$ = 6 to 12) with the ARO 12m telescope.

\subsubsection{\texorpdfstring{HCNS (59)}{}}

Thiofulminic acid was detected in the direction of TMC-1 using the Yebes 40m and 30m telescopes \citep{Cernicharo2024} with three lines (J = 3--2, J = 4--3 and J = 6--5) .

\subsubsection{\texorpdfstring{HOCS$^{+}$ (61)}{}}

Thioxyhydroxymethylium was detected towards the Galactic Center molecular cloud G+0.693--0.027 with the Yebes 40m and IRAM 30m telescopes with transitions covering 34 to 161 GHz with J$_{\rm up}$ $\le$ 14 and K$_{a}$ = 0 \citep{SanzNovo2024a}.

\subsubsection{\texorpdfstring{HNSO (63)}{}}

Thionylimide was detected towards the Galactic Center molecular cloud G+0.693--0.027 with the Yebes 40m and IRAM 30m telescopes. The a-type transitions cover 34 to 171 GHz with 1 $\le$ J $\le$ 10 and K$_{a}$ $\le$ 2 \citep{SanzNovo2024b}.

\subsubsection{\texorpdfstring{c-SiC$_3$ (64)}{}}

The first detection of silicon tricarbide in interstellar material (seven transitions with J$_{\rm up}$=7 to 9) was made by \citet{Apponi1999} with NRAO 12m observations of IRC+10216. 

\subsubsection{\texorpdfstring{H$_2$O$_2$ (64)}{}}

\citet{Bergman2011} reported the first detection of hydrogen peroxide in the ISM towards the star-forming region $\rho$-Ophiuchi A (transitions J=3$_{0,3}$--2$_{1,1}$, J=5$_{0,5}$--4$_{1,3}$, and J=6$_{1,5}$--5$_{0,5}$), obtained with the APEX telescope.

\subsubsection{\texorpdfstring{C$_3$S (68)}{}}

The presence of tricarbon monosulfide radical was revealed the first time in an interstellar source (TMC-1) by \citet{Yamamoto1987a}, who identified three transitions (J=4--3, 7--6, and 8--7) of its rotational spectrum to unidentified lines in the observations obtained with the Nobeyama 45m telescope by \citet{Kaifu1987}.

\subsection{Five atoms}

\subsubsection{\texorpdfstring{CH$_4$ (16)}{}}

Methane was detected in the gas phase and probably detected in the solid phase using the 3m NASA Infrared Telescope Facility (IRTF) observations toward NGC 7538 IRS 9 in absorption at 7.6 $\mu$m \citet{Lacy1991}. 


\subsubsection{\texorpdfstring{H$_2$CNH (29)}{}}

Methanimine (formaldimine) has first been detected using Parkes 64m towards SgrB2 \citep{Godfrey1973} with its 1$_{1,0}$--1$_{1,1}$ transition at 5.290 GHz. It was later detected in absorption toward the quasar PKS 1830-211, in a foreground galaxy at z=0.89 using ATCA \citep{Muller2011}.

\subsubsection{\texorpdfstring{H$_2$COH$^+$ (31)}{}}

Hydroxymethylium, also known as protonated formaldehyde, was detected towards SgrB2(M) and (N), Orion KL, W51, and possibly in NGC 7538 and DR21(OH) by \citet{Ohishi1996} using the Nobeyama 45m and the Kitt Peak 12m with six transitions between 31 and 174 GHz.

\subsubsection{\texorpdfstring{CH$_3$O (31)}{}}

Methoxy radical was detected \citep{Cernicharo2012} towards B1-b using the IRAM 30m telescope through two components of the N=1--0, K=0, J=3/2--1/2, F=2--1 transition ($\Lambda$=1 at 82.458 and $\Lambda$=1 at 82.472 GHz).

\subsubsection{\texorpdfstring{SiH$_4$ (32)}{}}

Silane was first detected towards IRC+10216 through 13 rovibrational transitions of the $\nu_4$ band around 917 cm$^{-1}$ using the IRTF telescope \citep{Goldhaber1984}.

\subsubsection{\texorpdfstring{NH$_2$OH (33)}{}}

Hydroxylamine was detected towards SgrB2(N) with the IRAM 30m \citep{Rivilla2020} with a-type transitions covering J=2--1, 3--2, and 4--3 near 100.7, 151.1, and 201.5 GHz. 

\subsubsection{\texorpdfstring{c-C$_3$H$_2$ (38)}{}}

Cyclopropenylidene was first detected with its ortho ground state lines 2$_{1,2}$--1$_{0,1}$ in emission at 85.3389 GHz towards Ori A, SgrB2(OH), and TMC-1 \citet{Thaddeus1981} although as an unidentified feature. It was later identified as c-C$_3$H$_2$ by \citet{Thaddeus1985} with eleven other ortho and para transitions between 18 and 266 GHz. It was later detected at 18.343 GHz in absorption against the nuclear continuum of the nearby radio galaxy NGC 5128 (= Centaurus A) using the NRAO 42.7m \citep{Seaquist1986}. 

\subsubsection{\texorpdfstring{l-C$_3$H$_2$ (38)}{}}

Propadienylidene is a higher energy isomer of the previous species. It has been detected in TMC-1 and possibly IRC + 10216 with the IRAM 30m through the ortho transitions 5$_{1,5}$--4$_{1,4}$, 5$_{1,4}$--4$_{1,3}$, and 7$_{1,6}$--6$_{1,5}$ near 103.0, 104.9, and 146.9 GHz, respectively. The para ground state transition 1$_{0,1}$--0$_{0,0}$ near 20.8 GHz was also  identified from a previously unidentified line with the Effelsberg 100m telescope \citep{Cernicharo1987}. It was later detected in absorption toward the quasar PKS 1830-211, in a foreground galaxy at z=0.89 using ATCA \citep{Muller2011}.

\subsubsection{\texorpdfstring{H$_2$CCN (40)}{}}

Cyanomethyl radical was first detected towards TMC-1 (20.12 and 40.24 GHz) and SgrB2 (40, 80, and 100 GHz, although blended) by \citet{Irvine1988} using FCRAO 14m, NRAO 43m, Onsala 20m and Nobeyama 45m. Some of the lines were previously observed before , towards SgrB2 using the Bell Laboratories 7m telescope, and attributed to some unknown insterstellar molecule \citep{Cummins1986}. It was later detected in absorption toward the quasar PKS 1830-211, in a foreground galaxy at z=0.89 using ATCA \citep{Muller2011}.

\subsubsection{\texorpdfstring{H$_2$NCN (42)}{}}

Cyanamide was first detected through its 4$_{1,3}$--3$_{1,2}$ and 5$_{1,4}$--4$_{1,3}$ transitions at at 80.5045 and 100.6295 GHz respectively using the NRAO 11m telescope towards SgrB2 by \citet{Turner1975}.

\subsubsection{\texorpdfstring{HNCNH (42)}{}}

Carbodiimide was first detected towards SgrB2 using the Green Bank Telescope (GBT) \citep{McGuire2012} in the 1–46 GHz line survey. It is a high energy isomer ($\sim$ 2000 K) of the cyanamide  (H$_2$NCN) species typically found in hot cores.

\subsubsection{\texorpdfstring{H$_2$C$_2$O (42)}{}}

Ethenone (also called ketene) was first detected toward SgrB2(OH) by \citet{Turner1977} using the NRAO 11m telescope which identified three K$_a$=1 transitions with J$_{up}$=4 and 5. The fourth transition and the two K$_a$=0 transitions were detected tentatively around 81 and 101 GHz.  It was later detected in absorption toward the quasar PKS 1830-211, in a foreground galaxy at z=0.89 using ATCA \citep{Muller2011}.

\subsubsection{\texorpdfstring{H$_2$NCO$^+$ (44)}{}}

Protonated isocyanic acid was tentatively detected towards SgrB2 using the GBT \citep{Gupta2013} with weak absorption features of the para 1$_{0,1}$--0$_{0,0}$ (20.228 GHz) and ortho 2$_{1,1}$--1$_{1,0}$ (40.783 GHz). 

\subsubsection{\texorpdfstring{HCOOH (46)}{}}

Formic acid was first identified with its 1$_{1,1}$--1$_{1,0}$ transition 1638.805 MHz towards SgrB2 using NRAO 43m \citep{Zuckerman1971}. It was later confirmed by \citet{Winnewisser1975} with the detection towards the same source of a second transition, the 2$_{1,1}$--2$_{1,2}$ at 4.9 GHz. The first detection in an extragalactic source was made much later by \citet{Tercero2020} in absorption in the spiral arm of a galaxy located at z = 0.89 on the line of sight to the quasar PKS 1830-211 using Yebes 40m.

\subsubsection{\texorpdfstring{C$_4$H (49)}{}}

Butadiynyl radical has been detected using the NRAO 11m towards IRC+10216 in its 2$\Sigma$ ground vibrational state between 85 and 115 GHz  in eight fine structure lines of four rotational transitions (N=9--8 to 12--11) by \citet{Guelin1978}. 

\subsubsection{\texorpdfstring{C$_4$H$^-$ (49)}{}}

Butadiynyl anion has been detected toward the carbon rich star IRC+10216 in 5 rotational transitions from J=9--8 to 15--14 between 83.8 and 139.6 GHz by \citet{Cernicharo2007} using the IRAM 30m.

\subsubsection{\texorpdfstring{CH$_3$Cl (50)}{}}

Methyl chloride (chloromethan) is a trace constituent in Earth's atmosphere and is largely produced by industrial processes. It was detected with ALMA near 345.4 GHz through the J=13--12 transitions with K = 0 to 4 \citep{Fayolle2017} towards IRAS16293. 

\subsubsection{\texorpdfstring{HC$_3$N (51)}{}}

Cyanoacetylene was first detected towards SgrB2 using the NRAO 43m by \citet{Turner1971} through the F=2--1 and 1--1 hyperfine components of the J=1--0 transition at 9.098 and 9.097 GHz, respectively. Six millimeter transitions have later been detected toward the nucleus of the starburst galaxy NGC 253 \citet{Mauersberger1990}, confirming an earlier tentative detection towards M82 and IC 342 \citep{Henkel1988}.

\subsubsection{\texorpdfstring{HC$_3$N$^{+}$ (51)}{}}
The Yebes 40m detected  six strong hyperfine structure components of the J = 7/2--5/2 near 31.63 GHz and 9/2--7/2 near 40.66 GHz in its $^{2}\Pi_{3/2}$ lower spin ladder, towards TMC-1 \citep{Cabezas2024}.

\subsubsection{\texorpdfstring{HCCNC (51)}{}}

Isocyanoacetylene was first detected towards TMC-1 using the  Nobeyama 45m to detect the J = 4--3, 5--4, and 9--8 transitions near 39.7, 49.7, and 89.4 GHz \citep{Kawaguchi1992a}.  

\subsubsection{\texorpdfstring{HNC$_3$ (51)}{}}

Iminopropadienylidene has first been detected by \citet{Kawaguchi1992} towards TMC-1 using the Nobeyama 45m telescope through the J=3--2, 4--3, and 5--4 transitions at 28.0, 37.3, and 46.7 GHz.

\subsubsection{\texorpdfstring{HC$_3$O$^+$ (53)}{}}

Protonated tricarbon monoxide (ethynyloxomethylium) was only detected, using the IRAM 30m and the Yebes 40m towards TMC-1 \citep{Cernicharo2020} J$_{up}$=3 and 4 near 35.7 and 44.6 GHz  J$_{up}$ = 9 and 10 near 89.2 and 98.1 GHz.

\subsubsection{\texorpdfstring{HCCCO (53)}{}}

Propynonyl was identified \citep{Cernicharo2021a} through two rotational transitions (3$_{0,1}$--2$_{0,2}$ and 4$_{0,4}$--3$_{0,3}$) in the course of a molecular line survey of the prototypical cold dark molecular cloud TMC-1 carried out with the Yebes 40 m radio telescope between 31.0 and 50.3 GHz.

\subsubsection{\texorpdfstring{NCCNH$^+$ (53)}{}}

Protonated cyanogen was first detected in emission toward the dark cloud TMC-1 (cyanopolyyne peak) and the young stellar object IRAS 18148-0440 in the L483 dense core \citep{Agundez2015b} using the IRAM 30m and Yebes 40m (J=10--9 and 5--4 transitions near 88.8 and 44.4 GHz respectively). 

\subsubsection{\texorpdfstring{CNCHO (55)}{}}

Formyl cyanide (cyanoformaldehyde) was first detected towards SgrB2 using the GBT \citep{Remijan2008} by means of four P-branch rotational transitions in emission, the 7$_{0,7}$--6$_{1,6}$ at 8.6 GHz, the 8$_{0,8}$--7$_{1,7}$ at 19.4 GHz, the 9$_{0,9}$--8$_{1,8}$ at 30.3 GHz, and the 10$_{0,10}$--9$_{1,9}$ at 41.3 GHz, and one P-branch transition in absorption, the 5$_{1,5}$--6$_{0,6}$ at 2.1 GHz. 

\subsubsection{\texorpdfstring{H$_2$C$_2$S (58)}{}}

Ethenthione (thioketene) was detected through 6 a-type transitions between 33.4 and 44.8 GHz in the course of a spectral survey of the cold dark molecular cloud TMC-1 carried out with the Yebes 40m telescope \citep{Cernicharo2021}.

\subsubsection{\texorpdfstring{C$_5$ (60)}{}}

Pentatetraenediylidene has been detected towards IRC+20216 by \citet{Bernath1989} before it was even detected in the laboratory. More than a dozen P and R branch transitions of the $\nu_3$ asymmetric stretching mode near 2164 cm$^{-1}$ were detected using the KPNO 4m telescope. It was later detected in absorption toward the quasar PKS 1830-211, in a foreground galaxy at z = 0.89, identified by its N=6--5 to 9--8 transitions with rest frequencies between 57.0 and 85.7 GHz using ATCA \citep{Muller2011}.

\subsubsection{\texorpdfstring{CHOSH (62)}{}}

Monothioformic acid was detected towards the quiescent giant molecular cloud G+0.693-0.027, about 1$^{\prime}$ north-east of Sagittarius B2(N) using the Yebes 40m and the IRAM 30m \citep{Rodriguez2021}. Nine rotational transitions with J from 2 to 8 and K$_a$=0 or 1 were reported, seven of which were not blended or marginally blended. 

\subsubsection{\texorpdfstring{HC$_3$S$^+$ (69)}{}}

Ethynylthioxomethylium was detected through J$_{up}$=5 to 8 transitions between 32.8 and 49.2 GHz in the course of a spectral survey of the cold dark molecular cloud TMC-1 carried out with the Yebes 40m telescope \citep{Cernicharo2021b}.

\subsubsection{\texorpdfstring{HC$_3$S (69)}{}}

Propadienethionyl was detected towards TMC-1 with the Yebes 40m telescope through three consecutive transitions with 13/2 $\le$ J$_{up}$ $\le$ 17/2 between 34.8 and 45.6 GHz \citep{Cernicharo2024b}.

\subsubsection{\texorpdfstring{HC(S)CN (71)}{}}

Cyanothioformaldehyde (thioformyl cyanide) was detected towards TMC-1 with Yebes 40m, between 30.1 and 50.4 GHz covering 10 unblended (or only slightly blended) a-type transitions with J$_{up}$ from 4 to 8 and K$_a$=0 or 1 \citep{Cernicharo2021c}

\subsubsection{\texorpdfstring{NaC$_3$N (73)}{}}

Sodium cyanoacetylide has been detected with a low S/N towards IRC +10216 using the Yebes 40 m telescope \citep{Cabezas2023}. 

\subsubsection{\texorpdfstring{MgC$_3$N (74)}{}}

Magnesium monocyanoacetylides has been detected with a low S/N using the Yebes 40m and the IRAM 30m towards IRC+10216 with 22 rotational transitions between N$_{up}$ = 12 to 40 and between 33 and 111 GHz \citep{Cernicharo2019a}.

\subsubsection{\texorpdfstring{MgC$_3$N$^+$ (74)}{}}

MgC$_3$N$^+$ has been detected (13--12 at 37.62 GHz, 14-13 at 40.51 GHz and 15--14 at 43.41 GHz, non blended with other features ) by \citet{Cernicharo2023} in the spectrum of IRC+10216 using the Yebes 40m.

\subsubsection{\texorpdfstring{C$_4$Si (76)}{}}

Butadiynylidenesilylidyne (or silicon tetracarbide) has been detected towards IRC+10216 with J transitions between 11 and 27 using the Nobeyama 45m between 36 and 86 GHz \citep{Ohishi1989}. 

\subsubsection{\texorpdfstring{C$_4$S (80)}{}}

Tetracarbon monosulfide was detected through the fine structure transitions with N$_{up}$=9 to 12 and J=N+1 and with moderate to good S/N ratio, between 32.8 and 49.2 GHz in the course of a spectral survey of the cold dark molecular cloud TMC-1 carried out with the Yebes 40m telescope \citep{Cernicharo2021}.

\subsubsection{\texorpdfstring{NC$_3$S (82)}{}}

Cyanoethynylsulfanyl was detected it $^{2}\Pi_{3/2}$ lower spin ladder towards TMC-1 with the Yebes 40m telescope through five consecutive transitions with 23/2 $\le$ J$_{up}$ $\le$ 33/2 between 33.0 and 47.5 GHz \citep{Cernicharo2024b}.

\section{Radiative and collisional excitation of molecules}
\label{sec:rad}

The observations of molecular spectral lines are crucial to determine the physical and chemical conditions of the observed object. Nowadays, the technological advances offer a high spatial and spectral resolution and sensitivity, with a new spectral window for millimeter and submillimeter observations where most molecules emit their rotational transitions. Hundreds of transitions are now being accessible for some species and we sometimes need detailed radiative transfer modelling code to characterise the physical and chemical conditions of the object, sometimes with some background and foreground contamination. We will present in the following the basics for understanding the radiative transfer modelling (Sec. \ref{sec:RT}), in order to compute the column densities of the observed species (Sec. \ref{sec:CD}) and their chemical abundances (Sec. \ref{sec:AB}).

\subsection{Radiative transfer}
\label{sec:RT}

\subsubsection{Population levels}

\begin{figure}[h!]
  \centering
  \caption{Example of a 2-level system with an upper level energy E$_{u}$ and a lower level energy E$_{\ell}$.}
  \includegraphics[width=0.85\textwidth]{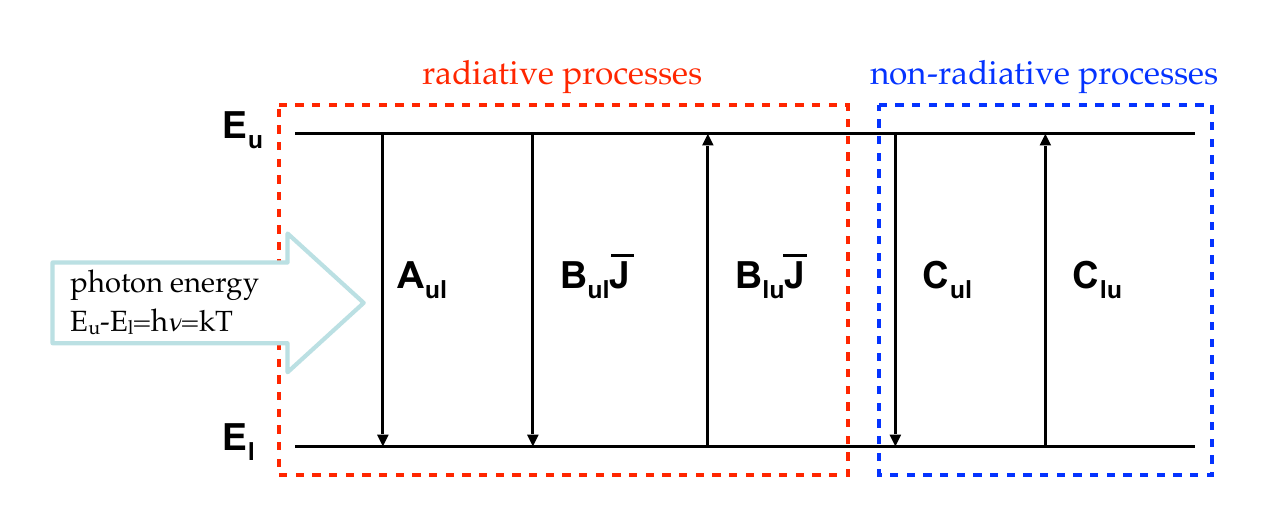}
  \label{fig:2level}
\end{figure}

When considering a multi energy level system of a considered species, we first need to evaluate the rate of transitions populating a given energy level. Figure \ref{fig:2level} shows the example of a 2-level system with an upper level energy E$_{u}$ and a lower level energy E$_{\ell}$. The A$_{u\ell}$, B$_{u\ell}$ and B$_{\ell u}$ parameters represent the so-called Einstein coefficients, respectively describing the spontaneous radiative de-excitation, the stimulated radiative de-excitation and the radiative excitation. The C$_{u\ell}$ and C$_{\ell u}$ parameters represent the collisional de-excitation and excitation respectively (non radiative processes as they are independent of the photon interaction with the 2-level system). These rates are the collision rates per second per molecule of the species of interest and they depend on the density of the collision partner. They can be expressed as:
\begin{equation}
C_{ij} = \gamma_{ij} \times n_{collider} \, ,
\end{equation}
where $n_{collider}$ is the density of the collision partner which can be H$_{2}$, Helium, electrons, depending on the properties of the observed ISM and participates to the level population. The collisional rate coefficients $\gamma_{ij}$ (in cm$^{3}$~s$^{-1}$) are the velocity-integrated collisional cross sections, and depend on the kinetic temperature (T$_k$) through the relative velocity of the colliding molecules and possibly also through the collisional cross sections directly. The downward collisional rate coefficients are tabulated in various databases such as LAMDA\footnote{\url{https://home.strw.leidenuniv.nl/~moldata/}} and Basecol\footnote{\url{https://basecol.vamdc.eu/}}. They represent the Maxwellian average of the collisional cross section ($\sigma$), depending on the collision energy ($E$), the kinetic temperature ($T_k$) and the reduced mass ($\mu$) of the system:
\begin{equation}
\gamma_{u\ell} = \sqrt{\frac{8kT_k}{\pi\mu}}\left(\frac{1}{kT_k}\right)^2\int \sigma_{u\ell} E \exp\left(\frac{-E}{kT_k}\right) dE \, ,
\end{equation}
where k is the Boltzmann constant. The upward and downward rates are related through the following:
\begin{equation}
\label{eq:gamma}
\gamma_{\ell u} = \gamma_{u\ell}\frac{g_u}{g_\ell}e^{-h\nu/kT_k} \, ,    
\end{equation}
where $g_i$ is the statistical weight of the $i$ level and T$_k$ is the kinetic temperature. 
The radiation field is noted in Fig. \ref{fig:2level} as $\bar{J}$=$\int_{0}^{\infty}J_{\nu}\Phi(\nu) \,d\nu$, where J$_{\nu}$ is defined as the integral of the specific intensity I$_{\nu}$ over the source of emission and $\Phi(\nu)$ is the line profile function (Gaussian, Lorentzian...): \\
\begin{equation}
J_{\nu} = \frac{1}{4\pi} \int I_{\nu}\,d\Omega \, .
\end{equation}

We need to solve the population rate for each level:
\begin{equation}
\label{eq:pop}
\frac{dn_{i}}{dt} = -n_{i}\left[ \sum_{k<i} A_{ik} + \sum_{k \neq i}(B_{ik}\bar{J} + C_{ik}) \right] + \sum_{k > i} n_{k}A_{ki} + \sum_{k \neq i} n_{k}(B_{ki}\bar{J} + C_{ki}) \, ,
\end{equation}

where $n_{i}$ is the population of the energy level i.  To solve these equations we need to know the radiation field which is the amount of radiation “inside” the source and which we do not know.

When the energy levels of a molecule are in statistical equilibrium, the rate of transition populating a given energy level is balanced by the rate of transitions which depopulates that same energy level and $\frac{dn_{i}}{dt} = 0$.\\
We can derive the Einstein coefficients by considering a 2-level system and only radiation excitation in Eq. \ref{eq:pop} with $C_{\ell u} = C_{u\ell} = 0$:
\begin{equation}
\label{eq:pop-eq}
n_{u}A_{u\ell} + n_{u}B_{u\ell}\bar{J}   = n_{\ell} B_{\ell u}\bar{J}
\end{equation}
For a system in thermal equilibrium, the relative level populations follow the Boltzmann distribution:
\begin{equation}
\label{eq:boltz}
\frac{n_{u}}{n_{\ell}} = \frac{g_{u}}{g_{\ell}}\exp\left(-\frac{h\nu}{kT}\right) \, ,
\end{equation}
where h is the Planck constant, g$_{u}$ and g$_{\ell}$ are the statistical weights of levels up and low respectively and $T$ is the temperature of the region. In non-LTE (Local Thermodynamic Equilibrium), T can be replaced by the so-called excitation temperature. T$_{ex}$ is not a {\it real} temperature (such as T$_k$) and corresponds to the temperature for a Boltzmann population in system made of these two levels. This definition is broader than the case of thermal equilibrium and remains valid even if the level population is not at equilibrium. T$_{ex}$ then depends on levels ($\ell$,$u$). \\
In thermodynamic equilibrium, the radiation field $J_{\nu}$ can be described by the Planck function $B_{\nu}(T)$:
\begin{equation}
\label{eq:planck}
B_{\nu}(T) = \frac{2h\nu^{3}}{c^{2}}\frac{1}{\exp\left(\frac{h\nu}{kT} \right) -1} \hspace{1cm}(erg~s^{-1}~cm^{-2}~Hz^{-1}~sr^{-1}) \, .
\end{equation}
We can now substitute Eq. \ref{eq:boltz} into Eq. \ref{eq:pop-eq} to obtain:
\begin{equation}
\label{eq:JnuTex}
    \bar{J} = \frac{A_{u\ell}/B_{u\ell}}{\frac{N_lB_{\ell u}}{N_uB_{u\ell}}-1} = \frac{A_{u\ell}/B_{u\ell}}{\frac{g_\ell B_{\ell u}}{g_uB_{u\ell}}\exp(\frac{h\nu}{kT})-1} 
\end{equation}
Comparing Eq. \ref{eq:planck} with Eq. \ref{eq:JnuTex} we can now get simplified relationships that allows to express Eq. \ref{eq:pop} as a function of $A_{u\ell}$ only:
\begin{align}
g_{u}B_{u\ell} = g_{\ell}B_{\ell u} \, , \\
A_{u\ell} = \frac{2h\nu^{3}}{c^{2}}B_{u\ell}\, .
\end{align}
Note that some authors use the energy blackbody radiation $U_{\nu}$ (=$I_{\nu}(T)\times4\pi/c$) instead of $J_{\nu}$ through $\bar{J}$ in Eq. \ref{eq:pop-eq}. A different expression relating the $A$ and $B$ Einstein coefficients is then used: $A_{u\ell} = B_{u\ell} \times 8\pi h\nu^3/c^3$. Note however that, in the later definition, only the B coefficients vary, and A is unchanged from one definition to the other. The Einstein coefficient A$_{u\ell}$ (which are proportional to $\nu^3$) can be found tabulated in the spectroscopic databases such as CDMS\footnote{\url{https://cdms.astro.uni-koeln.de/}}, JPL\footnote{\url{https://spec.jpl.nasa.gov/}} and  NIST\footnote{\url{https://www.nist.gov/pml/observed-interstellar-molecular-microwave-transitions/}}.

\subsubsection{Radiative transfer equation}

\begin{figure}[h]
\begin{center}
\includegraphics[width=0.85\textwidth]{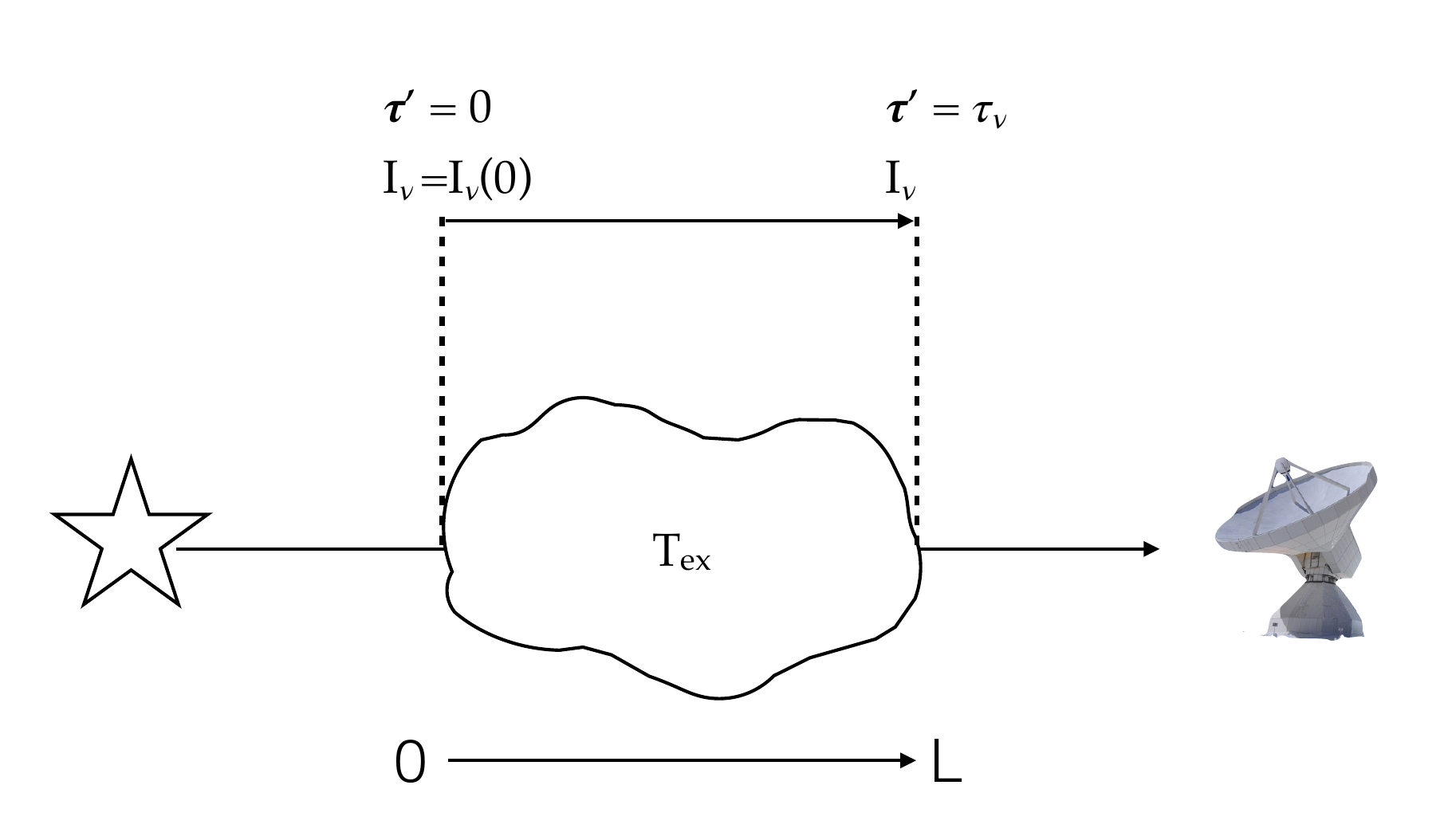}
\caption{\label{fig:opacity_cloud}{Passage of a beam through a gaseous object of length L, excitation temperature T$_{ex}$.}}
\end{center}
\end{figure}

The intensity of a source emitting in the ISM along a line of sight $I_{\nu}$, will change 
if the radiation is absorbed or emitted, and this change can be described by the equation of transfer\,:
\begin{equation}
\label{eq:dnu}
\frac{dI_{\nu}}{ds} = -\alpha_{\nu}I_{\nu} + j_{\nu} \, ,
\end{equation}
where $\frac{dI_{\nu}}{ds}$ represents the change of the intensity $I_{\nu}$ at the corresponding 
frequency $\nu$ through a slab of material of thickness {\it s}. It depends on the absorption coefficient 
$\alpha_{\nu}$ and the emissivity $j_{\nu}$ \citep[see e.g][]{Spitzer1978}.  
The expressions of $\alpha_{\nu}$ and j$_{\nu}$ are defined as:
\begin{align}
\label{eq:em_abs}
\alpha_{\nu} & = \frac{h\nu}{4\pi}(n_{\ell}B_{\ell u}-n_{u}B_{u\ell})\Phi(\nu) \, ,\\
	             & = \frac{c^{2}}{8\pi\nu^{2}}\frac{g_{u}}{g_{\ell}}n_{\ell}A_{u\ell}\left(1 - \frac{g_{\ell}n_{u}}{g_{u}n_{\ell}}\right)\Phi(\nu)   \hspace{1cm}(cm^{-1}) \, ,\\
\label{eq:abs_em}
 j_{\nu} & = \frac{h\nu}{4\pi}A_{u\ell}n_{u}\Phi(\nu)\hspace{1cm} (erg~s^{-1}~cm^{-3}~Hz^{-1}~sr^{-1}) \, .	             
\end{align}
Since we do not know the path of propagation $s$, it is convenient to define a new variable called {\it optical depth}, or {\it opacity} of a line at the frequency $\nu$ such as:
\begin{equation}
\label{eq:taunu}
d\tau_{\nu} = \alpha_{\nu} ds \, ,
\end{equation}
and define the so-called source function $S_{\nu}$ (Kirchhoff's law of thermal radiation) as:
\begin{equation}
\label{eq:Snu}
S_{\nu} = \frac{j_{\nu}}{\alpha_{\nu}} = \frac{n_u A_{ul}}{\left(n_l B_{lu}-n_u B_{ul}\right)}\, .
\end{equation}
Then we get\,:
\begin{align}
\frac{dI_{\nu}}{d\tau_{\nu}} = -I_{\nu} + S_{\nu}\, ,\\
\frac{dI_{\nu}}{d\tau_{\nu}}e^{\tau} + I_{\nu}e^{\tau} = S_{\nu}e^{\tau}\, ,\\
\frac{d}{d\tau_{\nu}}\left(I_{\nu}e^{\tau}\right) = S_{\nu}e^{\tau} \, .
\end{align}
We can then integrate this equation between 0 and $\tau_{\nu}$ (cf. Fig. \ref{fig:opacity_cloud})\,:
\begin{equation}
\int_{0}^{\tau_{\nu}}\frac{d}{d\tau_{\nu}}\left(I_{\nu}e^{\tau}\right)\,d\tau_{\nu} = \int_{0}^{\tau_{\nu}}S_{\nu}e^{\tau}\,d\tau_{\nu} \, ,
\end{equation}
\begin{equation}
I_{\nu}e^{\tau}-I_{\nu}(0) = \int_{0}^{\tau_{\nu}}S_{\nu}e^{\tau}\,d\tau^{\prime}_{\nu} \, ,
\end{equation}
\begin{equation}
I_{\nu} = I_{\nu}(0)e^{-\tau}+\int_{0}^{\tau_{\nu}}S_{\nu}\exp[-({\tau_{\nu}-\tau^{\prime}})]\,d\tau^{\prime} \, ,
\end{equation}
where I$_{\nu}$(0) represents the background radiation, i.e the cosmic microwave background (CMB) at 2.7 K. 

Assuming that the source function does not vary in the observed medium at a constant temperature (it does not vary as a function of the opacity), we then get\,:
\begin{equation}
I_{\nu} = I_{\nu}(0)e^{-\tau_{\nu}} + S_{\nu}(1-e^{-\tau_{\nu}})  \, .
\end{equation}
From this equation we can consider two cases depending on the optical depth of the medium:
\begin{itemize}
    \item $\tau \ll 1$, hence I$_{\nu}$ = I$_{\nu}$(0), the total emission is equal to the background emission,
    \item $\tau \gg 1$ hence I$_{\nu}$ = S$_{\nu}$, the total emission is equal to the source function.
\end{itemize}
In order to compare the intensity of the observed signal with the original intensity of the emitting source in absence of the intervening ISM ($I_{\nu}(0)$), we get\,:
\begin{equation}
\label{eq:Inuobs}
I_{\nu_{obs}}(s) = I_{\nu}(s)-I_{\nu}(0) = (S_{\nu}(T)-I_{\nu}(0))(1-e^{-\tau_{\nu}}) \, .
\end{equation} 
The Source function is equivalent to the Planck function at the temperature $T_{ex}$: $S_{\nu}$ = $B_{\nu}$($T_{ex}$) (see Eq. \ref{eq:planck} and Eq. \ref{eq:Snu}). Equation \ref{eq:Inuobs} can then be rewritten as:
\begin{equation}
\label{eq:final}
I_{\nu_{obs}} = \frac{2h\nu^{3}}{c^{2}} \left[\frac{1}{e^{h\nu/kT_{ex}}-1} -\frac{1}{e^{h\nu/kT_{CMB}}-1}\right](1-e^{-\tau_{\nu}}) \, .
\end{equation} 
The radiation $I_{\nu_{obs}}$ is defined by the Planck's function at $T_b$ ($I_{\nu_{obs}}$=$B_{\nu}(T_b)$), the brightness temperature of the source (in K). 
In the Rayleigh-Jeans (RJ) limit (namely $\frac{T_0}{T}$ $\ll$ 1 where $T_0$ = $h\nu/k$), 
\begin{equation}
\label{eq:RJ}
I_{\nu} = \frac{2k\nu^2T_b}{c^2} \, .
\end{equation}
It is the custom in radioastronomy to measure the brightness of a source by its brightness temperature, $T_{b}$ (see Eq.(\ref{eq:TB}). The RJ limit stands for frequencies $\nu$ (in GHz) $\ll$ 20.84 $\times$ T (in K), and it is valid for radio emission except perhaps for the low temperatures (cold cores of about 10 K). When $\nu$ is so high that RJ does not stand, Eq. \ref{eq:RJ} can still be used but in this case but it should be understood that in this case, $T_b$ is different than the thermodynamic temperature of a blackbody. \\
The  Equation \ref{eq:final} can be expressed as \,:
\begin{equation}
\label{eq:tbv}
T_b(v) = T_0\left(\frac{1}{e^{T_0/T_{ex}}-1}-\frac{1}{e^{T_0/T_{CMB}}-1}\right)(1-e^{-\tau(v)}) \, ,
\end{equation}
where $T_{0}=h\nu/k$. One might apply the filling factor correction $\eta$ (lying between 0 and unity) as a multiplicative factor to Eq. \ref{eq:final} \citep[see][]{Ulich1976}:
\begin{equation}
\eta = \frac{\Omega_{source}}{\Omega_{observed}} \, ,
\end{equation}
where $\Omega$ is the solid angle. We assume for single dish observations, that the telescope beam Full Width Half maximum (FWHM) size is related to the diameter of the telescope by the diffraction limit:
\begin{equation}
\theta_b = 1.22 \frac{\lambda}{D} \, ,
\end{equation}
where $\theta_b$ is the angular resolution (radians), $\lambda$ is the wavelength of light, and $D$ is the diameter of the telescope. The gaussian beam is frequently used in formula deduction for single dish. The size of a gaussian beam is characterized by Half Power Beam Width of the main lobe: $\theta_b$. The solid angle of such a gaussian beam is:
\begin{equation}
\Omega_b = \int{e^{-4 \times \ln(2) \times \left(\frac{\theta^2}{\theta_{b}^2}\right)} \times 2\pi\times\theta} d(\theta) = \frac{\pi}{4\times \ln(2)} \times \theta_b^2 =~ 1.133 \times \theta_b^2  \, .
\end{equation}
Therefore, the filling factor can be expressed as:
\begin{equation}
\eta = \frac{\theta^2_{source}}{\theta^2_{source}+\theta^2_{beam}}
\end{equation}
where $\theta_{source}$ and $\theta_b$ are the circular 2D Gaussian sizes of the source (if centered in the telescope beam) and half-power telescope beam respectively. For example, with a 10$^{\prime\prime}$ source size and a 10$^{\prime\prime}$ telescope beam, we still get a factor 1/2 for the filling factor to be applied to Eq. \ref{eq:tbv}. A non-gaussian intensity distribution requires to work out more closely the relation between the telescope response (beam) and the source intensity distribution. Indeed, if not centered, the measured intensity is attenuated with respect to the intrinsic intensity. 

\begin{figure}[h]
\begin{center}
\includegraphics[width=0.65\textwidth]{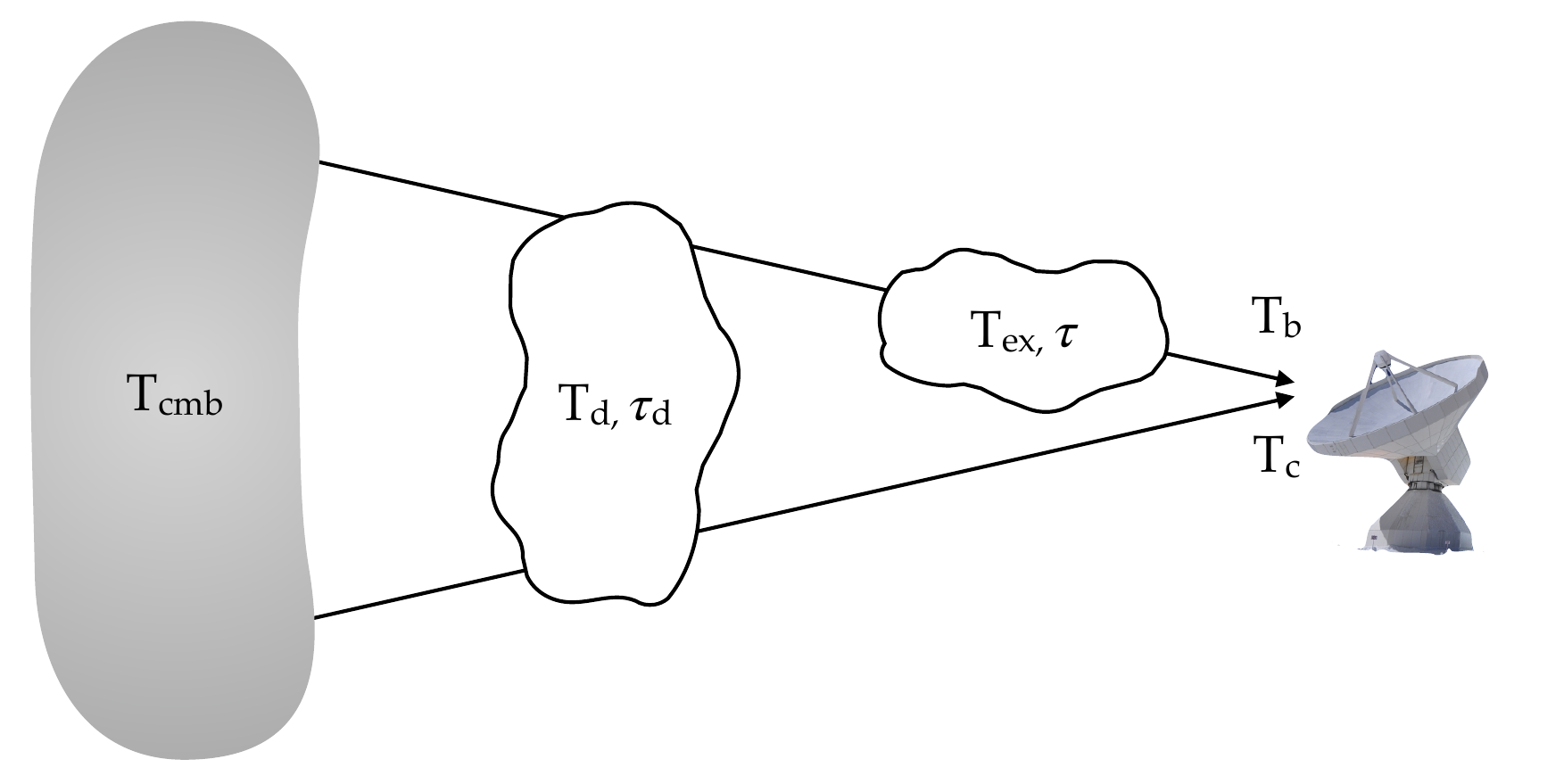}
\caption{\label{fig:sketch_cloud}{Sketch of multiple clouds along the line of sight. The brightness temperature of the cloud is $T_b$ and the one of the continuum is $T_c$.}}
\end{center}
\end{figure}

When a background continuum source ($T_{dust}$, $\tau_{dust}$) is coupled to the molecular/atomic cloud ($T_{ex}$, $\tau$) along the line of sight (see Fig. \ref{fig:sketch_cloud}), the previous equation must take into account the dust temperature and opacity, as well as the cosmic microwave background (CMB).
In an ON-OFF observation (see Sect.~\ref{subsec:mapping}), the resulting brightness temperature obtained from the telescope is\,:
\begin{equation}
T_b = J_{\nu}(T_{CMB})e^{-\tau_{dust}}e^{-\tau} + \eta_{dust}J_{\nu}(T_{dust})(1 - e^{-\tau_{dust}})e^{-\tau} + \eta J_{\nu}(T_{ex})(1 - e^{-\tau}) - J_{\nu}(T_{CMB}) \, ,
\end{equation}
where $J_{\nu}(T) =(h\nu/k) \times 1/(e^{h\nu/kT}-1)$ is the radiation temperature, $\eta_{dust}$ represents the filling factor for the continuum source and $\eta$ represents the dilution factor for the molecular/atomic cloud.
In the case where $\tau_{dust}$ = 0 and $\eta = 1$, the equation becomes\,:
\begin{equation}
T_b = ( J_{\nu}(T_{ex})-J_{\nu}(T_{CMB}))(1 - e^{-\tau}) \, .
\end{equation}
In the case where $\eta$=$\eta_{dust}$=1, then:
\begin{equation}
\label{eq:tb_full}
T_b = (J_{\nu}(T_{CMB})e^{-\tau_{dust}} + J_{\nu}(T_{dust})(1 - e^{-\tau_{dust}}))e^{-\tau} + J_{\nu}(T_{ex})(1 - e^{-\tau}) - J_{\nu}(T_{CMB}) \, .
\end{equation}
Outside the line, towards the continuum source, the continuum obtained in the ON-OFF observation is defined by:
\begin{equation}
\label{eq:tc_full}
T_c = (J_{\nu}(T_{CMB})e^{-\tau_{dust}} + J_{\nu}(T_{dust})(1 - e^{-\tau_{dust}})) - J_{\nu}(T_{CMB}) \, .
\end{equation}
Combining Eq. \ref{eq:tb_full} and \ref{eq:tc_full} give a resulting brightness temperature\,:
\begin{equation}
\label{tb_CASSIS}
\rm T_b = T_c \times e^{-\tau} + (1 - e^{-\tau})\left(J_{\nu}(T_{ex})-J_{\nu}(T_{CMB})\right) \, .
\end{equation}

\begin{itemize}
\item For emission lines:
\begin{equation}
T_b - T_c = \Delta T = (1 - e^{-\tau})\left(J_{\nu}(T_{ex}) - J_{\nu}(T_{CMB}) - T_c\right) \, .
\end{equation}
\item For absorption lines:
\begin{equation}
T_c - T_b = T_{abs} = (1 - e^{-\tau})\left(T_c - J_{\nu}(T_{ex}) + J_{\nu}(T_{CMB})\right) \, .
\end{equation}
\end{itemize}

Several components (spatially distributed, with different $V_{LSR}$ and/or temperature/density) and molecules can be modelled at the same time. For each molecules, several transitions can be modelled within the spectrum. The spectra of these components are computed separately, and then added iteratively. In the following equations, the indices i, j and k correspond to components, molecules, and lines, respectively. The spectrum is computed in a first iteration with the first component:
\begin{equation}
\label{equation_Tb}
T_{b,0} (\nu) = T_c(\nu) e^{-\sum\limits_{j,k} \tau_{0,j,k}(\nu)}  + \sum\limits_{j} \eta_{0,j} \left(1-e^{-\sum\limits_{k} \tau_{0,j,k}}\right)\left(J_{\nu}(T_{ex,0,j,k})-J_{\nu}(T_{bg})\right) \, .
\end{equation}
The other components are then added iteratively in the case of an onion-like structure:
\begin{equation}
T_{b,i=[1,N-1]} (\nu)  = T_{b,0} (\nu) e^{-\sum\limits_{j,k} \tau_{i,j,k}(\nu)}  + \sum\limits_{j} \eta_{i,j} \left(1-e^{-\sum\limits_{k} \tau_{i,j,k}}\right)\left(J_{\nu}(T_{ex,i,j,k})-J_{\nu}(T_{bg})\right) \, .
\end{equation}
The continuum emission is here assumed to be optically thin (i.e. transparent to the CMB) and spatially uniform, to fill the beam of the single-dish telescope or the synthesised beam of the interferometer. In LTE, $T_{ex,i,j,k}$ has the same value within each component ($i$) for each transition ($k$).

\subsection{Molecular column densities}
\label{sec:CD}

\subsubsection{Opacity}
\label{lte-form}
In order to derive the physical conditions of the observed medium, it is useful to measure the number of molecules per unit area along the targeted line of sight. This quantity is called the molecular column density and is the first step before measuring the molecular abundances (in LTE) or the kinetic temperature, collider density and molecular abundances (in non-LTE). We can express the column density in $i$ level, $N_{i}$ as a function of the number of molecules in the energy level $i$ ($n_{i}$: number per cm$^{-3}$):
\begin{equation}
\label{eq:N}
N_{i} = \int_0^L n_{i} ds \, ,
\end{equation}
L being the size of the source along the line-of-sight, and {\it ds} the infinitesimal element of length along the line-of-sight. The line opacity can be expressed as a function of the column density and the excitation temperature, that we assume to be constant on the line of sight. For that we can integrate $\tau$ along the line profile\,: 
\begin{align}
\label{eq:taudv}
\int{}{}\tau_{\nu} d\nu &=~\int{}{}\frac{h\nu\Phi_{\nu}}{4\pi}(B_{lu}N_{l}-B_{ul}N_{u})d\nu\\
                      & =~\frac{A_{ul}c^{3}N_{u}}{8\pi\nu^{3}}(\exp(h\nu/kT_{ex})-1) \, ,
\end{align}
Where $\Phi_{\nu}$ is the line profile with $\int{}{}\Phi(\nu)d\nu$=1.
For a gaussian line shape, we can express the opacity (Eq. \ref{eq:taunu}) as a function of the cloud's depth\,:
\begin{equation}
\tau_{ul}(z) = \frac{A_{ul}c^2}{8\pi\nu_{ul}^3\Delta \varv \sqrt{\pi}/2\sqrt{\ln2}}\int_{0}^{z} n_u\left(\frac{n_lg_u}{n_ug_l}-1\right) \, dz'  \, ,
\end{equation}
where $\Delta \varv$ (velocity units) is the full width at half maximum of the observed line. Integrating on the line of sight (see equation \ref{eq:N}), we get, at the line center\,:
\begin{equation}
\tau_{0} = \frac{g_u}{g_l}\frac{c^2}{8\pi\nu^2\Delta \nu\sqrt{\pi}/2\sqrt{\ln2}}A_{ul}N_l\left(1-e^{-h\nu/kT_{ex}}\right) \, ,
\end{equation}
where $\Delta \nu$ (frequency units) is the FWHM of the observed line, and $N_{l}$ is the column density in the lower state.\\
This equation can also be expressed as:
\begin{equation}
\label{eq:tau0}
\tau_{0} = \frac{c^2A_{ul}N_u}{8\pi\nu^2\Delta \nu\sqrt{\pi}/2\sqrt{\ln2}}\left(e^{h\nu/kT_{ex}}-1\right) \, .
\end{equation}

\begin{figure}[h]
\begin{center}
\includegraphics[width=0.75\textwidth]{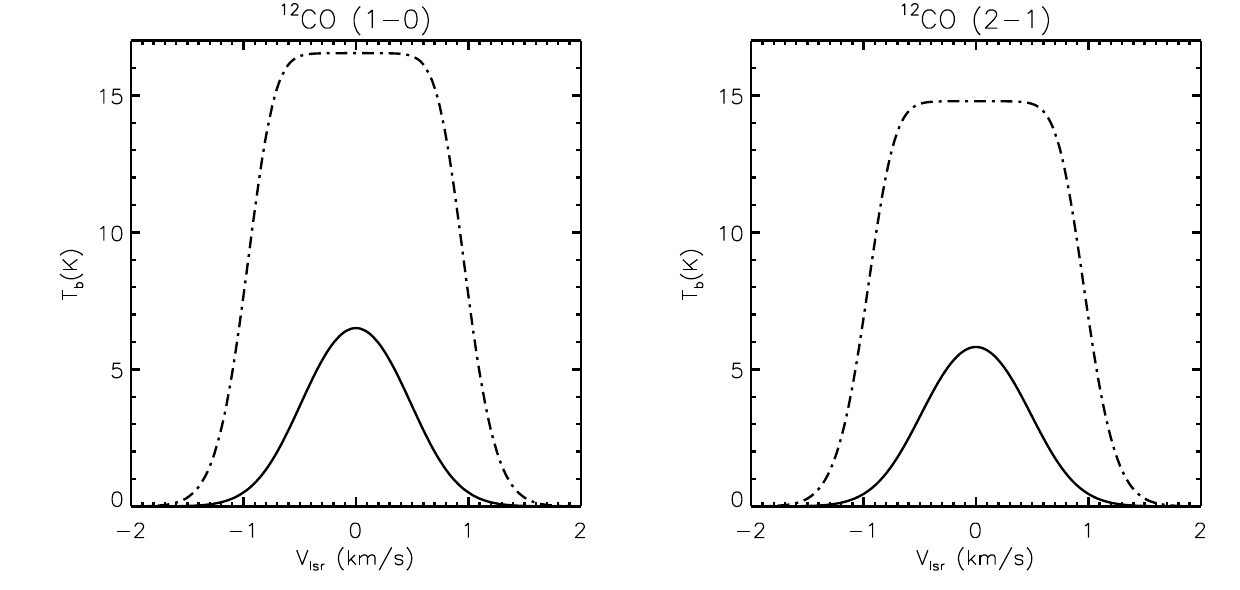}
\caption{\label{fig:coopacity}{Line profiles for the $^{12}$CO transitions 1 $\rightarrow$ 0 and 2 $\rightarrow$ 1 
at $T_{ex}$ = 20 K, $\Delta \varv$ = 1 km~s$^{-1}$ and the opacity $\tau$ = 0.5 (plain lines) and 10 (dot-dashed lines) at the line centre, using the CASSIS software.}}
\end{center}
\end{figure}

The next step is now to estimate the total column density of the observed molecular species, not just the column density in an energy level. Statistical mechanics states that, when the gas exchange energy with the ambient medium, the partition fonction Q describes the relative population of states in the gas as:
\begin{equation}
\label{eq:Q}
Q(T) = \sum_{i}{}g_{i}\exp\left(-\frac{E_{i}}{kT}\right) \, .
\end{equation}
The partition function is a function of the nuclear spin, rotational, vibrational, electronic states of the molecule: $Q=Q_{n}Q_{r}Q_{v}Q_{e}$ (see \citet{Gordy1984} and \citet{Mangum2015}). \\
The total column density N$_{tot}$ can then be computed:
\begin{equation}
N_{tot} = \sum_{n=0}^{\infty}N_{i} \, .
\end{equation}
Using equation \ref{eq:Q} we can now express the total column density: 
\begin{equation}
\label{eq:ntot}
N_{tot}=\frac{N_{lowest}Q(T_{ex})}{g_{lowest}}=\frac{N_{u}Q(T_{ex})e^{E_u/kT_{ex}}}{g_{u}}  \, ,
\end{equation}
where $Q(T_{ex})$ is the partition function for an excitation temperature $T_{ex}$ and the index \textit{lowest} represents 
the lowest level associated to the molecule and its form (ortho, para, ...). Indeed, in the case of the water molecule, the lowest energy level for the para form is 0 K and 34.23 K for ortho.  $E_u$ is the upper level of the transition compared to the ground 
level (different from zero when ortho, para or meta forms). \\
Combining Eq. \ref{eq:tau0} and Eq. \ref{eq:ntot}, we finally obtain:
\begin{equation}
N_{tot} = \frac{Q(T_{ex})e^{E_u/kT_{ex}}}{g_{u}} \times \frac{\tau_0 8\pi\left(\frac{\sqrt{\pi}}{2\sqrt{\ln(2)}}\right)\nu^2\Delta\nu}{c^2A_{ul}(e^{h\nu/kT_{ex}}-1)}
\end{equation}
where $\tau_0$ is the opacity at the line centre, derived from the maximum intensity at the line peak emission (see Eq. \ref{eq:tbv}). \\
When the collision rate between molecules is high, $T_{ex}$ $\rightarrow$ $T_{b}$ $\rightarrow$ $T_{K}$. By choosing 
a column density such that $\tau$~$\gg$~1 at the line center, the line will therefore saturate, the temperature at the line center will become constant, and will result in a non-negligible line broadening (see Eq. \ref{eq:tbv}). 
\\
Figure \ref{fig:coopacity} presents the modelled line profiles of the $^{12}$CO molecule for transitions 1 $\rightarrow$ 0 and 2 $\rightarrow$ 1 at $T_{ex}$ = 20 K, FWHM = 1 km~s$^{-1}$, $\tau$ = 10 and $\tau$ = 0.5 at the line center, assuming a gaussian profile: \\
\begin{equation}
\label{tau}
\tau(\varv)~=~\tau_{0} \exp\left(-\frac{(\varv-\varv_{0})^{2}}{2\sigma^{2}}\right)  \, ,
\end{equation}
where $\varv_{0}$ is the velocity in the local standard of rest (V$_{LSR}$), $\sigma$(km/s)~=~$\Delta \varv$ (km/s)/(2$\sqrt{2\ln2}$), 
where $\Delta \varv$ is the FWHM. Using Eq. \ref{eq:tbv} we get $T_b$~=~16.5~K for the 1 $\rightarrow$ 0 transition and $T_b$~=~14.8~K for the 2 $\rightarrow$ 1 transition. Note that in LTE, $T_{ex}$ = $T_{k}$, then in the RJ limit with $\tau$ $\gg$ 1, $T_{b}$ $\approx$ $T_{k}$ - $T_{CMB}$. Therefore T$_{b}$ measured at the peak (Fig. \ref{fig:coopacity}) gives a direct measure of T$_{k}$. In this case, at 115 GHz, $h\nu/kT$ $\ll$ 1, but not at 230 GHz, and $T_{ex}$ = $T_{k}$=16.7+2.73 $\approx$ 20 K, which is consistent with the value given as an input for the model. We can see that for low opacity the profile follows a Gaussian profile, and increasing the opacity tends to saturate the line profile, deviating from a Gaussian one. \\
The CASSIS software (\url{https://cassis.irap.omp.eu/}) developed at IRAP/France \citep{Vastel2015} for quick line identification with access to spectroscopic and collisional databases and modelling in LTE and non-LTE uses the same formalism as explained above and can be exploited to better understand radiative transfer modelling. 

In the LTE regime, proper estimation of $T_{ex}$ requires the observation of multiple transitions (many upper energy levels covering at least the temperature range of the observed source) coupled with modelling of the statistical equilibrium and radiative transfer. The gas excitation temperature varies between the background radiation temperature ($T_{CMB}$ when no background is present) at low density and the gas kinetic temperature at high density. At low densities, collisions are not the dominant excitation mechanisms and $T_{ex}$ is in equilibrium with the radiation temperature. It can be demonstrated from Eq. \ref{eq:pop-eq}, combined with Eq. \ref{eq:boltz} and Eq. \ref{eq:planck} with $B_{\nu}(T_{CMB})$. At high densities, collisions dominate in setting the level population and T$_{ex}$ is equal to  the gas kinetic temperature of the dominant collisional partner. It can be demonstrated from Eq. \ref{eq:pop}, neglecting the radiative terms:
\begin{equation}
n_u(A_{u\ell} + B_{u\ell}\bar{J} + C_{u\ell}) = n_\ell(B_{\ell u}\bar{J} + C_{\ell u}) \, .
\end{equation}
Therefore, 
\begin{equation}
n_uC_{u\ell} = n_{\ell}C_{\ell u}\, .
\end{equation}
Using Eq. \ref{eq:gamma} and \ref{eq:boltz}, the excitation temperature tends to reach the kinetic temperature and the transition is {\it thermalised} (i.e LTE). We say that the line is sub-thermally excited when $T_{ex}$ is less than $T_k$. Note that some transitions of the same molecule can be thermalised (for example the CO 1--0 transition) while higher energy levels are sub-thermally excited. \\

We can now introduce a new parameter, called the {\it critical density} ($n_{cr}$), which has traditionally been used as a measure of the density at which a particular transition is excited and is observed at radio wavelengths. The definition of the critical density is unfortunately not consistent throughout the literature. Some definitions only consider the two energy levels involved in the transition (2-level approximation):
\begin{equation}
\label{eq:ncr}
n_{cr} (u\ell) = \frac{A_{u\ell}}{\gamma_{u\ell}} \, ,
\end{equation}
where $\gamma_{u\ell}$ depends on the kinetic temperature. It defines the density of the gas required for the collisions to dominate over the radiative processes. Other definitions use the multi-level nature of collisions to sum over all collisions out of the upper energy level or only from the upper energy level to lower energy levels \citep[see for example][]{Shirley2015}. \\
From Eq. \ref{eq:ncr}, the critical density is defined for each transition and is proportional to $\nu^3$, so the higher the upper energy is, the higher the frequency, therefore the Einstein coefficient and the critical density. As an example, using the LAMDA catalog, the critical density of HC$_3$N J = 9--8 at 81.88 GHz is $n_{cr} = [2.8-7.0] \times 10^5$ cm$^{-3}$ (E$_{u}$ = 19.6 K, A$_{9-8}$ = 4.2 $\times$ 10$^{-5}$ s$^{-1}$, $\gamma_{9-8}$ (10 - 300 K)=[5.97 $\times$ 10$^{-11}$ -- 1.51 $\times$ 10$^{-10}$]~cm$^3$~s$^{-1}$). At 336.52 GHz, the critical density of HC$_3$N J = 37--36 is $n_{cr} = [8.4-27.7] \times 10^6$ cm$^{-3}$ (E$_{u}$ = 306.9 K, A$_{37-36}$ = 3.05 $\times$ 10$^{-3}$ s$^{-1}$, $\gamma_{37-36}$ (10 - 300 K)=[1.11 $\times$ 10$^{-10}$ -- 3.62 $\times$ 10$^{-10}$]~cm$^3$~s$^{-1}$). Therefore, the higher-frequency transitions are more readily sub-thermally excited. In contrast, the CO 1--0 and 2--1 transitions have much lower critical densities but are easily observed from the ground and detected (high Einstein coefficients) in the ISM, as the second most abundant molecule. These CO transitions are therefore good tracers of molecular gas in our Galaxy and beyond \citep[see e.g.][]{Liszt1998,Dame2001,Goldsmith2008}. However it should be noted that the use of CO has some caveats: 1) the low-J CO lines become easily optically thick (the 1--0 transition becomes optically thick beyond 10$^{16}$ cm$^{-2}$, which corresponds to a few visual extinction \citep{Liszt1998} so that they cannot directly probe the highest density regions; 2) at low column densities, CO is rapidly photodissociated by the interstellar radiation field \citep{vanDishoeck1988}. That is why there have been a lot of effort to detect higher density tracers such as HCO$^+$, N$_2$H$^+$, HCN, HC$_3$N, NH$_3$, although with much weaker line intensities compared to CO or even CS. Note that the high-J CO lines detected in the submm/infrared may be optically thin. \\

Going back to the Fig. \ref{fig:coopacity} line profiles, several mechanisms can broaden spectral lines. The dynamical structure of the source, if unresolved, may contribute to such broach lines such as outflows, collapsing envelopes, stellar winds, etc...Also, individual atoms in a gaseous medium are in random, chaotic motion: the hotter the gas, the faster the random thermal motions of the atoms. When 
a photon is emitted by an atom in motion, the frequency of the detected photon is changed by the Doppler effect. The 
photon is then not recorded at the precise frequency predicted by atomic physics but rather at a slightly shifted. 
Throughout the whole cloud, atoms move in every possible direction, resulting in a broadening of the line. 
From \citet{Fuller1992}\,:
\begin{equation}
(\Delta \varv)^{2}~=~8\ln(2) \frac{kT}{m} \, ,
\end{equation}
where $\Delta \varv$ is the FWHM, $T$ is temperature in the gas, and $m$ is the mass of the atom (or molecule). For example, in the case of the D$_{2}$H$^{+}$ molecule, at a temperature of 8 K, the thermal linewidth should be 0.27 km~s$^{-1}$. Note that turbulence can also result in the broadening of a spectral line.\\

By defining a column density and an excitation temperature, we get from \ref{eq:taudv} $\int \tau d\varv$. Then, using a linewidth (before broadening due to optical depth), we get $\tau(\varv)$ (as a function of velocity), which has a Gaussian profile (Eq. \ref{tau}). Then from \ref{eq:tbv} we get $T_{b}$ as a function of velocity. Different line profiles can therefore be used.

\subsubsection{LTE Rotational diagram analysis}
\label{rotd}

This analysis refers to a plot of the column density per statistical weight of a number of molecular energy levels, as a function of their energy above the ground state (see \citet{Goldsmith1999}). In LTE, this corresponds to a Boltzmann distribution, so a plot of the natural 
logarithm of $N_{u}$/g$_{u}$ versus $E_{u}$/k yields a straight line with a slope of 1/$T_{rot}$. The temperature inferred is called the rotation temperature.\\
We can rewrite Eq. \ref{eq:tau0} as:
\begin{equation}
\label{eq:tau}
\tau = \frac{c^3 A_{ul}N_{u}}{8\pi\nu^3\Delta \varv \sqrt{\pi}/2\sqrt{\ln2}}[e^{h\nu/kT_{rot}}-1] \, .
\end{equation}
Neglecting $J_{\nu}(T_{CMB})$ from Eq. \ref{eq:tbv} compared to $J_{\nu}(T_{rot})$, which means $T_{CMB}$ $\ll$ $T_{rot}$, we can express the main beam temperature as:
\begin{equation}
\label{eq:tbreduced}
T_{b} = \frac{h\nu}{k}\frac{1}{\exp(h\nu/kT_{rot})-1} \frac{1-e^{-\tau}}{\tau}\times \tau  \, .
\end{equation}
Therefore, we can compute the column density in the upper state as:
\begin{equation}
N_{u} = \int T_{b}dv \times \frac{8\pi k\nu^2}{hc^3A_{u\ell}} \times \frac{\tau}{1-e^{-\tau}}  \, .
\end{equation}
\begin{equation}
N_{u} = W \times \frac{8\pi k\nu^2}{hc^3 A_{u\ell}} \times C_{\tau}  \, ,
\end{equation}
where W in the integrated area and $C_{\tau}$ is the optical depth correction factor. When the line is optically thin, $C_{\tau}$ is equal to unity. \\
For a molecule in LTE, all excitation temperatures are the same, and the population of each level is given by:
\begin{equation}
N_{u}=\frac{N_{tot}}{Q(T_{rot})}g_{u}e^{-E_u/kT_{rot}} \, .
\end{equation}
We can rewrite this equation to obtain:
\begin{equation}
\label{eq:rot}
\ln\frac{N_u}{g_u} = \ln \frac{N_{tot}}{Q(T_{rot})} -\frac{E_u}{kT_{rot}} \, .
\end{equation}

\noindent A rotational diagram can be useful to determine whether the emission is optically thick or thin, whether the level populations are described by LTE, and to determine what temperature describe the population distribution in the event that LTE applies.
Equation \ref{eq:rot}  can be written in terms of the observed integrated area W (K~km~s$^{-1}$):
\begin{equation}
\ln\frac{8\pi k\nu^2 W}{hc^3A_{u\ell}g_u} = \ln \frac{N_{tot}}{Q(T_{rot})} - \ln C_{\tau} -\frac{E_u}{kT_{rot}} \, .
\end{equation}

\noindent If we do not take into account the $C_{\tau}$ factor, each of the upper level populations would be underestimated 
by a factor $C_{\tau}$, different for each transition. Therefore, the ordinate of the rotation diagram would be 
below its correct value by the factor $\ln C_{\tau}$. A change in the temperature for lines of different excitation 
might indicate that the source has different temperature components or that the lines considered are not 
optically thin and cannot be easily used to obtain a meaningful excitation temperature.\\
Note that the error bars should be taken into account for the order 1 polynomial fit, in order to obtain a reliable 
value for the uncertainty on the rotational temperature as well as the total column density. The uncertainty of the integrated area is computed though the following formula: 
\begin{equation}
\Delta W = \sqrt{(cal/100 \times W)^2 + (rms \sqrt{2 \times FWHM \times \delta \varv})^2} \, ,
\end{equation}
where cal is the instrumental calibration uncertainty ($\%$), W is the integrated area (in K~km~s$^{-1}$), rms is the noise around the selected species (in K), FWHM is expressed in km~s$^{-1}$ and $\delta \varv$ is the bin size (in km~s$^{-1}$).  We assumed that the number of channels in the line is $2 \times FWHM/\delta \varv$
For undetected transitions we estimated the upper limit of 3$\sigma$:
\begin{equation}
\label{eq: upper_limit}
W \ (\mathrm{K~km~s^{-1}}) \le  3(rms \times FWHM) \times \sqrt{(\rm 2\times\alpha)^2 + (2\times\delta V/FWHM)} \, .
\end{equation}

\noindent Therefore, the plotted uncertainties are simply:
\begin{equation}
\Delta \left( \ln\frac{N_u}{g_u} \right) = \frac{\Delta W}{W} \, .
\end{equation}

\noindent Now, how do we estimate the uncertainty on the values of $T_{rot}$ and $N_{tot}$? \\
From the fitted straight line (y = ax+b) the slope a is related to the rotational excitation temperature as  $T_{rot}$ = -1/a. 
Then $\Delta T_{rot} = \Delta a/a^2$. The intercept b is related to the total column density as $N_{tot}$ = Q(rot) $\times$ e$^b$. 
Therefore $\Delta N_{tot} = Q(rot) \times \Delta b \times e^b$.\\

\begin{figure}[h]
\begin{center}
\includegraphics[width=0.85\textwidth]{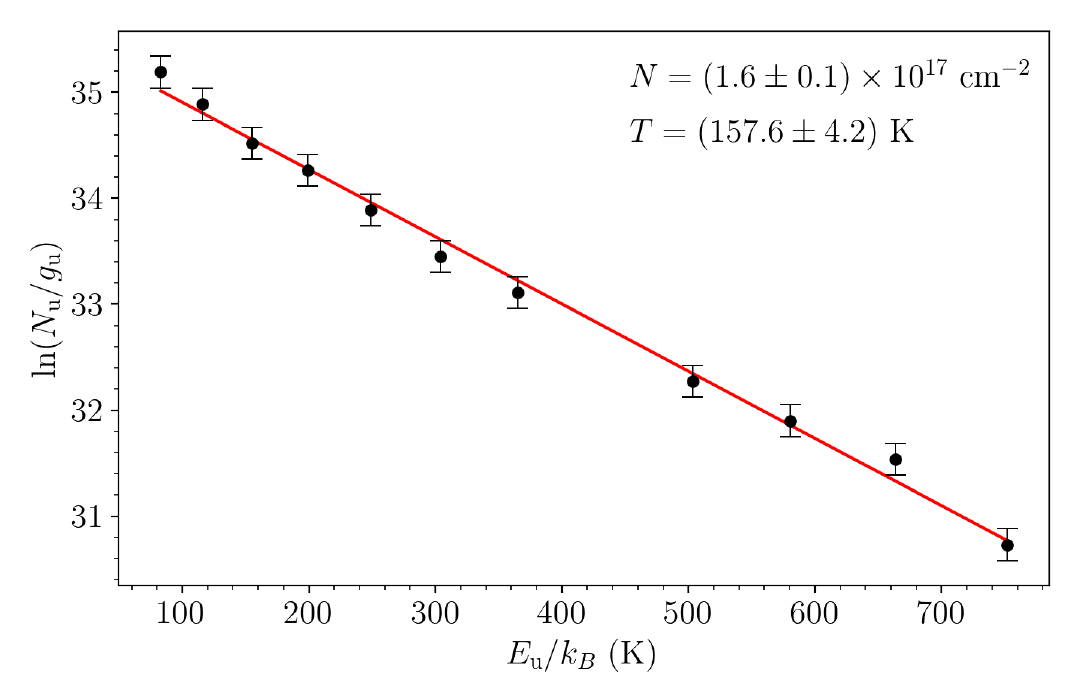}
\caption{\label{fig:ccs}{Rotational diagram analysis for the CO detected transitions using Herschel/HIFI towards the Orion Bar using the CASSIS software. The opacity correction has been applied. The CO column density and rotational temperature are quoted in the upper right corner. No beam dilution has been taken into account.}}
\end{center}
\end{figure}

\noindent We can iteratively apply the $C_{\tau}$ correction to the rotational diagram analysis until a solution for $T_{rot}$ and $N_{tot}$ 
has converged (when the last result has not changed by a small value). For the first iteration we use Equation \ref{eq:rot} and obtain values 
for the transitions opacity. In the second iteration we add the $C_{\tau}$ correction to the linear equation: 
\begin{equation}
\ln\frac{N_u}{g_u} = \ln \frac{N_{tot}}{Q(T_{rot})} -\frac{E_u}{kT_{rot}} - \ln C_{\tau} \, .
\end{equation}
The iterations go on until a convergence has been obtained.\\
As Goldsmith and Langer (1999) nicely said, this method requires quite a large number of transitions spread over a range of upper state energies.\\
Please note that between Eq. \ref{eq:tau} and Eq. \ref{eq:tbreduced}, $J_{\nu}(T_{CMB})$ has been neglected from Eq. \ref{eq:tbv} compared to $J_{\nu}(T_{rot})$. Therefore the above analysis does not stand when $T_{CMB}$ is not negligible compared to T$_{rot}$.\\
An example is presented in Fig. \ref{fig:ccs} using the CASSIS software.

\subsubsection{Non-LTE formalism}

When the LTE conditions are not fulfilled, the $C_{u\ell}$ and $C_{\ell u}$ collisional coefficients cannot be neglected and Eq. \ref{eq:pop} must be solved. Some simplifications must be done. The problem is how to {\it decouple} the radiative transfer calculations from the calculations of the level populations. A popular approach for this is the so-called escape probability method (described by \citet{Sobolev1960}). A factor ($\beta$) that determines the probability that a photon at some position in the cloud can escape the system is introduced in the equations \ref{eq:pop}: $\bar{J_{\nu}} = S_{\nu}(1 - \beta)$. 
Indeed, locally, the number of photons used for absorptions ($(n_\ell B_{\ell u}-n_uB_{u\ell})\bar{J}$) is equal to the number of photons available for local absorption and therefore not escaping the cloud ($n_u(1-\beta(\tau_{u\ell}))A_{u\ell}$). Now the statistical equilibrium equations can take a much  easier form:
\begin{equation}
\label{eq:nonlte}
\frac{dn_u}{dt} = n_\ell C_{\ell u} - n_uC_{u\ell}- \beta n_uA_{u\ell}     \, .
\end{equation}
So now we can solve the level populations and the radiation field separately as they are now decoupled. We can then estimate the escape probability value. \\

\noindent A first expression of $\beta$ has been derived for an expanding spherical sphere by \citet{Sobolev1960} and also applies to moderate velocity gradients. It is called the Large Velocity Gradient (LVG) approximation (also called Sobolev) in which \citet{Castor1970} and \citet{Elitzur1992} (Chapter 2) derived:
\begin{equation}
\beta = \frac{1-e^{-\tau}}{\tau} \, .
\end{equation}

\noindent For a uniform sphere, \citet{Osterbrock2006} derived:
\begin{equation}
\beta = \frac{1.5}{\tau}\left[1-\frac{2}{\tau^2} -\left(\frac{2}{\tau}+\frac{2}{\tau^2}\right)e^{-\tau}\right] \, .
\end{equation}
For a homogeneous slab geometry, also applicable to shocks, \citet{Dejong1975} derived:
\begin{equation}
\beta = \frac{1-e^{-3\tau}}{3\tau} \, .
\end{equation}

\noindent When the gas becomes optically thick ($\tau \gg 1$), the probability for a photon to escape the medium is considerably reduced, because of the {\it trapping} of emitted photons. In this case, the effective rate of spontaneous emission has to be reduced by the number of photons leaving the system: $A_{u\ell}^{eff}$ = $A_{u\ell}\beta(\tau)$. Hence the critical density is now described by:
\begin{equation}
\label{eq:ncr_eff}
    n_{cr}^{eff}(u\ell) = \frac{A_{u\ell}\beta(\tau)}{\gamma_{u\ell}}\, .
\end{equation}
This situation leads to more easily thermalised molecular levels since the $n_{cr}$ leading to thermalisation is reduced ($\beta(\tau)<1$). The resolution of the LVG method is quite similar to the LTE method except that the term $A_{ul}$ is replaced by  $A_{ul}\beta(\tau)$ and the term $\bar{J_{\nu}}$ is replaced by $S_{\nu}(1 - \beta)$ in the equations. We can write:
\begin{equation}
    n_\ell n_{collider}\gamma_{\ell u} = n_u(n_{collider}\gamma_{u\ell} + \beta A_{u\ell})\, .
\end{equation}
Combined with Eq. \ref{eq:gamma} we obtain:
\begin{equation}
    \frac{n_u}{n_\ell} = \frac{g_u}{g_\ell}\frac{e^{-h\nu_0/kT_k}}{\frac{\beta A_{u\ell}}{n_{collider}\gamma_{u\ell}} + 1} = \frac{g_u}{g_\ell}\frac{e^{-h\nu_0/kT_k}}{\frac{n_{cr}^{eff}}{n_{collider}} + 1} \, .
\end{equation}
Taking into account the Boltzmann equation (Eq. \ref{eq:boltz}) we obtain:
\begin{equation}
\label{eq:tex_lvg}
    T_{ex} = \frac{T_k}{1 + \frac{kT_k}{h\nu_0}\ln\left(\frac{n_{cr}^{eff}}{n_{collider}} + 1\right)}\, .
\end{equation}
If $n \gg n_{cr}^{eff}$, hence $T_{ex}$=$T_k$ and the line is thermalised (LTE case).\\

Equation \ref{eq:nonlte} can then be solved for each level assuming equilibrium, computing $T_{ex}$ for each level (Eq. \ref{eq:tex_lvg}), then the opacity (Eq. \ref{eq:tau0}) then the integrated line intensity (from Eq. \ref{eq:tbv}). As a first guess, we consider the level populations in the optically thin case and we solve $n_i$. Then we compute $\tau$ and then $\beta$, which we re-inject into the equilibrium equations to solve $n_i$ (and therefore $N_i$) and $T_{ex}$ for each transition. We can then iterate the procedure and stop when the values do not change. The unknown parameters are therefore the kinetic temperature of the medium, the number density of the collider, the column density of the molecular species and the width of the transitions (assumed to be same for all the transitions of the same species). These can be constrained when a few transitions of the same species have been detected. Large modelling grids can be computed and a $\chi^2$ minimisation can be used to match the integrated intensities of the observed transitions:\\ 
\begin{equation}
  \chi^2 = \sum_{i=1}^{N}\frac{(W_i^{obs}-W_i^{mod})^2}{(cal/100 \times W_i^{obs})^2 + (rms \sqrt{2 \times fwhm \times \delta \varv})^2}
\end{equation}
where $N$ is the number of observed lines, $W_i^{obs}$ is the observed integrated line intensity, $W_i^{mod}$ is the modelled integrated line intensity such as provided with RADEX, cal is the instrumental calibration uncertainty ($\%$), rms is the noise around the selected transitions (in K), FWHM is expressed in km~s$^{-1}$ and $\delta \varv$ is the bin size (in km~s$^{-1}$).

\subsection{Abundances}
\label{sec:AB}

Molecules can be used as probes of the physical (kinetic temperature and H$_2$ density as well as motions such as collapse or rotation) and chemical (abundances) information on the gas. In order to properly determine the abundances of the detected species, it is mandatory to determine the H$_2$ column density. 
In molecular clouds, H$_2$ produced on the dust grain surfaces and ejected in the gas-phase cannot be directly traced. Indeed, it is a homonuclear linear molecule with no permanent dipole moment, and all of the low-lying energy levels are quadrupole transitions with small transition probabilities ($A_{ul}$ values) and relatively high excitation energies. These transitions are therefore only excited at high temperatures or in strong UV radiation fields (i.e., fluorescence). The generally high energies of the first excited states of H$_2$ mean that we expect negligible H$_2$ emission unless we are looking at
unusually warm (500--1000K) gas in proximity to hot stars or in regions of active star formation. Consequently the most abundant molecule in the ISM, carrying most of the mass and playing a key role for the thermal balance and gas-phase chemistry of the ISM, is virtually invisible to direct observation. In this section, we review the methodology leading to  the determination of the abundances of the observed molecular species.\\

\subsubsection{\texorpdfstring{H$_2$ column density from CO observations}{}}

Cold molecular clouds are primarily traced by the second most abundant molecule, CO, which is asymmetric. The first rotational lines are the most commonly observed transitions at 115, 230 and 345 GHz, with the first lying only $\sim$ 5 K above ground, a relatively low effective density ($\sim$10$^{2-3}$ cm$^{-3}$) and a wavelength (3 mm) which is readily observable from the ground. It has therefore historically been one of the most commonly used tracers of physical conditions in the molecular ISM. A CO-to-H$_2$ conversion factor (also called the X factor) can be established based on the 115 GHz line intensity ($T_{mb}(CO)$):
\begin{equation}
N(H_2) \sim X(CO) \times T_{mb}(CO)
\end{equation}
In the galactic molecular clouds X(CO) $\sim$ $2 \times 10^{20}$ cm$^{-2}$~(K.km/s)$^{-1}$, but this value is dependent on the metallicity \citep{Wilson1995,Boselli2002,Bolatto2013} with a factor of 2–20 lower towards starburst galaxies \citep{Downes2003}. The exact value of the conversion factor between CO integrated line intensity and mass, X, is however a matter of some dispute.\\
An alternative estimate of N(H$_2$) using $^{13}$CO (or even C$^{18}$O) emission requires several steps and assumptions. One can assume optically thin $^{13}$CO emission or derive the $^{13}$CO opacity ($\tau_{13}$), using the T$_{mb}$ ($^{12}$CO) which is optically thick, as a measure of the $^{13}$CO excitation temperature (see Sec. \ref{lte-form} and discussion of Fig. \ref{fig:coopacity}) and then correct the optically thin column density value by the factor $\rm \tau_{13}(V)dV/[1-\exp(\tau_{13}(V))]dV$. 
We can rewrite Eq. \ref{eq:final} into:
\begin{equation}
\tau_{\nu} = -\ln\left[1-\frac{T_{mb}}{J_{\nu}(T_{ex})-J_{\nu}(T_{CMB})}\right]
\end{equation}
and insert this equation in Eq. \ref{eq:ntot}.
In the case of optically thin line (cf Eq. \ref{eq:tbv}):
\begin{equation}
\label{eq:tbvthin}
T_b(v) = [J_{\nu}(T_{ex})-J_{\nu}(T_{CMB})] \times \tau(v) \, .
\end{equation}
Then:
\begin{equation}
N_{tot} = N_{tot}^{thin} \times \frac{\tau}{1-\exp(-\tau)}
\end{equation}
where the fraction corresponds to the {\it optical depth correction factor} (see Sec. \ref{rotd}). 

Note that a constant T$_{ex}$ is assumed to estimate the fractional population in all J levels of $^{13}$CO. In the low-density regime of molecular clouds, the J = 1--0 T$_{ex}$ of $^{13}$CO is often smaller than the value determined from $^{12}$CO. By adopting the excitation temperature of $^{12}$CO, the resultant $^{13}$CO column density is therefore underestimated. Finally, to derive the H$_2$ column density, an isotopic ratio of $^{12}$CO/$^{13}$CO and H$_2$/$^{12}$CO abundance ratio are assumed. Based on H$_2$ and CO IR absorption lines, the H$_2$/CO abundance ratio ranges from 4,000 to 7,000 \citep{Lacy1994,Kulesa2002}. However, both isotopic and CO abundance ratios can vary within clouds and from cloud to cloud owing to isotopic fractionation and local UV fields. For example, at the center of dense cores, the [CO]/][H$_2$] ratio is expected to be reduced by up to five orders of magnitude \citep{Bergin1997}. This so-called {\it depletion} is dependent on the the temperature, the density and the timescale. Other tools must therefore be used for the determination of H$_2$ in the cold and dense regions of the ISM. 

\subsubsection{\texorpdfstring{H$_2$ column density from dust measurements}{}}

Dust grains are made up of metals such as carbon and silicon (with a mass fraction in metals is 1$\%$), so a (more or less) constant gas-to-dust ratio is expected in the ISM. The observations of the dust column density are therefore often used to estimate the total H$_2$ column density in the gas phase. At millimetric wavelengths, in the Rayleigh-Jeans domain, dust emission depends linearly on temperature, and its great advantage is its optical thinness. \\
The observed flux density F$_{\nu}$ (Jy~beam$^{-1}$) is approximated with a modified black body curve. For optically thin emission:
\begin{equation}
\label{eq:fnu_thin}
F_{\nu} = B_{\nu}(T_{dust})(1-e^{-\tau}) = B_{\nu}(T_{dust}) \times \tau_{\nu}
\end{equation}
In the (sub)millimetre regime (emission of cold dust), the dust emission is rarely optically
thick, except possibly at high resolution or towards high-mass star forming regions.
Having F$_{\nu}$ we can deduce $\tau$. 
The dust opacity is defined as:
\begin{equation}
\tau(\nu) = \rho_{dust} \times \kappa(\nu) \times L 
\end{equation}
where $\rho_{dust}$ is the mass density (g~cm$^{-3}$), $\kappa(\nu)$ is the mass absorption coefficient in cm$^2$~g$^{-1}$, L the thickness of structure along the line-of-sight. The so-called dust opacity, $\kappa(\nu)$, which expresses the effective surface area for extinction per unit mass, depends on the chemical composition and structure of dust grains, but not the size for particles. However, $\kappa$ will be modified in the dense regions of the ISM, with an increase by a factor of 2--3, leading to a higher uncertainty of the mass and abundances determinations \citep{Juvela2015}. The dust opacity is usually described as a power law $\kappa(\nu)$ = $\kappa_0(\nu/\nu_0)^{\beta}$ \citep{Hildebrand1983,Compiegne2011}, where $\kappa_0$ is the emission cross-section at a reference frequency $\nu_0$. The fit is possible when observations consist of at least three wavelengths covering frequencies before and after the maximum value of the intensity. For example, for cold cores, observations at long wavelengths (beyond 200 $\mu$m) are needed to constrain the spectral index, while shorter wavelengths are better for determining colour temperature. For isothermal clouds in the millimeter wavelength range, the spectral index value can be derived using the ratio of the surface brightness at for example 1.2 and 3 mm:
\begin{equation}
\label{eq:beta}
\beta = \frac{log(I_{1.2mm}/I_{3mm})-log(B_{1.2mm}(T_{dust})/B_{3mm}(T_{dust}))}{log(\nu_{1.2mm}/\nu_{3mm})}
\end{equation}

So the fit of the modified blackbody involves three free parameters: the spectral index ($\beta$), the colour temperature (T), the intensity (I$_0$) at a reference frequency $\nu_0$ and B$_{\nu}$ is the Planck function.  


We are interested in the H$_2$ column density:
\begin{equation}
\label{eq:nh2_dust}
N(H_2) = \int n_{H_2} ds =\int \frac{N_{b}(H_2)}{Volume} ds = \int \frac{Mass}{Volume \times \mu(H_2)m_H} ds = \int \frac{\rho_{gas}}{\mu(H_2)m_H} ds = \frac{\tau(\nu)}{\mu(H_2)m_H\kappa(\nu)}
\end{equation}

where m$_H$ is the hydrogen atom mass, $N_{b}(H_2)$ is the number of H$_2$ molecules,  $\mu(H_2)$ is the total mass (Mass) relative to the H$_2$ molecule ($N_{b}(H_2) \times m_H = N_{b}(H)/2 \times m_H$ as $N_{b}(H)=2N_{b}(H_{2})$ in the cold regions of the ISM). As hydrogen represents 71\% of the total mass of metals in the ISM we can approximate $\mu(H_2)$ as 2.8 ($2M(H)/0.71/M(H)$).\\
Note that in the above equation, an assumption on the dust opacity has been made, as the gas-to-dust mass ratio of 0.1 in our Galaxy \citep{Beckwith1990}, has been taken into account:
\begin{equation}
\label{eq:kappa}
\kappa(\nu) = 0.1(\nu/1000 GHz)^{\beta} \, ,
\end{equation}
assuming a value for the spectral index $\beta$ (for example, 1.8 is appropriate for dense regions \citep{Juvela2015}).\\
In conclusion, when you measure a flux density at frequency $\nu$ you can deduce $\tau$ from Eq. \ref{eq:fnu_thin}. You can now determine the H$_2$ column density using Eq. \ref{eq:nh2_dust}, after computing $\beta$ (Eq. \ref{eq:beta}) or assuming a value, to determine $\kappa_{\nu}$ (Eq. \ref{eq:kappa}). The beam averaged column density can be expressed as:
\begin{equation}
    N(H_2) = \frac{F(\nu)}{\Omega\mu(H_2)m_H\kappa(\nu)B_{\nu}(T_{dust})}
\end{equation}
$\Omega$ being the beam solid angle. We can then express the above equation in useful units:
\begin{equation}
    N(H_2) = 2.02 \times 10^{20} ( cm^{-2})\left(e^{1.439/[\lambda(mm)(T/10K)]}-1\right)(\lambda(mm))^3\left(\frac{0.1}{\kappa_{\nu} (cm^2~g^{-1})}\right)\left(\frac{10}{\theta_b(^{\prime\prime})}\right)^2\left(F(\nu) (mJy~beam^{-1}) \right)
\end{equation}

\subsubsection{Comparison with chemical models}

\begin{figure}
	\centering
	\includegraphics[width=6.8cm]{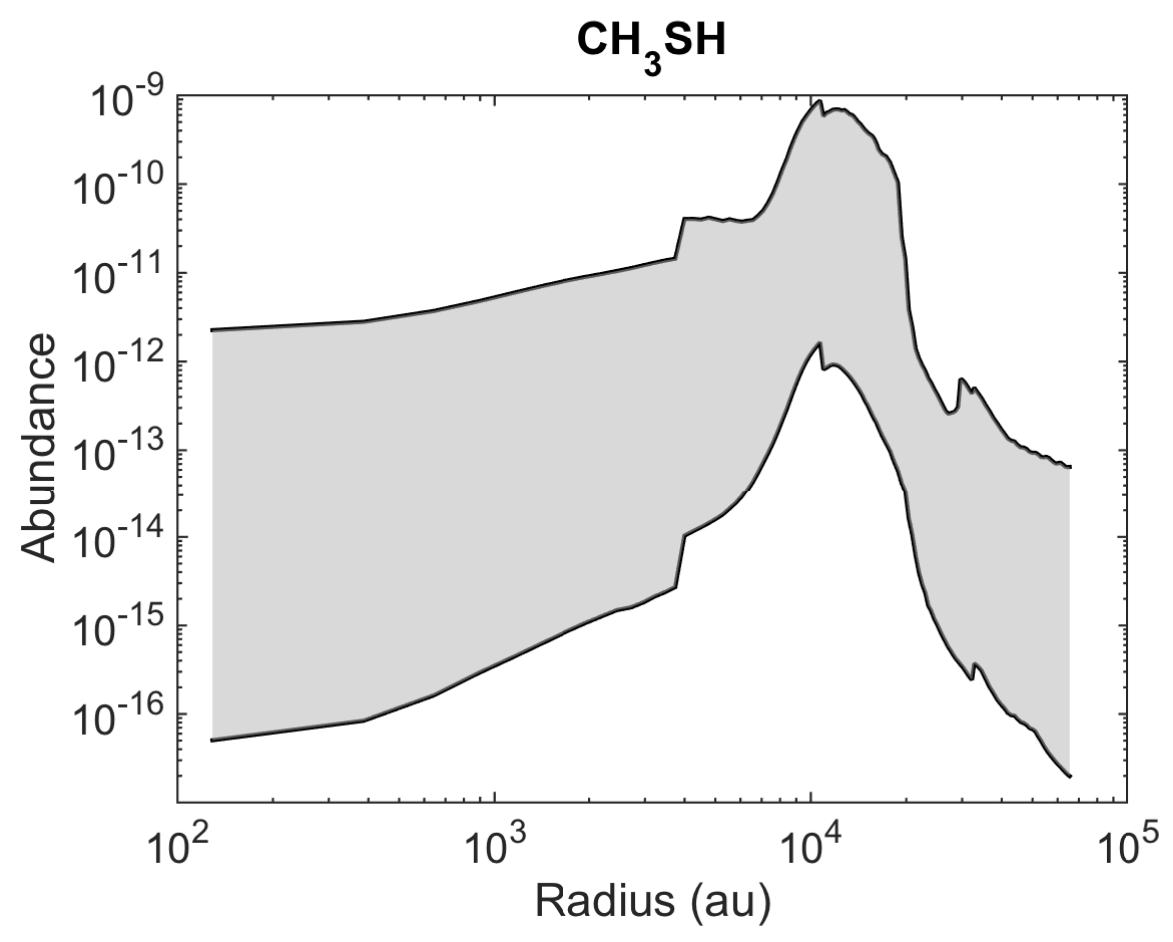}
	\includegraphics[width=6.8cm]{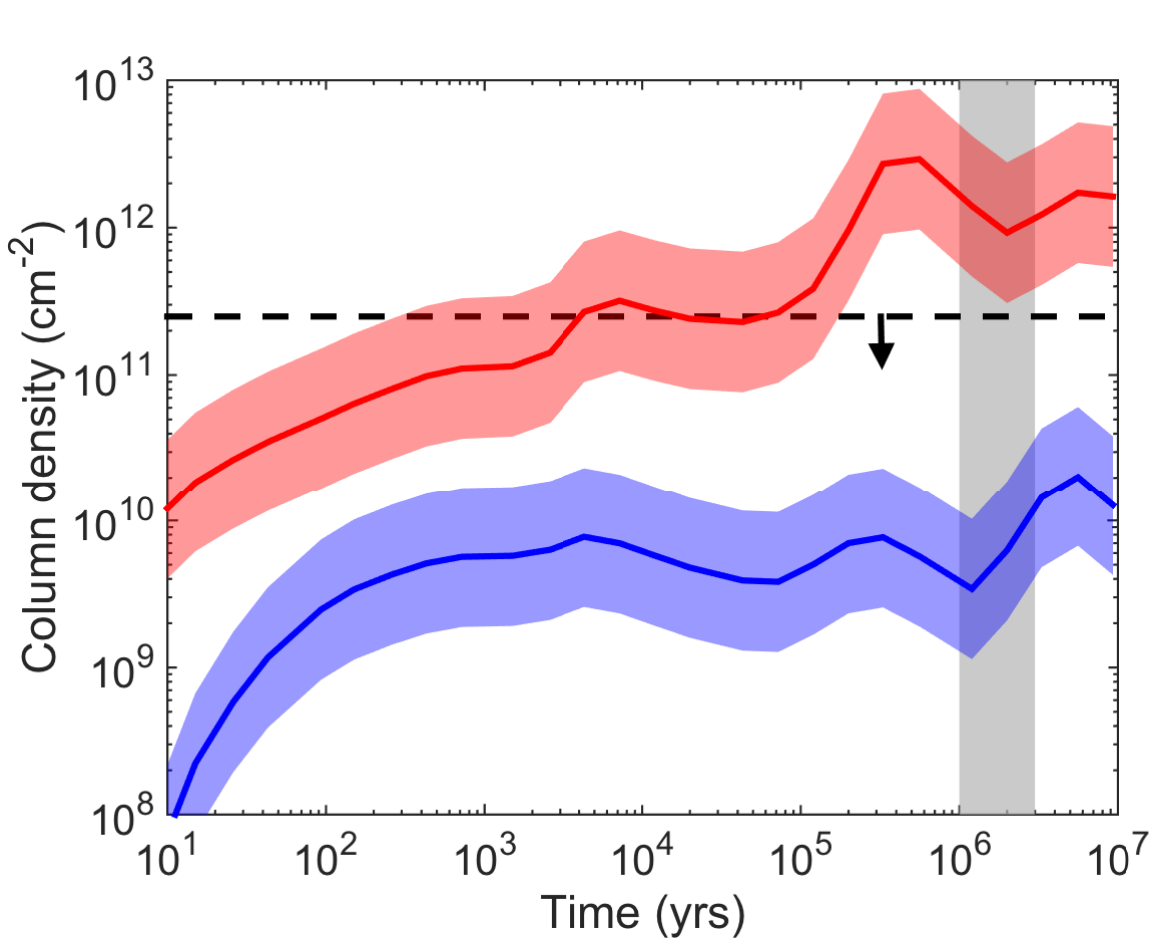}
	\caption{Results from a chemical model using the KIDA (\url{https://kida.astrochem-tools.org/}) database and the  NAUTILUS (\url{https://kida.astrochem-tools.org/codes.html}) chemical code.}
	\label{ch3sh_model}
\end{figure}

Modellings and simulations of the chemistry, in which the abundances are calculated based on the rates of their formation and destruction, can now be compared to the observations. Since the calculated abundances are function of time as well as the initial conditions of the modelled source, the modelling can provide information about the history of the source. A lot of efforts have been made during the past 10 years to feed the chemical databases based on the astronomical discoveries, but it should be emphasised that models are still imperfects as our knowledge must be improved based on laboratory work and theoretical calculations. \\
To easily compare the abundances as a function of time, derived from the chemical models, with the observations, we first need to calculate the column density of the modelled molecular species along the radius of the observed object. We therefore must convert the modelled abundance $[X]$ (with respect to H) of a species $X$ into column densities $N(X)$. The most simple example for a typical prestellar core with a 1D symmetry gives, for a beam smaller than the core:
\begin{equation}
\label{eq_Ncol}
	N(X) = 2\,\times\,\sum_{i=2}^{n}\left(\frac{n(\textrm{H})_i[X]_i + n(\textrm{H})_{i-1}[X]_{i-1}}{2}\right)\,\times\,(R_{i-1} - R_{i})
\end{equation}
where $R$ is the radius from the centre and $i$ the position in the grid along the line of sight ($i=1$ being the outermost position) composed of $n$ shells. $n(\textrm{H})_i$ is the gas density at radial point $i$, $[X]_i$ the abundance of the species. Different column densities are derived depending on the observed position towards the core.
The different column densities obtained using Eq. (\ref{eq_Ncol}) must then be convolved with the beam size of the telescope used at the frequency of the observations. The H$_2$ column density can also be derived using the same method if the density profile of the source is known.\\
We therefore obtain, from the chemical models, a variation in time of the column density that can be compared with the observed column density to constrain the age using as many species as possible. For that, in order to find the {\it best-fit} model, we can use the distance of disagreement computation, applied on the column density, which is computed as follows:
\begin{equation}
D(t) = \frac{1}{n_{obs}} \sum_{i}|log(N(X))_{obs,i}-log(N(X))_{i}(t)|
\end{equation}
where $N(X)_{obs,i}$ is the observed column density, $N(X)_{i}$(t) is the modelled column density at a specific age and n$_{obs}$ is the total number of observed species considered in this computation. The distances can be compared with many models with different initial conditions (atomic abundances, temperature, density, etc...) to obtain the lowest value and therefore the {\it best-fit} age and initial conditions. 
Fig. \ref{ch3sh_model} shows an example of the radial distribution of the modelled CH$_3$SH for an age between 10$^6$ et 3 $\times$ 10$^6$ years already constrained using many species (gray vertical area), for different models (red: non-sulphur depletion and blue: sulphur depletion in the initial atomic abundances) for the prestellar core L1544. The dashed line correspond to the column density computed by \citet{Vastel2018}. 

\section{Most used molecular tracers in star-forming regions}

Molecular lines of simple species give us the opportunity of identifying different types of objects in molecular clouds.
The best tracers of diffuse ($\rm \sim 10^2-10^3~cm^{-3}$) and extended molecular clouds where star-forming cores are embedded are low-excitation lines (e.g. $J=1-0$ and $2-1$) of CO and its isotopologues, owing to the high abundance of CO, to the relatively low critical density ($\rm \sim 10^3~cm^{-3}$) of these transitions, and to the fact that CO is not frozen on to dust grains at such low densities.
In denser ($\rm \geq 10^3~cm^{-3}$) star-forming regions, the most appropriate tracers depend both on the evolutionary stage of the cores and on the physical conditions inside them. 
In this section, we briefly summarise some of the most commonly used tracers of molecular cores hosting star formation. For a detailed discussion, we refer to \citet{Williams2014} and \citet{Ceccarelli2023}.

{\bf Pre-stellar cores:} Pre-stellar cores represent the earliest stage of the formation process of stars and planets. In such stage, the core is characterised by a low temperature ($T\sim 10$~K), and a density profile increasing towards the centre, with an inner nucleus (a few thousand au) with an almost flat density profile. In the so-called molecular zone, the best gas-phase tracers are CO and their optically thin less abundant isotopologues ($^{13}$CO and C$^{18}$O) in the lower density envelope ($\rm \sim 10^3-10^4~cm^{-3}$, 7000--15000 au from the centre of the core), as well as HCO$^+$, formed from the reaction between CO and H$_3^+$. Complex organic molecules (starting with methanol) are also detected in this external layer due to non-thermal desorption processes. In the so-called depletion zone ($\sim 2000-7000$ au), density increases up to $\rm 10^5~cm^{-3}$\ and N-bearing species such as NH$_3$ and N$_2$H$^+$ appear to maintain high abundances where CO molecules are mainly in the solid phase due to freeze-out.
The CO freeze-out also boosts the formation of deuterated molecules, such as N$_2$D$^+$ and NH$_2$D, which are also good tracers of this intermediate inner region. 
In the central high-density nucleus ($\rm \geq 10^6~cm^{-3}$, $\sim$ 2000 au), an almost complete freeze-out is expected where even NH$_{2}$D starts to deplete, and a few molecular ions (H$_3^+$, H$_2$D$^+$, D$_2$H$^+$ and possibly D$_3^+$) should remain in the gas. 

{\bf Protostellar cores:} The gravitational collapse of a rotating dense core forms a protostar at core centre surrounded by an infalling envelope and a rotating disc. 
The protostar heats up the environment, triggering desorption of dust grains and endothermic reactions. The gas and dust temperatures decrease with increasing distance from the central protostar, going from $\sim 100$~K in the inner 100 au to $\sim 10$~K in the outer envelope, extended up to 5000--10000 au.
The chemical composition changes accordingly, depending on the desorption energies of the species on ice mantles, and on the endothermicity of the gas-phase reactions.
In the innermost, warmer regions, the chemistry is dominated by the sublimation of CO at $\sim$ 20 K (and its hydrogenated species up to methanol), CH$_4$ at $\sim$ 30 K, and H$_2$O at $\sim$ 100 K from grain mantles, which increase the formation of complex organic molecules and carbon-chain molecules. Among the simple species, good tracers of infall motions are the low-$J$ transitions of HCO$^+$, CS, and HNC, as well as the inversion transitions of NH$_3$.
Finally, in the outer zone, the same chemical composition as that of the natal molecular cloud is expected.

{\bf Outflows:} In the protostellar stage, collimated outflows are observed to emerge from the protostellar core, with velocities attaining $\rm \sim 100~km~s^{-1}$. Ro-vibrational transitions of H$_2$, especially at the head of these jets where the high-velocity material impacts violently the quiescent gas, are detected in the infrared. The shock due to the passage of the outflow can destroy both the ice mantles and the refractory cores of dust grains (the so-called sputtering mechanism for the former and shattering for the latter).
Molecules efficiently formed on ice mantles, such as H$_2$O and OH, are indeed greatly enhanced in abundance.
Disruption of the refractory core causes the release of refractory elements in the gas, so that many other simple species containing refractory elements, are substantially enhanced in abundance along the shocked gas, such as SiO, SiS, H$_2$S, SO, and SO$_2$. 
Phosphorus-bearing species such as PN and PO are also suggested to be excellent outflow tracers, but it is not yet clear whether the phosphorus main carrier is in the refractory core or in the ice mantles of the grains \citep[e.g.][]{Fontani2024}. Complex organic molecules are also good tracers of shocks in outflowing gas.


\bibliographystyle{elsarticle-harv} 
\bibliography{cas-refs.bib}

@ARTICLE{Adams1941,
       author = {{Adams}, Walter S.},
        title = "{Some Results with the COUD{\'E} Spectrograph of the Mount Wilson Observatory.}",
      journal = {\apj},
         year = 1941,
        month = jan,
       volume = {93},
        pages = {11},
          doi = {10.1086/144237},
       adsurl = {https://ui.adsabs.harvard.edu/abs/1941ApJ....93...11A}
}

@ARTICLE{Agundez2007,
       author = {{Ag{\'u}ndez}, Marcelino and {Cernicharo}, Jos{\'e} and {Gu{\'e}lin}, Michel},
        title = "{Discovery of Phosphaethyne (HCP) in Space: Phosphorus Chemistry in Circumstellar Envelopes}",
      journal = {\apjl},
         year = 2007,
        month = jun,
       volume = {662},
       number = {2},
        pages = {L91-L94},
          doi = {10.1086/519561},
       adsurl = {https://ui.adsabs.harvard.edu/abs/2007ApJ...662L..91A}
}

@ARTICLE{Agundez2008,
       author = {{Ag{\'u}ndez}, M. and {Cernicharo}, J. and {Pardo}, J.~R. and {Gu{\'e}lin}, M. and {Phillips}, T.~G.},
        title = "{Tentative detection of phosphine in IRC +10216}",
      journal = {\aap},
         year = 2008,
        month = jul,
       volume = {485},
       number = {3},
        pages = {L33-L36},
          doi = {10.1051/0004-6361:200810193},
archivePrefix = {arXiv},
       eprint = {0805.4297},
 primaryClass = {astro-ph},
       adsurl = {https://ui.adsabs.harvard.edu/abs/2008A&A...485L..33A}
}

@ARTICLE{Agundez2010,
       author = {{Ag{\'u}ndez}, M. and {Cernicharo}, J. and {Gu{\'e}lin}, M. and et al.},
        title = "{Astronomical identification of CN$^{-}$, the smallest observed molecular anion}",
      journal = {\aap},
         year = 2010,
        month = jul,
       volume = {517},
          eid = {L2},
        pages = {L2},
          doi = {10.1051/0004-6361/201015186},
archivePrefix = {arXiv},
       eprint = {1007.0662},
 primaryClass = {astro-ph.GA},
       adsurl = {https://ui.adsabs.harvard.edu/abs/2010A&A...517L...2A}
}

@ARTICLE{Agundez2014,
       author = {{Ag{\'u}ndez}, Marcelino and {Cernicharo}, Jos{\'e} and {Gu{\'e}lin}, Michel},
        title = "{New molecules in IRC +10216: confirmation of C$_{5}$S and tentative identification of MgCCH, NCCP, and SiH$_{3}$CN}",
      journal = {\aap},
         year = 2014,
        month = oct,
       volume = {570},
          eid = {A45},
        pages = {A45},
          doi = {10.1051/0004-6361/201424542},
archivePrefix = {arXiv},
       eprint = {1408.6306},
 primaryClass = {astro-ph.GA},
       adsurl = {https://ui.adsabs.harvard.edu/abs/2014A&A...570A..45A}
}

@ARTICLE{Agundez2015a,
       author = {{Ag{\'u}ndez}, Marcelino and {Cernicharo}, Jos{\'e} and {Gu{\'e}lin}, Michel},
        title = "{Discovery of interstellar ketenyl (HCCO), a surprisingly abundant radical}",
      journal = {\aap},
         year = 2015,
        month = may,
       volume = {577},
          eid = {L5},
        pages = {L5},
          doi = {10.1051/0004-6361/201526317},
archivePrefix = {arXiv},
       eprint = {1504.05721},
 primaryClass = {astro-ph.GA},
       adsurl = {https://ui.adsabs.harvard.edu/abs/2015A&A...577L...5A}
}

@ARTICLE{Agundez2015b,
       author = {{Ag{\'u}ndez}, M. and {Cernicharo}, J. and {de Vicente}, P. and et al.},
        title = "{Probing non-polar interstellar molecules through their protonated form: Detection of protonated cyanogen (NCCNH$^{+}$)}",
      journal = {\aap},
         year = 2015,
        month = jul,
       volume = {579},
          eid = {L10},
        pages = {L10},
          doi = {10.1051/0004-6361/201526650},
archivePrefix = {arXiv},
       eprint = {1506.07043},
 primaryClass = {astro-ph.GA},
       adsurl = {https://ui.adsabs.harvard.edu/abs/2015A&A...579L..10A}
}

@ARTICLE{Agundez2018,
       author = {{Ag{\'u}ndez}, M. and {Marcelino}, N. and {Cernicharo}, J. and {Tafalla}, M.},
        title = "{Detection of interstellar HCS and its metastable isomer HSC: new pieces in the puzzle of sulfur chemistry}",
      journal = {\aap},
         year = 2018,
        month = mar,
       volume = {611},
          eid = {L1},
        pages = {L1},
          doi = {10.1051/0004-6361/201832743},
archivePrefix = {arXiv},
       eprint = {1802.09401},
 primaryClass = {astro-ph.GA},
       adsurl = {https://ui.adsabs.harvard.edu/abs/2018A&A...611L...1A}
}

@ARTICLE{Agundez2018a,
       author = {{Ag{\'u}ndez}, M. and {Marcelino}, N. and {Cernicharo}, J.},
        title = "{Discovery of Interstellar Isocyanogen (CNCN): Further Evidence that Dicyanopolyynes Are Abundant in Space}",
      journal = {\apjl},
         year = 2018,
        month = jul,
       volume = {861},
       number = {2},
          eid = {L22},
        pages = {L22},
          doi = {10.3847/2041-8213/aad089},
archivePrefix = {arXiv},
       eprint = {1806.10328},
 primaryClass = {astro-ph.GA},
       adsurl = {https://ui.adsabs.harvard.edu/abs/2018ApJ...861L..22A}
}

@ARTICLE{Aladro2015,
       author = {{Aladro}, R. and {Mart{\'\i}n}, S. and {Riquelme}, D. and {Henkel}, C. and {Mauersberger}, R. and {Mart{\'\i}n-Pintado}, J. and {Wei{\ss}}, A. and {Lefevre}, C. and {Kramer}, C. and {Requena-Torres}, M.~A. and {Armijos-Abenda{\~n}o}, R.~J.},
        title = "{Lambda = 3 mm line survey of nearby active galaxies}",
      journal = {\aap},
         year = 2015,
        month = jul,
       volume = {579},
          eid = {A101},
        pages = {A101},
          doi = {10.1051/0004-6361/201424918},
archivePrefix = {arXiv},
       eprint = {1504.03743},
 primaryClass = {astro-ph.GA},
       adsurl = {https://ui.adsabs.harvard.edu/abs/2015A&A...579A.101A}
}

@ARTICLE{Anderson2014,
       author = {{Anderson}, J.~K. and {Ziurys}, L.~M.},
        title = "{Detection of CCN (x $^{2}${\ensuremath{\pi}}$_{r}$) in IRC+10216: Constraining Carbon-chain Chemistry}",
      journal = {\apjl},
         year = 2014,
        month = nov,
       volume = {795},
       number = {1},
          eid = {L1},
        pages = {L1},
          doi = {10.1088/2041-8205/795/1/L1},
       adsurl = {https://ui.adsabs.harvard.edu/abs/2014ApJ...795L...1A}
}

@ARTICLE{Apponi1999,
       author = {{Apponi}, A.~J. and {McCarthy}, M.~C. and {Gottlieb}, C.~A. and {Thaddeus}, P.},
        title = "{Astronomical Detection of Rhomboidal SiC$_{3}$}",
      journal = {\apjl},
         year = 1999,
        month = may,
       volume = {516},
       number = {2},
        pages = {L103-L106},
          doi = {10.1086/311998},
       adsurl = {https://ui.adsabs.harvard.edu/abs/1999ApJ...516L.103A}
}

@ARTICLE{Barlow2013,
       author = {{Barlow}, M.~J. and {Swinyard}, B.~M. and {Owen}, P.~J. and et al.},
        title = "{Detection of a Noble Gas Molecular Ion, $^{36}$ArH$^{+}$, in the Crab Nebula}",
      journal = {Science},
         year = 2013,
        month = dec,
       volume = {342},
       number = {6164},
        pages = {1343-1345},
          doi = {10.1126/science.1243582},
archivePrefix = {arXiv},
       eprint = {1312.4843},
 primaryClass = {astro-ph.GA},
       adsurl = {https://ui.adsabs.harvard.edu/abs/2013Sci...342.1343B}
}

@ARTICLE{Beckwith1990,
       author = {{Beckwith}, Steven V.~W. and {Sargent}, Anneila I. and {Chini}, Rolf S. and {Guesten}, Rolf},
        title = "{A Survey for Circumstellar Disks around Young Stellar Objects}",
      journal = {\aj},
         year = 1990,
        month = mar,
       volume = {99},
        pages = {924},
          doi = {10.1086/115385},
       adsurl = {https://ui.adsabs.harvard.edu/abs/1990AJ.....99..924B}
}

@ARTICLE{Bergin1997,
       author = {{Bergin}, E.~A. and {Langer}, W.~D.},
        title = "{Chemical Evolution in Preprotostellar and Protostellar Cores}",
      journal = {\apj},
         year = 1997,
        month = sep,
       volume = {486},
       number = {1},
        pages = {316-328},
          doi = {10.1086/304510},
       adsurl = {https://ui.adsabs.harvard.edu/abs/1997ApJ...486..316B}
}

@ARTICLE{Bergman2011,
       author = {{Bergman}, P. and {Parise}, B. and {Liseau}, R. and {Larsson}, B. and {Olofsson}, H. and {Menten}, K.~M. and {G{\"u}sten}, R.},
        title = "{Detection of interstellar hydrogen peroxide}",
      journal = {\aap},
         year = 2011,
        month = jul,
       volume = {531},
          eid = {L8},
        pages = {L8},
          doi = {10.1051/0004-6361/201117170},
archivePrefix = {arXiv},
       eprint = {1105.5799},
 primaryClass = {astro-ph.GA},
       adsurl = {https://ui.adsabs.harvard.edu/abs/2011A&A...531L...8B}
}

@ARTICLE{Bernath1989,
       author = {{Bernath}, Peter F. and {Hinkle}, Kenneth H. and {Keady}, John J.},
        title = "{Detection of C5 in the circumstellar shell of IRC +10216.}",
      journal = {Science},
         year = 1989,
        month = may,
       volume = {244},
        pages = {562-564},
          doi = {10.1126/science.244.4904.562},
       adsurl = {https://ui.adsabs.harvard.edu/abs/1989Sci...244..562B}
}

@ARTICLE{Berne2014,
       author = {{Bern{\'e}}, O. and {Marcelino}, N. and {Cernicharo}, J.},
        title = "{IRAM 30 m Large Scale Survey of $^{12}$CO(2-1) and $^{13}$CO(2-1) Emission in the Orion Molecular Cloud}",
      journal = {\apj},
         year = 2014,
        month = nov,
       volume = {795},
       number = {1},
          eid = {13},
        pages = {13},
          doi = {10.1088/0004-637X/795/1/13},
archivePrefix = {arXiv},
       eprint = {1408.2999},
 primaryClass = {astro-ph.GA},
       adsurl = {https://ui.adsabs.harvard.edu/abs/2014ApJ...795...13B}
}

@ARTICLE{Berne2023,
       author = {{Bern{\'e}}, O. and {Martin-Drumel}, M.A. and {Schroetter}, I. et al.},
        title = "{Formation of the Methyl Cation by Photochemistry in a Protoplanetary Disk}",
      journal = {Nature},
         year = 2023,
        month = Jun,
          doi = {10.1038/s41586-023-06307-x},
}

@ARTICLE{Blake1985,
       author = {{Blake}, G.~A. and {Keene}, J. and {Phillips}, T.~G.},
        title = "{Chlorine in dense interstellar clouds : the abundance of HCl in OMC-1.}",
      journal = {\apj},
         year = 1985,
        month = aug,
       volume = {295},
        pages = {501-506},
          doi = {10.1086/163394},
       adsurl = {https://ui.adsabs.harvard.edu/abs/1985ApJ...295..501B}
}

@ARTICLE{Bogey1984,
       author = {{Bogey}, M. and {Demuynck}, C. and {Destombes}, J.~L.},
        title = "{Laboratory detection of the protonated carbon dioxide by submillimeter wave spectroscopy}",
      journal = {\aap},
         year = 1984,
        month = sep,
       volume = {138},
       number = {1},
        pages = {L11},
       adsurl = {https://ui.adsabs.harvard.edu/abs/1984A&A...138L..11B}
}

@ARTICLE{Bolatto2013,
       author = {{Bolatto}, Alberto D. and {Wolfire}, Mark and {Leroy}, Adam K.},
        title = "{The CO-to-H$_{2}$ Conversion Factor}",
      journal = {\araa},
         year = 2013,
        month = aug,
       volume = {51},
       number = {1},
        pages = {207-268},
          doi = {10.1146/annurev-astro-082812-140944},
archivePrefix = {arXiv},
       eprint = {1301.3498},
 primaryClass = {astro-ph.GA},
       adsurl = {https://ui.adsabs.harvard.edu/abs/2013ARA&A..51..207B}
}

@ARTICLE{Boselli2002,
       author = {{Boselli}, Alessandro and {Lequeux}, James and {Gavazzi}, Giuseppe},
        title = "{The CO to H$_{2}$ Conversion Factor in Normal Late-Type Galaxies}",
      journal = {\apss},
         year = 2002,
        month = jul,
       volume = {281},
       number = {1},
        pages = {127-128},
          doi = {10.1023/A:1019599512232},
       adsurl = {https://ui.adsabs.harvard.edu/abs/2002Ap&SS.281..127B}
}

@ARTICLE{Brazier1983,
       author = {{Brazier}, C.~R. and {Brown}, J.~M.},
        title = "{The microwave spectrum of the CH free radical}",
      journal = {\jcp},
         year = 1983,
        month = feb,
       volume = {78},
       number = {3},
        pages = {1608-1610},
          doi = {10.1063/1.444853},
       adsurl = {https://ui.adsabs.harvard.edu/abs/1983JChPh..78.1608B}
}

@ARTICLE{Bruenken2009,
       author = {{Br{\"u}nken}, S. and {Gottlieb}, C.~A. and {McCarthy}, M.~C. and {Thaddeus}, P.},
        title = "{Laboratory Detection of HOCN and Tentative Identification in Sgr B2}",
      journal = {\apj},
         year = 2009,
        month = may,
       volume = {697},
       number = {1},
        pages = {880-885},
          doi = {10.1088/0004-637X/697/1/880},
       adsurl = {https://ui.adsabs.harvard.edu/abs/2009ApJ...697..880B}
}

@ARTICLE{Buhl1970,
       author = {{Buhl}, D. and {Snyder}, L.~E.},
        title = "{Unidentified Interstellar Microwave Line}",
      journal = {\nat},
         year = 1970,
        month = oct,
       volume = {228},
       number = {5268},
        pages = {267-269},
          doi = {10.1038/228267a0},
       adsurl = {https://ui.adsabs.harvard.edu/abs/1970Natur.228..267B}
}

@ARTICLE{Cabezas2013,
       author = {{Cabezas}, C. and {Cernicharo}, J. and {Alonso}, J.~L. and {Ag{\'u}ndez}, M. and {Mata}, S. and {Gu{\'e}lin}, M. and {Pe{\~n}a}, I.},
        title = "{Laboratory and Astronomical Discovery of HydroMagnesium Isocyanide}",
      journal = {\apj},
         year = 2013,
        month = oct,
       volume = {775},
       number = {2},
          eid = {133},
        pages = {133},
          doi = {10.1088/0004-637X/775/2/133},
archivePrefix = {arXiv},
       eprint = {1309.0371},
 primaryClass = {astro-ph.GA},
       adsurl = {https://ui.adsabs.harvard.edu/abs/2013ApJ...775..133C}
}

@ARTICLE{Cabezas2021,
       author = {{Cabezas}, C. and {Ag{\'u}ndez}, M. and {Marcelino}, N. and {Tercero}, B. and {Cuadrado}, S. and {Cernicharo}, J.},
        title = "{Interstellar detection of the simplest aminocarbyne H$_{2}$NC: an ignored but abundant molecule}",
      journal = {\aap},
         year = 2021,
        month = oct,
       volume = {654},
          eid = {A45},
        pages = {A45},
          doi = {10.1051/0004-6361/202141491},
archivePrefix = {arXiv},
       eprint = {2107.08389},
 primaryClass = {astro-ph.GA},
       adsurl = {https://ui.adsabs.harvard.edu/abs/2021A&A...654A..45C}
}

@ARTICLE{Cabezas2022,
       author = {{Cabezas}, C. and {Ag{\'u}ndez}, M. and {Marcelino}, N. and {Tercero}, B. and {Endo}, Y. and {Fuentetaja}, R. and {Pardo}, J.~R. and {de Vicente}, P. and {Cernicharo}, J.},
        title = "{Discovery of the elusive thioketenylium, HCCS$^{+}$, in TMC-1}",
      journal = {\aap},
         year = 2022,
        month = jan,
       volume = {657},
          eid = {L4},
        pages = {L4},
          doi = {10.1051/0004-6361/202142815},
archivePrefix = {arXiv},
       eprint = {2112.11855},
 primaryClass = {astro-ph.GA},
       adsurl = {https://ui.adsabs.harvard.edu/abs/2022A&A...657L...4C}
}

@ARTICLE{Cabezas2023,
       author = {{Cabezas}, C. and {Pardo}, J.~R. and {Ag{\'u}ndez}, M. and {Tercero}, B. and {Marcelino}, N. and {Endo}, Y. and {de Vicente}, P. and {Gu{\'e}lin}, M. and {Cernicharo}, J.},
        title = "{Discovery of two metallic cyanoacetylides in IRC +10216: HMgCCCN and NaCCCN}",
      journal = {\aap},
         year = 2023,
        month = apr,
       volume = {672},
          eid = {L12},
        pages = {L12},
          doi = {10.1051/0004-6361/202346462},
archivePrefix = {arXiv},
       eprint = {2304.01066},
 primaryClass = {astro-ph.GA},
       adsurl = {https://ui.adsabs.harvard.edu/abs/2023A&A...672L..12C}
}

@ARTICLE{Cabezas2024,
       author = {{Cabezas}, C. and {Ag{\'u}ndez}, M. and {Endo}, Y. and {Tercero}, B. and {Marcelino}, N. and {de Vicente}, P. and {Cernicharo}, J.},
        title = "{Discovery of the interstellar cyanoacetylene radical cation HC$_{3}$N$^{+}$}",
      journal = {\aap},
         year = 2024,
        month = jul,
       volume = {687},
          eid = {L22},
        pages = {L22},
          doi = {10.1051/0004-6361/202451081},
archivePrefix = {arXiv},
       eprint = {2407.02121},
 primaryClass = {astro-ph.GA},
       adsurl = {https://ui.adsabs.harvard.edu/abs/2024A&A...687L..22C}
}

@ARTICLE{Carruthers1970,
       author = {{Carruthers}, George R.},
        title = "{Rocket Observation of Interstellar Molecular Hydrogen}",
      journal = {\apjl},
         year = 1970,
        month = aug,
       volume = {161},
        pages = {L81},
          doi = {10.1086/180575},
       adsurl = {https://ui.adsabs.harvard.edu/abs/1970ApJ...161L..81C}
}

@ARTICLE{Castor1970,
       author = {{Castor}, J.~I.},
        title = "{Spectral line formation in Wolf-Rayet envelopes.}",
      journal = {\mnras},
         year = 1970,
        month = jan,
       volume = {149},
        pages = {111},
          doi = {10.1093/mnras/149.2.111},
       adsurl = {https://ui.adsabs.harvard.edu/abs/1970MNRAS.149..111C}
}

@INPROCEEDINGS{Ceccarelli2023,
       author = {{Ceccarelli}, C. and {Codella}, C. and {Balucani}, N. and {Bockelee-Morvan}, D. and {Herbst}, E. and {Vastel}, C. and {Caselli}, P. and {Favre}, C. and {Lefloch}, B. and {Oberg}, K. and {Yamamoto}, S.},
        title = "{Organic Chemistry in the First Phases of Solar-Type Protostars}",
    booktitle = {Protostars and Planets VII},
         year = 2023,
       editor = {{Inutsuka}, S. and {Aikawa}, Y. and {Muto}, T. and {Tomida}, K. and {Tamura}, M.},
       series = {Astronomical Society of the Pacific Conference Series},
       volume = {534},
        month = jul,
        pages = {379},
       adsurl = {https://ui.adsabs.harvard.edu/abs/2023ASPC..534..379C}
}

@ARTICLE{Cernicharo1987,
       author = {{Cernicharo}, J. and {Guelin}, M. and {Menten}, K.~M. and {Walmsley}, C.~M.},
        title = "{C6-H : astronomical study of its fine and hyperfine structure.}",
      journal = {\aap},
         year = 1987,
        month = jul,
       volume = {181},
        pages = {L1-L4},
       adsurl = {https://ui.adsabs.harvard.edu/abs/1987A&A...181L...1C}
}

@ARTICLE{Cernicharo1987b,
       author = {{Cernicharo}, J. and {Guelin}, M.},
        title = "{Metals in IRC +10216 : detection of NaCl, AlCl, and KCl, and tentative detection of alf.}",
      journal = {\aap},
         year = 1987,
        month = sep,
       volume = {183},
        pages = {L10-L12},
       adsurl = {https://ui.adsabs.harvard.edu/abs/1987A&A...183L..10C}
}

@ARTICLE{Cernicharo1989,
       author = {{Cernicharo}, J. and {Gottlieb}, C.~A. and {Guelin}, M. and {Thaddeus}, P. and {Vrtilek}, J.~M.},
        title = "{Astronomical and Laboratory Detection of the SiC Radical}",
      journal = {\apjl},
         year = 1989,
        month = jun,
       volume = {341},
        pages = {L25},
          doi = {10.1086/185449},
       adsurl = {https://ui.adsabs.harvard.edu/abs/1989ApJ...341L..25C}
}

@ARTICLE{Cernicharo1997,
       author = {{Cernicharo}, J. and {Liu}, X. -W. and {Gonz{\'a}lez-Alfonso}, E. and et al.},
        title = "{Discovery of Far-Infrared Pure Rotational Transitions of CH$^{+}$ in NGC 7027}",
      journal = {\apjl},
         year = 1997,
        month = jul,
       volume = {483},
       number = {1},
        pages = {L65-L68},
          doi = {10.1086/310729},
       adsurl = {https://ui.adsabs.harvard.edu/abs/1997ApJ...483L..65C}
}

@ARTICLE{Cernicharo2007,
       author = {{Cernicharo}, J. and {Gu{\'e}lin}, M. and {Ag{\'u}ndez}, M. and et al.},
        title = "{Astronomical detection of C$_4H^-$, the second interstellar anion}",
      journal = {\aap},
         year = 2007,
        month = may,
       volume = {467},
       number = {2},
        pages = {L37-L40},
          doi = {10.1051/0004-6361:20077415},
       adsurl = {https://ui.adsabs.harvard.edu/abs/2007A&A...467L..37C}
}

@ARTICLE{Cernicharo2012,
       author = {{Cernicharo}, J. and {Marcelino}, N. and {Roueff}, E. and et al.},
        title = "{Discovery of the Methoxy Radical, CH$_{3}$O, toward B1: Dust Grain and Gas-phase Chemistry in Cold Dark Clouds}",
      journal = {\apjl},
         year = 2012,
        month = nov,
       volume = {759},
       number = {2},
          eid = {L43},
        pages = {L43},
          doi = {10.1088/2041-8205/759/2/L43},
       adsurl = {https://ui.adsabs.harvard.edu/abs/2012ApJ...759L..43C}
}

@ARTICLE{Cernicharo2015,
       author = {{Cernicharo}, J. and {McCarthy}, M.~C. and {Gottlieb}, C.~A. and et al.},
        title = "{Discovery of SiCSi in IRC+10216: A Missing Link between Gas and Dust Carriers of Si\&ndashC Bonds}",
      journal = {\apjl},
         year = 2015,
        month = jun,
       volume = {806},
       number = {1},
          eid = {L3},
        pages = {L3},
          doi = {10.1088/2041-8205/806/1/L3},
archivePrefix = {arXiv},
       eprint = {1505.01633},
 primaryClass = {astro-ph.GA},
       adsurl = {https://ui.adsabs.harvard.edu/abs/2015ApJ...806L...3C}
}

@ARTICLE{Cernicharo2018,
       author = {{Cernicharo}, J. and {Lefloch}, B. and {Ag{\'u}ndez}, M. and et al.},
        title = "{Discovery of the Ubiquitous Cation NS$^{+}$ in Space Confirmed by Laboratory Spectroscopy}",
      journal = {\apjl},
         year = 2018,
        month = feb,
       volume = {853},
       number = {2},
          eid = {L22},
        pages = {L22},
          doi = {10.3847/2041-8213/aaa83a},
archivePrefix = {arXiv},
       eprint = {1801.05559},
 primaryClass = {astro-ph.GA},
       adsurl = {https://ui.adsabs.harvard.edu/abs/2018ApJ...853L..22C}
}

@ARTICLE{Cernicharo2019,
       author = {{Cernicharo}, J. and {Velilla-Prieto}, L. and {Ag{\'u}ndez}, M. and et al.},
        title = "{Discovery of the first Ca-bearing molecule in space: CaNC}",
      journal = {\aap},
         year = 2019,
        month = jul,
       volume = {627},
          eid = {L4},
        pages = {L4},
          doi = {10.1051/0004-6361/201936040},
archivePrefix = {arXiv},
       eprint = {1906.09352},
 primaryClass = {astro-ph.SR},
       adsurl = {https://ui.adsabs.harvard.edu/abs/2019A&A...627L...4C}
}

@ARTICLE{Cernicharo2019a,
       author = {{Cernicharo}, J. and {Cabezas}, C. and {Pardo}, J.~R. and et al.},
        title = "{Discovery of two new magnesium-bearing species in IRC+10216: MgC$_{3}$N and MgC$_{4}$H}",
      journal = {\aap},
         year = 2019,
        month = oct,
       volume = {630},
          eid = {L2},
        pages = {L2},
          doi = {10.1051/0004-6361/201936372},
       adsurl = {https://ui.adsabs.harvard.edu/abs/2019A&A...630L...2C}
}

@ARTICLE{Cernicharo2020,
       author = {{Cernicharo}, J. and {Marcelino}, N. and {Ag{\'u}ndez}, M. and et al.},
        title = "{Discovery of HC$_{3}$O$^{+}$ in space: The chemistry of O-bearing species in TMC-1}",
      journal = {\aap},
         year = 2020,
        month = oct,
       volume = {642},
          eid = {L17},
        pages = {L17},
          doi = {10.1051/0004-6361/202039351},
archivePrefix = {arXiv},
       eprint = {2010.04419},
 primaryClass = {astro-ph.GA},
       adsurl = {https://ui.adsabs.harvard.edu/abs/2020A&A...642L..17C}
}

@ARTICLE{Cernicharo2021,
       author = {{Cernicharo}, J. and {Cabezas}, C. and {Ag{\'u}ndez}, M. and et al.},
        title = "{TMC-1, the starless core sulfur factory: Discovery of NCS, HCCS, H$_{2}$CCS, H$_{2}$CCCS, and C$_{4}$S and detection of C$_{5}$S}",
      journal = {\aap},
         year = 2021,
        month = apr,
       volume = {648},
          eid = {L3},
        pages = {L3},
          doi = {10.1051/0004-6361/202140642},
archivePrefix = {arXiv},
       eprint = {2103.12431},
 primaryClass = {astro-ph.GA},
       adsurl = {https://ui.adsabs.harvard.edu/abs/2021A&A...648L...3C}
}

@ARTICLE{Cernicharo2021a,
       author = {{Cernicharo}, J. and {Ag{\'u}ndez}, M. and {Cabezas}, C. and et al.},
        title = "{Discovery of HCCCO and C$_{5}$O in TMC-1 with the QUIJOTE line survey}",
      journal = {\aap},
         year = 2021,
        month = dec,
       volume = {656},
          eid = {L21},
        pages = {L21},
          doi = {10.1051/0004-6361/202142634},
archivePrefix = {arXiv},
       eprint = {2112.01130},
 primaryClass = {astro-ph.GA},
       adsurl = {https://ui.adsabs.harvard.edu/abs/2021A&A...656L..21C}
}

@ARTICLE{Cernicharo2021b,
       author = {{Cernicharo}, J. and {Cabezas}, C. and {Endo}, Y. and et al.},
        title = "{Space and laboratory discovery of HC$_{3}$S$^{+}$}",
      journal = {\aap},
         year = 2021,
        month = feb,
       volume = {646},
          eid = {L3},
        pages = {L3},
          doi = {10.1051/0004-6361/202040013},
archivePrefix = {arXiv},
       eprint = {2101.05163},
 primaryClass = {astro-ph.GA},
       adsurl = {https://ui.adsabs.harvard.edu/abs/2021A&A...646L...3C}
}

@ARTICLE{Cernicharo2021c,
       author = {{Cernicharo}, J. and {Cabezas}, C. and {Endo}, Y. and et al.},
        title = "{The sulphur saga in TMC-1: Discovery of HCSCN and HCSCCH}",
      journal = {\aap},
         year = 2021,
        month = jun,
       volume = {650},
          eid = {L14},
        pages = {L14},
          doi = {10.1051/0004-6361/202141297},
archivePrefix = {arXiv},
       eprint = {2105.12996},
 primaryClass = {astro-ph.GA},
       adsurl = {https://ui.adsabs.harvard.edu/abs/2021A&A...650L..14C}
}

@ARTICLE{Cernicharo2023,
       author = {{Cernicharo}, J. and {Cabezas}, C. and {Pardo}, J.~R. and {Ag{\'u}ndez}, M. and {Roncero}, O. and {Tercero}, B. and {Marcelino}, N. and {Gu{\'e}lin}, M. and {Endo}, Y. and {de Vicente}, P.},
        title = "{The magnesium paradigm in IRC +10216: Discovery of MgC$_{4}$H$^{+}$, MgC$_{3}$N$^{+}$, MgC$_{6}$H$^{+}$, and MgC$_{5}$N$^{+}$}",
      journal = {\aap},
         year = 2023,
        month = apr,
       volume = {672},
          eid = {L13},
        pages = {L13},
          doi = {10.1051/0004-6361/202346467},
archivePrefix = {arXiv},
       eprint = {2304.05117},
 primaryClass = {astro-ph.GA},
       adsurl = {https://ui.adsabs.harvard.edu/abs/2023A&A...672L..13C}
}

@ARTICLE{Cernicharo2024,
       author = {{Cernicharo}, J. and {Ag{\'u}ndez}, M. and {Cabezas}, C. and {Tercero}, B. and {Fuentetaja}, R. and {Marcelino}, N. and {de Vicente}, P.},
        title = "{Discovery of thiofulminic acid with the QUIJOTE line survey: A study of the isomers of HNCS and HNCO in TMC-1}",
      journal = {\aap},
         year = 2024,
        month = feb,
       volume = {682},
          eid = {L4},
        pages = {L4},
          doi = {10.1051/0004-6361/202349105},
archivePrefix = {arXiv},
       eprint = {2401.11785},
 primaryClass = {astro-ph.GA},
       adsurl = {https://ui.adsabs.harvard.edu/abs/2024A&A...682L...4C}
}

@ARTICLE{Cernicharo2024b,
       author = {{Cernicharo}, J. and {Cabezas}, C. and {Ag{\'u}ndez}, M. and {Fuentetaja}, R. and {Tercero}, B. and {Marcelino}, N. and {de Vicente}, P.},
        title = "{More sulphur in TMC-1: Discovery of the NC$_{3}$S and HC$_{3}$S radicals with the QUIJOTE line survey}",
      journal = {\aap},
         year = 2024,
        month = aug,
       volume = {688},
          eid = {L13},
        pages = {L13},
          doi = {10.1051/0004-6361/202451256},
archivePrefix = {arXiv},
       eprint = {2407.15275},
 primaryClass = {astro-ph.GA},
       adsurl = {https://ui.adsabs.harvard.edu/abs/2024A&A...688L..13C}
}

@ARTICLE{Changala2022,
       author = {{Changala}, P.~B. and {Gupta}, H. and {Cernicharo}, J. and {Pardo}, J.~R. and {Ag{\'u}ndez}, M. and {Cabezas}, C. and {Tercero}, B. and {Gu{\'e}lin}, M. and {McCarthy}, M.~C.},
        title = "{Laboratory and Astronomical Discovery of Magnesium Dicarbide, MgC$_{2}$}",
      journal = {\apjl},
         year = 2022,
        month = dec,
       volume = {940},
       number = {2},
          eid = {L42},
        pages = {L42},
          doi = {10.3847/2041-8213/aca144},
archivePrefix = {arXiv},
       eprint = {2210.17348},
 primaryClass = {astro-ph.SR},
       adsurl = {https://ui.adsabs.harvard.edu/abs/2022ApJ...940L..42C}
}

@ARTICLE{Cheung1968,
       author = {{Cheung}, A.~C. and {Rank}, D.~M. and {Townes}, C.~H. and {Thornton}, D.~D. and {Welch}, W.~J.},
        title = "{Detection of NH$_{3}$ Molecules in the Interstellar Medium by Their Microwave Emission}",
      journal = {\prl},
         year = 1968,
        month = dec,
       volume = {21},
       number = {25},
        pages = {1701-1705},
          doi = {10.1103/PhysRevLett.21.1701},
       adsurl = {https://ui.adsabs.harvard.edu/abs/1968PhRvL..21.1701C}
}

@ARTICLE{Cheung1969,
       author = {{Cheung}, A.~C. and {Rank}, D.~M. and {Townes}, C.~H. and {Thornton}, D.~D. and {Welch}, W.~J.},
        title = "{Detection of Water in Interstellar Regions by its Microwave Radiation}",
      journal = {\nat},
         year = 1969,
        month = feb,
       volume = {221},
       number = {5181},
        pages = {626-628},
          doi = {10.1038/221626a0},
       adsurl = {https://ui.adsabs.harvard.edu/abs/1969Natur.221..626C}
}

@ARTICLE{Churchwell1977,
       author = {{Churchwell}, E. and {Witzel}, A. and {Huchtmeier}, W. and {Pauliny-Toth}, I. and {Roland}, J. and {Sieber}, W.},
        title = "{Detection of H$_{2}$O maser emission in the galaxy M33.}",
      journal = {\aap},
         year = 1977,
        month = feb,
       volume = {54},
        pages = {969-971},
       adsurl = {https://ui.adsabs.harvard.edu/abs/1977A&A....54..969C}
}

@ARTICLE{Compiegne2011,
       author = {{Compi{\`e}gne}, M. and {Verstraete}, L. and {Jones}, A. and et al.},
        title = "{The global dust SED: tracing the nature and evolution of dust with DustEM}",
      journal = {\aap},
         year = 2011,
        month = jan,
       volume = {525},
          eid = {A103},
        pages = {A103},
          doi = {10.1051/0004-6361/201015292},
archivePrefix = {arXiv},
       eprint = {1010.2769},
 primaryClass = {astro-ph.GA},
       adsurl = {https://ui.adsabs.harvard.edu/abs/2011A&A...525A.103C}
}

@ARTICLE{Coutens2019,
       author = {{Coutens}, A. and {Ligterink}, N.~F.~W. and {Loison}, J. -C. and {Wakelam}, V. and {Calcutt}, H. and {Drozdovskaya}, M.~N. and {J{\o}rgensen}, J.~K. and {M{\"u}ller}, H.~S.~P. and {van Dishoeck}, E.~F. and {Wampfler}, S.~F.},
        title = "{The ALMA-PILS survey: First detection of nitrous acid (HONO) in the interstellar medium}",
      journal = {\aap},
         year = 2019,
        month = mar,
       volume = {623},
          eid = {L13},
        pages = {L13},
          doi = {10.1051/0004-6361/201935040},
archivePrefix = {arXiv},
       eprint = {1903.03378},
 primaryClass = {astro-ph.SR},
       adsurl = {https://ui.adsabs.harvard.edu/abs/2019A&A...623L..13C}
}

@ARTICLE{Cummins1986,
       author = {{Cummins}, S.~E. and {Linke}, R.~A. and {Thaddeus}, P.},
        title = "{A Survey of the Millimeter-Wave Spectrum of Sagittarius B2}",
      journal = {\apjs},
         year = 1986,
        month = mar,
       volume = {60},
        pages = {819},
          doi = {10.1086/191102},
       adsurl = {https://ui.adsabs.harvard.edu/abs/1986ApJS...60..819C}
}

@ARTICLE{Dame2001,
       author = {{Dame}, T.~M. and {Hartmann}, Dap and {Thaddeus}, P.},
        title = "{The Milky Way in Molecular Clouds: A New Complete CO Survey}",
      journal = {\apj},
         year = 2001,
        month = feb,
       volume = {547},
       number = {2},
        pages = {792-813},
          doi = {10.1086/318388},
archivePrefix = {arXiv},
       eprint = {astro-ph/0009217},
 primaryClass = {astro-ph},
       adsurl = {https://ui.adsabs.harvard.edu/abs/2001ApJ...547..792D}
}

@ARTICLE{Dejong1975,
       author = {{de Jong}, T. and {Chu}, S. and {Dalgarno}, A.},
        title = "{Carbon monoxide in collapsing interstellar clouds.}",
      journal = {\apj},
         year = 1975,
        month = jul,
       volume = {199},
        pages = {69-78},
          doi = {10.1086/153665},
       adsurl = {https://ui.adsabs.harvard.edu/abs/1975ApJ...199...69D}
}

@ARTICLE{DeLuca2012,
       author = {{De Luca}, M. and {Gupta}, H. and {Neufeld}, D. and et al.},
        title = "{Herschel/HIFI Discovery of HCl$^{+}$ in the Interstellar Medium}",
      journal = {\apjl},
         year = 2012,
        month = jun,
       volume = {751},
       number = {2},
          eid = {L37},
        pages = {L37},
          doi = {10.1088/2041-8205/751/2/L37},
       adsurl = {https://ui.adsabs.harvard.edu/abs/2012ApJ...751L..37D}
}

@ARTICLE{DHendecourt1989,
       author = {{D'Hendecourt}, L.~B. and {Jourdain de Muizon}, M.},
        title = "{The discovery of interstellar carbon dioxide.}",
      journal = {\aap},
         year = 1989,
        month = oct,
       volume = {223},
        pages = {L5-L8},
       adsurl = {https://ui.adsabs.harvard.edu/abs/1989A&A...223L...5D}
}

@ARTICLE{Douglas1941,
       author = {{Douglas}, A.~E. and {Herzberg}, G.},
        title = "{Note on CH\^\{+\} in Interstellar Space and in the Laboratory.}",
      journal = {\apj},
         year = 1941,
        month = sep,
       volume = {94},
        pages = {381},
          doi = {10.1086/144342},
       adsurl = {https://ui.adsabs.harvard.edu/abs/1941ApJ....94..381D}
}

@ARTICLE{Downes2003,
       author = {{Downes}, D. and {Solomon}, P.~M.},
        title = "{Molecular Gas and Dust at z=2.6 in SMM J14011+0252: A Strongly Lensed Ultraluminous Galaxy, Not a Huge Massive Disk}",
      journal = {\apj},
         year = 2003,
        month = jan,
       volume = {582},
       number = {1},
        pages = {37-48},
          doi = {10.1086/344594},
archivePrefix = {arXiv},
       eprint = {astro-ph/0210040},
 primaryClass = {astro-ph},
       adsurl = {https://ui.adsabs.harvard.edu/abs/2003ApJ...582...37D}
}

@ARTICLE{Dunham1937,
       author = {{Dunham}, T., Jr.},
        title = "{Interstellar Neutral Potassium and Neutral Calcium}",
      journal = {\pasp},
         year = 1937,
        month = feb,
       volume = {49},
       number = {287},
        pages = {26-28},
          doi = {10.1086/124759},
       adsurl = {https://ui.adsabs.harvard.edu/abs/1937PASP...49...26D}
}

@BOOK{Elitzur1992,
       author = {{Elitzur}, Moshe},
        title = "{Astronomical masers}",
         year = 1992,
       volume = {170},
          doi = {10.1007/978-94-011-2394-5},
       adsurl = {https://ui.adsabs.harvard.edu/abs/1992ASSL..170.....E}
}

@ARTICLE{Fayolle2017,
       author = {{Fayolle}, Edith C. and {{\"O}berg}, Karin I. and {J{\o}rgensen}, Jes K. and  {Rosina Team}},
        title = "{Protostellar and cometary detections of organohalogens}",
      journal = {Nature Astronomy},
         year = 2017,
        month = oct,
       volume = {1},
        pages = {703-708},
          doi = {10.1038/s41550-017-0237-7},
       adsurl = {https://ui.adsabs.harvard.edu/abs/2017NatAs...1..703F}
}

@ARTICLE{Feuchtgruber2000,
       author = {{Feuchtgruber}, H. and {Helmich}, F.~P. and {van Dishoeck}, E.~F. and {Wright}, C.~M.},
        title = "{Detection of Interstellar CH$_{3}$}",
      journal = {\apjl},
         year = 2000,
        month = jun,
       volume = {535},
       number = {2},
        pages = {L111-L114},
          doi = {10.1086/312711},
archivePrefix = {arXiv},
       eprint = {astro-ph/0005273},
 primaryClass = {astro-ph},
       adsurl = {https://ui.adsabs.harvard.edu/abs/2000ApJ...535L.111F}
}

@ARTICLE{Fontani2024,
       author = {{Fontani}, Francesco},
        title = "{Observations of phosphorus-bearing molecules in the interstellar medium}",
      journal = {Frontiers in Astronomy and Space Sciences},
         year = 2024,
        month = aug,
       volume = {11},
          eid = {1451127},
        pages = {1451127},
          doi = {10.3389/fspas.2024.1451127},
archivePrefix = {arXiv},
       eprint = {2407.19006},
 primaryClass = {astro-ph.GA},
       adsurl = {https://ui.adsabs.harvard.edu/abs/2024FrASS..1151127F}
}

@ARTICLE{Frerking1979,
       author = {{Frerking}, M.~A. and {Linke}, R.~A. and {Thaddeus}, P.},
        title = "{Interstellar isothiocyanic acid.}",
      journal = {\apjl},
         year = 1979,
        month = dec,
       volume = {234},
        pages = {L143-L145},
          doi = {10.1086/183126},
       adsurl = {https://ui.adsabs.harvard.edu/abs/1979ApJ...234L.143F}
}

@ARTICLE{Friberg1980,
       author = {{Friberg}, P. and {Hjalmarson}, A. and {Guelin}, M. and {Irvine}, W.~M.},
        title = "{Interstellar C3N - Detection in Taurus dark clouds}",
      journal = {\apjl},
         year = 1980,
        month = oct,
       volume = {241},
        pages = {L99-L103},
          doi = {10.1086/183369},
       adsurl = {https://ui.adsabs.harvard.edu/abs/1980ApJ...241L..99F}
}

@ARTICLE{Fuente2006,
       author = {{Fuente}, A. and {Garc{\'\i}a-Burillo}, S. and {Gerin}, M. and et al.},
        title = "{Detection of CO$^{+}$ in the Nucleus of M82}",
      journal = {\apjl},
         year = 2006,
        month = apr,
       volume = {641},
       number = {2},
        pages = {L105-L108},
          doi = {10.1086/503605},
archivePrefix = {arXiv},
       eprint = {astro-ph/0602509},
 primaryClass = {astro-ph},
       adsurl = {https://ui.adsabs.harvard.edu/abs/2006ApJ...641L.105F}
}

@ARTICLE{Fuente2017,
       author = {{Fuente}, Asunci{\'o}n and {Goicoechea}, Javier R. and {Pety}, J{\'e}r{\^o}me and et al.},
        title = "{First Detection of Interstellar S$_{2}$H}",
      journal = {\apjl},
         year = 2017,
        month = dec,
       volume = {851},
       number = {2},
          eid = {L49},
        pages = {L49},
          doi = {10.3847/2041-8213/aaa01b},
archivePrefix = {arXiv},
       eprint = {1712.03036},
 primaryClass = {astro-ph.GA},
       adsurl = {https://ui.adsabs.harvard.edu/abs/2017ApJ...851L..49F}
}

@ARTICLE{Fuller1992,
       author = {{Fuller}, G.~A. and {Myers}, P.~C.},
        title = "{Dense Cores in Dark Clouds. VII. Line Width--Size Relations}",
      journal = {\apj},
         year = 1992,
        month = jan,
       volume = {384},
        pages = {523},
          doi = {10.1086/170894},
       adsurl = {https://ui.adsabs.harvard.edu/abs/1992ApJ...384..523F}
}

@ARTICLE{Garcia2002,
       author = {{Garc{\'\i}a-Burillo}, S. and {Mart{\'\i}n-Pintado}, J. and {Fuente}, A. and {Usero}, A. and {Neri}, R.},
        title = "{Widespread HCO Emission in the Nuclear Starburst of M82}",
      journal = {\apjl},
         year = 2002,
        month = aug,
       volume = {575},
       number = {2},
        pages = {L55-L58},
          doi = {10.1086/342743},
archivePrefix = {arXiv},
       eprint = {astro-ph/0207313},
 primaryClass = {astro-ph},
       adsurl = {https://ui.adsabs.harvard.edu/abs/2002ApJ...575L..55G}
}

@ARTICLE{Geballe1996,
       author = {{Geballe}, T.~R. and {Oka}, T.},
        title = "{Detection of H$^{+}_{3}$ in interstellar space}",
      journal = {\nat},
         year = 1996,
        month = nov,
       volume = {384},
       number = {6607},
        pages = {334-335},
          doi = {10.1038/384334a0},
       adsurl = {https://ui.adsabs.harvard.edu/abs/1996Natur.384..334G}
}

@ARTICLE{Geballe2006,
       author = {{Geballe}, T.~R. and {Goto}, M. and {Usuda}, T. and {Oka}, T. and {McCall}, B.~J.},
        title = "{The Interstellar Medium of IRAS 08572+3915 NW: H$^+_3$ and Warm High-Velocity CO}",
      journal = {\apj},
         year = 2006,
        month = jun,
       volume = {644},
       number = {2},
        pages = {907-913},
          doi = {10.1086/503763},
archivePrefix = {arXiv},
       eprint = {astro-ph/0603041},
 primaryClass = {astro-ph},
       adsurl = {https://ui.adsabs.harvard.edu/abs/2006ApJ...644..907G}
}

@ARTICLE{Gerin2010,
       author = {{Gerin}, M. and {de Luca}, M. and {Black}, J. and et al.},
        title = "{Interstellar OH$^{+}$, H$_{2}$O$^{+}$ and H$_{3}$O$^{+}$ along the sight-line to G10.6-0.4}",
      journal = {\aap},
         year = 2010,
        month = jul,
       volume = {518},
          eid = {L110},
        pages = {L110},
          doi = {10.1051/0004-6361/201014576},
archivePrefix = {arXiv},
       eprint = {1005.5653},
 primaryClass = {astro-ph.GA},
       adsurl = {https://ui.adsabs.harvard.edu/abs/2010A&A...518L.110G}
}

@ARTICLE{Ginsburg2019,
       author = {{Ginsburg}, Adam and {McGuire}, Brett and {Plambeck}, Richard and et al.},
        title = "{Orion SrcI{\textquoteright}s Disk Is Salty}",
      journal = {\apj},
         year = 2019,
        month = feb,
       volume = {872},
       number = {1},
          eid = {54},
        pages = {54},
          doi = {10.3847/1538-4357/aafb71},
archivePrefix = {arXiv},
       eprint = {1901.04489},
 primaryClass = {astro-ph.GA},
       adsurl = {https://ui.adsabs.harvard.edu/abs/2019ApJ...872...54G}
}

@ARTICLE{Godfrey1973,
       author = {{Godfrey}, P.~D. and {Brown}, R.~D. and {Robinson}, B.~J. and {Sinclair}, M.~W.},
        title = "{Discovery of Interstellar Methanimine (Formaldimine)}",
      journal = {\aplett},
         year = 1973,
        month = feb,
       volume = {13},
        pages = {119},
       adsurl = {https://ui.adsabs.harvard.edu/abs/1973ApL....13..119G}
}

@ARTICLE{Goldhaber1984,
       author = {{Goldhaber}, D.~M. and {Betz}, A.~L.},
        title = "{Silane in IRC +10216.}",
      journal = {\apjl},
         year = 1984,
        month = apr,
       volume = {279},
        pages = {L55-L58},
          doi = {10.1086/184255},
       adsurl = {https://ui.adsabs.harvard.edu/abs/1984ApJ...279L..55G}
}

@ARTICLE{Goldsmith1999,
       author = {{Goldsmith}, Paul F. and {Langer}, William D.},
        title = "{Population Diagram Analysis of Molecular Line Emission}",
      journal = {\apj},
         year = 1999,
        month = may,
       volume = {517},
       number = {1},
        pages = {209-225},
          doi = {10.1086/307195},
       adsurl = {https://ui.adsabs.harvard.edu/abs/1999ApJ...517..209G}
}

@ARTICLE{Goldsmith2002,
       author = {{Goldsmith}, P.~F. and {Li}, D. and {Bergin}, E.~A. and et al.},
        title = "{Tentative Detection of Molecular Oxygen in the {\ensuremath{\rho}} Ophiuchi Cloud}",
      journal = {\apj},
         year = 2002,
        month = sep,
       volume = {576},
       number = {2},
        pages = {814-831},
          doi = {10.1086/341809},
       adsurl = {https://ui.adsabs.harvard.edu/abs/2002ApJ...576..814G}
}

@ARTICLE{Goldsmith2008,
       author = {{Goldsmith}, Paul F. and {Heyer}, Mark and {Narayanan}, Gopal and {Snell}, Ronald and {Li}, Di and {Brunt}, Chris},
        title = "{Large-Scale Structure of the Molecular Gas in Taurus Revealed by High Linear Dynamic Range Spectral Line Mapping}",
      journal = {\apj},
         year = 2008,
        month = jun,
       volume = {680},
       number = {1},
        pages = {428-445},
          doi = {10.1086/587166},
archivePrefix = {arXiv},
       eprint = {0802.2206},
 primaryClass = {astro-ph},
       adsurl = {https://ui.adsabs.harvard.edu/abs/2008ApJ...680..428G}
}

@ARTICLE{Goldsmith2011,
       author = {{Goldsmith}, Paul F. and {Liseau}, Ren{\'e} and {Bell}, Tom A. and et al.},
        title = "{Herschel Measurements of Molecular Oxygen in Orion}",
      journal = {\apj},
         year = 2011,
        month = aug,
       volume = {737},
       number = {2},
          eid = {96},
        pages = {96},
          doi = {10.1088/0004-637X/737/2/96},
archivePrefix = {arXiv},
       eprint = {1108.0441},
 primaryClass = {astro-ph.GA},
       adsurl = {https://ui.adsabs.harvard.edu/abs/2011ApJ...737...96G}
}

@ARTICLE{Gonzalez-Alfonso2004,
       author = {{Gonz{\'a}lez-Alfonso}, Eduardo and {Smith}, Howard A. and {Fischer}, Jacqueline and {Cernicharo}, Jos{\'e}},
        title = "{The Far-Infrared Spectrum of Arp 220}",
      journal = {\apj},
         year = 2004,
        month = sep,
       volume = {613},
       number = {1},
        pages = {247-261},
          doi = {10.1086/422868},
archivePrefix = {arXiv},
       eprint = {astro-ph/0406427},
 primaryClass = {astro-ph},
       adsurl = {https://ui.adsabs.harvard.edu/abs/2004ApJ...613..247G}
}

@BOOK{Gordy1984,
       author = {{Gordy}, W. and {Cook}, R.~L.},
        title = "{Microwave Molecular Spectra (New York: Wiley)}",
         year = 1984
}

@ARTICLE{Gottlieb1973,
       author = {{Gottlieb}, C.~A. and {Ball}, John A.},
        title = "{Interstellar Sulfur Monoxide}",
      journal = {\apjl},
         year = 1973,
        month = sep,
       volume = {184},
        pages = {L59},
          doi = {10.1086/181288},
       adsurl = {https://ui.adsabs.harvard.edu/abs/1973ApJ...184L..59G}
}

@ARTICLE{Gottlieb1975,
       author = {{Gottlieb}, C.~A. and {Ball}, J.~A. and {Gottlieb}, E.~W. and {Lada}, C.~J. and {Penfield}, H.},
        title = "{Detection of interstellar nitrogen sulfide.}",
      journal = {\apjl},
         year = 1975,
        month = sep,
       volume = {200},
        pages = {L147-L149},
          doi = {10.1086/181918},
       adsurl = {https://ui.adsabs.harvard.edu/abs/1975ApJ...200L.147G}
}

@ARTICLE{Green1974,
       author = {{Green}, S. and {Montgomery}, J.~A., Jr. and {Thaddeus}, P.},
        title = "{Tentative Identification of U93.174 as the Molecular Ion N$_{2}$H$^+$}",
      journal = {\apjl},
         year = 1974,
        month = oct,
       volume = {193},
        pages = {L89},
          doi = {10.1086/181639},
       adsurl = {https://ui.adsabs.harvard.edu/abs/1974ApJ...193L..89G}
}

@ARTICLE{Guelin1977,
       author = {{Guelin}, M. and {Thaddeus}, P.},
        title = "{Tentative Detection of the C3N Radical}",
      journal = {\apjl},
         year = 1977,
        month = mar,
       volume = {212},
        pages = {L81},
          doi = {10.1086/182380},
       adsurl = {https://ui.adsabs.harvard.edu/abs/1977ApJ...212L..81G}
}

@ARTICLE{Guelin1978,
       author = {{Guelin}, M. and {Green}, S. and {Thaddeus}, P.},
        title = "{Detection of the C$_{4}$H radical toward IRC +10216.}",
      journal = {\apjl},
         year = 1978,
        month = aug,
       volume = {224},
        pages = {L27-L30},
          doi = {10.1086/182751},
       adsurl = {https://ui.adsabs.harvard.edu/abs/1978ApJ...224L..27G}
}

@ARTICLE{Guelin1986,
       author = {{Guelin}, M. and {Cernicharo}, J. and {Kahane}, C. and {Gomez-Gonzales}, J.},
        title = "{A new free radical in IRC +10216.}",
      journal = {\aap},
         year = 1986,
        month = mar,
       volume = {157},
        pages = {L17-L20},
       adsurl = {https://ui.adsabs.harvard.edu/abs/1986A&A...157L..17G}
}

@ARTICLE{Guelin1990,
       author = {{Guelin}, M. and {Cernicharo}, J. and {Paubert}, G. and {Turner}, B.~E.},
        title = "{Free CP in IRC +10216.}",
      journal = {\aap},
         year = 1990,
        month = apr,
       volume = {230},
        pages = {L9-L11},
       adsurl = {https://ui.adsabs.harvard.edu/abs/1990A&A...230L...9G}
}

@ARTICLE{Guelin1991,
       author = {{Guelin}, M. and {Cernicharo}, J.},
        title = "{Astronomical detection of the HCCN radical. Toward a new family of carbon-chain molecules ?}",
      journal = {\aap},
         year = 1991,
        month = apr,
       volume = {244},
        pages = {L21},
       adsurl = {https://ui.adsabs.harvard.edu/abs/1991A&A...244L..21G}
}

@ARTICLE{Guelin2000,
       author = {{Gu{\'e}lin}, M. and {Muller}, S. and {Cernicharo}, J. and {Apponi}, A.~J. and {McCarthy}, M.~C. and {Gottlieb}, C.~A. and {Thaddeus}, P.},
        title = "{Astronomical detection of the free radical SiCN}",
      journal = {\aap},
         year = 2000,
        month = nov,
       volume = {363},
        pages = {L9-L12},
       adsurl = {https://ui.adsabs.harvard.edu/abs/2000A&A...363L...9G}
}

@ARTICLE{Gupta2013,
       author = {{Gupta}, H. and {Gottlieb}, C.~A. and {Lattanzi}, V. and {Pearson}, J.~C. and {McCarthy}, M.~C.},
        title = "{Laboratory Measurements and Tentative Astronomical Identification of H$_{2}$NCO$^{+}$}",
      journal = {\apjl},
         year = 2013,
        month = nov,
       volume = {778},
       number = {1},
          eid = {L1},
        pages = {L1},
          doi = {10.1088/2041-8205/778/1/L1},
       adsurl = {https://ui.adsabs.harvard.edu/abs/2013ApJ...778L...1G}
}

@ARTICLE{Gupta2024,
       author = {{Gupta}, H. and {Changala}, P.~B. and {Cernicharo}, J. and {Pardo}, J.~R. and {Ag{\'u}ndez}, M. and {Cabezas}, C. and {Tercero}, B. and {Gu{\'e}lin}, M. and {McCarthy}, M.~C.},
        title = "{Calcium Chemistry in Carbon-rich Circumstellar Environments: The Laboratory and Astronomical Discovery of Calcium Dicarbide, CaC$_{2}$}",
      journal = {\apjl},
         year = 2024,
        month = may,
       volume = {966},
       number = {2},
          eid = {L28},
        pages = {L28},
          doi = {10.3847/2041-8213/ad3336},
       adsurl = {https://ui.adsabs.harvard.edu/abs/2024ApJ...966L..28G}
}

@ARTICLE{Gusten2019,
       author = {{G{\"u}sten}, Rolf and {Wiesemeyer}, Helmut and {Neufeld}, David and et al.},
        title = "{Astrophysical detection of the helium hydride ion HeH$^{+}$}",
      journal = {\nat},
         year = 2019,
        month = apr,
       volume = {568},
       number = {7752},
        pages = {357-359},
          doi = {10.1038/s41586-019-1090-x},
archivePrefix = {arXiv},
       eprint = {1904.09581},
 primaryClass = {astro-ph.GA},
       adsurl = {https://ui.adsabs.harvard.edu/abs/2019Natur.568..357G}
}

@ARTICLE{Haasler2022,
       author = {{Haasler}, D. and {Rivilla}, V.~M. and {Mart{\'\i}n}, S. and et al.},
        title = "{First extragalactic detection of a phosphorus-bearing molecule with ALCHEMI: Phosphorus nitride (PN)}",
      journal = {\aap},
         year = 2022,
        month = mar,
       volume = {659},
          eid = {A158},
        pages = {A158},
          doi = {10.1051/0004-6361/202142032},
archivePrefix = {arXiv},
       eprint = {2112.04849},
 primaryClass = {astro-ph.GA},
       adsurl = {https://ui.adsabs.harvard.edu/abs/2022A&A...659A.158H}
}

@ARTICLE{Halfen2008,
       author = {{Halfen}, D.~T. and {Clouthier}, D.~J. and {Ziurys}, L.~M.},
        title = "{Detection of the CCP Radical (X$^{2}${\ensuremath{\Pi}}$_{r}$) in IRC +10216: A New Interstellar Phosphorus-containing Species}",
      journal = {\apjl},
         year = 2008,
        month = apr,
       volume = {677},
       number = {2},
        pages = {L101},
          doi = {10.1086/588024},
       adsurl = {https://ui.adsabs.harvard.edu/abs/2008ApJ...677L.101H}
}

@ARTICLE{Halfen2009,
       author = {{Halfen}, D.~T. and {Ziurys}, L.~M. and {Br{\"u}nken}, S. and {Gottlieb}, C.~A. and {McCarthy}, M.~C. and {Thaddeus}, P.},
        title = "{Detection of a New Interstellar Molecule: Thiocyanic Acid HSCN}",
      journal = {\apjl},
         year = 2009,
        month = sep,
       volume = {702},
       number = {2},
        pages = {L124-L127},
          doi = {10.1088/0004-637X/702/2/L124},
       adsurl = {https://ui.adsabs.harvard.edu/abs/2009ApJ...702L.124H}
}

@ARTICLE{Heikkila1999,
       author = {{Heikkil{\"a}}, A. and {Johansson}, L.~E.~B. and {Olofsson}, H.},
        title = "{Molecular abundance variations in the Magellanic Clouds}",
      journal = {\aap},
         year = 1999,
        month = apr,
       volume = {344},
        pages = {817-847},
       adsurl = {https://ui.adsabs.harvard.edu/abs/1999A&A...344..817H}
}

@ARTICLE{Henkel1988,
       author = {{Henkel}, C. and {Mauersberger}, R. and {Schilke}, P.},
        title = "{Molecules in external galaxies : the detection of CN, C2H and HNC and the tentative detection of HC3N.}",
      journal = {\aap},
         year = 1988,
        month = jul,
       volume = {201},
        pages = {L23-L26},
       adsurl = {https://ui.adsabs.harvard.edu/abs/1988A&A...201L..23H}
}

@ARTICLE{Henkel1985,
       author = {{Henkel}, C. and {Bally}, J.},
        title = "{Detection of extragalactic CS.}",
      journal = {\aap},
         year = 1985,
        month = sep,
       volume = {150},
        pages = {L25-L27},
       adsurl = {https://ui.adsabs.harvard.edu/abs/1985A&A...150L..25H}
}

@ARTICLE{Herbst2009,
       author = {{Herbst}, Eric and {van Dishoeck}, Ewine F.},
        title = "{Complex Organic Interstellar Molecules}",
      journal = {\araa},
         year = 2009,
        month = sep,
       volume = {47},
       number = {1},
        pages = {427-480},
          doi = {10.1146/annurev-astro-082708-101654},
       adsurl = {https://ui.adsabs.harvard.edu/abs/2009ARA&A..47..427H}
}

@ARTICLE{Hildebrand1983,
       author = {{Hildebrand}, R.~H.},
        title = "{The determination of cloud masses and dust characteristics from submillimetre thermal emission.}",
      journal = {\qjras},
         year = 1983,
        month = sep,
       volume = {24},
        pages = {267-282},
       adsurl = {https://ui.adsabs.harvard.edu/abs/1983QJRAS..24..267H}
}

@ARTICLE{Hinkle1988,
       author = {{Hinkle}, Kenneth W. and {Keady}, John J. and {Bernath}, Peter F.},
        title = "{Detection of C$_{3}$ in the circumstellar shell of IRC +10216.}",
      journal = {Science},
         year = 1988,
        month = sep,
       volume = {241},
        pages = {1319-1322},
          doi = {10.1126/science.241.4871.1319},
       adsurl = {https://ui.adsabs.harvard.edu/abs/1988Sci...241.1319H}
}

@ARTICLE{Hollis1986,
       author = {{Hollis}, J.~M. and {Churchwell}, E.~B. and {Herbst}, E. and {De Lucia}, F.~C.},
        title = "{An interstellar line coincident with the P(2,l)transition of hydronium (H$_{3}$O$^{+}$)}",
      journal = {\nat},
         year = 1986,
        month = aug,
       volume = {322},
       number = {6079},
        pages = {524-526},
          doi = {10.1038/322524a0},
       adsurl = {https://ui.adsabs.harvard.edu/abs/1986Natur.322..524H}
}

@ARTICLE{Hollis1989,
       author = {{Hollis}, J.~M. and {Jewell}, P.~R. and {Lovas}, F.~J.},
        title = "{A Search for Methylene in the Orion Nebula}",
      journal = {\apj},
         year = 1989,
        month = nov,
       volume = {346},
        pages = {794},
          doi = {10.1086/168059},
       adsurl = {https://ui.adsabs.harvard.edu/abs/1989ApJ...346..794H}
}

@ARTICLE{Hollis1995,
       author = {{Hollis}, J.~M. and {Jewell}, P.~R. and {Lovas}, F.~J.},
        title = "{Confirmation of Interstellar Methylene}",
      journal = {\apj},
         year = 1995,
        month = jan,
       volume = {438},
        pages = {259},
          doi = {10.1086/175070},
       adsurl = {https://ui.adsabs.harvard.edu/abs/1995ApJ...438..259H}
}

@ARTICLE{Hyland1969,
       author = {{Hyland}, A.~R. and {Becklin}, E.~E. and {Neugebauer}, G. and {Wallerstein}, George},
        title = "{Observations of the Infrared Object, VY Canis Majoris}",
      journal = {\apj},
         year = 1969,
        month = nov,
       volume = {158},
        pages = {619},
          doi = {10.1086/150224},
       adsurl = {https://ui.adsabs.harvard.edu/abs/1969ApJ...158..619H}
}

@ARTICLE{Irvine1988,
       author = {{Irvine}, W.~M. and {Friberg}, P. and {Hjalmarson}, A. and et al.},
        title = "{Identification of the interstellar cyanomethyl radical (CH2CN) in themolecular clouds TMC-1 and Sagittarius B2.}",
      journal = {\apjl},
         year = 1988,
        month = nov,
       volume = {334},
        pages = {L107-L111},
          doi = {10.1086/185323},
       adsurl = {https://ui.adsabs.harvard.edu/abs/1988ApJ...334L.107I}
}

@ARTICLE{Jefferts1970,
       author = {{Jefferts}, K.~B. and {Penzias}, A.~A. and {Wilson}, R.~W.},
        title = "{Observation of the CN Radical in the Orion Nebula and W51}",
      journal = {\apjl},
         year = 1970,
        month = aug,
       volume = {161},
        pages = {L87},
          doi = {10.1086/180576},
       adsurl = {https://ui.adsabs.harvard.edu/abs/1970ApJ...161L..87J}
}

@INPROCEEDINGS{Johansson1991,
       author = {{Johansson}, L.~E.~B.},
        title = "{Interstellar Gas in the Magellanic Clouds: SEST Observations of CO and Other Molecules}",
    booktitle = {Dynamics of Galaxies and Their Molecular Cloud Distributions},
         year = 1991,
       editor = {{Combes}, F. and {Casoli}, Fabienne},
       volume = {146},
        month = jan,
        pages = {1},
       adsurl = {https://ui.adsabs.harvard.edu/abs/1991IAUS..146....1J}
}

@ARTICLE{Juvela2015,
       author = {{Juvela}, M. and {Ristorcelli}, I. and {Marshall}, D.~J. and {et} al.},
        title = "{Galactic cold cores. V. Dust opacity}",
      journal = {\aap},
         year = 2015,
        month = dec,
       volume = {584},
          eid = {A93},
        pages = {A93},
          doi = {10.1051/0004-6361/201423788},
archivePrefix = {arXiv},
       eprint = {1501.07092},
 primaryClass = {astro-ph.GA},
       adsurl = {https://ui.adsabs.harvard.edu/abs/2015A&A...584A..93J}
}

@ARTICLE{Kaifu1987,
       author = {{Kaifu}, Norio and {Suzuki}, Hiroko and {Ohishi}, Masatoshi and et al.},
        title = "{Detection of Intense Unidentified Lines in TMC-1}",
      journal = {\apjl},
         year = 1987,
        month = jun,
       volume = {317},
        pages = {L111},
          doi = {10.1086/184922},
       adsurl = {https://ui.adsabs.harvard.edu/abs/1987ApJ...317L.111K}
}

@ARTICLE{Kaminski2013,
       author = {{Kami{\'n}ski}, T. and {Gottlieb}, C.~A. and {Menten}, K.~M. and et al.},
        title = "{Pure rotational spectra of TiO and TiO$_{2}$ in VY Canis Majoris}",
      journal = {\aap},
         year = 2013,
        month = mar,
       volume = {551},
          eid = {A113},
        pages = {A113},
          doi = {10.1051/0004-6361/201220290},
archivePrefix = {arXiv},
       eprint = {1301.4344},
 primaryClass = {astro-ph.SR},
       adsurl = {https://ui.adsabs.harvard.edu/abs/2013A&A...551A.113K}
}

@ARTICLE{Kawaguchi1992a,
       author = {{Kawaguchi}, Kentarou and {Ohishi}, Masatoshi and {Ishikawa}, Shin-Ichi and {Kaifu}, Norio},
        title = "{Detection of Isocyanoacetylene HCCNC in TMC-1}",
      journal = {\apjl},
         year = 1992,
        month = feb,
       volume = {386},
        pages = {L51},
          doi = {10.1086/186290},
       adsurl = {https://ui.adsabs.harvard.edu/abs/1992ApJ...386L..51K}
}

@ARTICLE{Kawaguchi1992,
       author = {{Kawaguchi}, Kentarou and {Takano}, Shuro and {Ohishi}, Masatoshi and et al.},
        title = "{Detection of HNCCC in TMC-1}",
      journal = {\apjl},
         year = 1992,
        month = sep,
       volume = {396},
        pages = {L49},
          doi = {10.1086/186514},
       adsurl = {https://ui.adsabs.harvard.edu/abs/1992ApJ...396L..49K}
}

@ARTICLE{Kawaguchi1993,
       author = {{Kawaguchi}, Kentarou and {Kagi}, Eriko and {Hirano}, Tsuneo and {Takano}, Shuro and {Saito}, Shuji},
        title = "{Laboratory Spectroscopy of MgNC: The First Radioastronomical Identification of Mg-bearing Molecule}",
      journal = {\apjl},
         year = 1993,
        month = mar,
       volume = {406},
        pages = {L39},
          doi = {10.1086/186781},
       adsurl = {https://ui.adsabs.harvard.edu/abs/1993ApJ...406L..39K}
}

@ARTICLE{Klemperer1970,
       author = {{Klemperer}, William},
        title = "{Carrier of the Interstellar 89.190 GHz Line}",
      journal = {\nat},
         year = 1970,
        month = sep,
       volume = {227},
       number = {5264},
        pages = {1230},
          doi = {10.1038/2271230a0},
       adsurl = {https://ui.adsabs.harvard.edu/abs/1970Natur.227.1230K}
}

@ARTICLE{Knauth2004,
       author = {{Knauth}, David C. and {Andersson}, B. -G. and {McCandliss}, Stephan R. and {Warren Moos}, H.},
        title = "{The interstellar N$_{2}$ abundance towards HD 124314 from far-ultraviolet observations}",
      journal = {\nat},
         year = 2004,
        month = jun,
       volume = {429},
       number = {6992},
        pages = {636-638},
          doi = {10.1038/nature02614},
       adsurl = {https://ui.adsabs.harvard.edu/abs/2004Natur.429..636K}
}

@ARTICLE{Koelemay2022,
       author = {{Koelemay}, L.~A. and {Burton}, M.~A. and {Singh}, A.~P. and {Sheridan}, P.~M. and {Bernal}, J.~J. and {Ziurys}, L.~M.},
        title = "{Laboratory and Astronomical Detection of the SiP Radical (X$^{2}${\ensuremath{\Pi}}$_{ i }$): More Circumstellar Phosphorus}",
      journal = {\apjl},
         year = 2022,
        month = nov,
       volume = {940},
       number = {1},
          eid = {L11},
        pages = {L11},
          doi = {10.3847/2041-8213/ac9d9b},
       adsurl = {https://ui.adsabs.harvard.edu/abs/2022ApJ...940L..11K}
}

@ARTICLE{Koelemay2023,
       author = {{Koelemay}, L.~A. and {Ziurys}, L.~M.},
        title = "{Elusive Iron: Detection of the FeC Radical (X $^{3}${\ensuremath{\Delta}}$_{ i }$) in the Envelope of IRC+10216}",
      journal = {\apjl},
         year = 2023,
        month = nov,
       volume = {958},
       number = {1},
          eid = {L6},
        pages = {L6},
          doi = {10.3847/2041-8213/ad0899},
       adsurl = {https://ui.adsabs.harvard.edu/abs/2023ApJ...958L...6K}
}

@ARTICLE{Kuiper1975,
       author = {{Kuiper}, T.~B.~H. and {Zuckerman}, B. and {Kakar}, R.~K. and {Rodriguez Kuiper}, E.~N.},
        title = "{Detection of 2.6-millimeter radiation probably due to nitrogen sulfide.}",
      journal = {\apjl},
         year = 1975,
        month = sep,
       volume = {200},
        pages = {L151-L153},
          doi = {10.1086/181919},
       adsurl = {https://ui.adsabs.harvard.edu/abs/1975ApJ...200L.151K}
}

@PHDTHESIS{Kulesa2002,
       author = {{Kulesa}, Craig Alan},
        title = "{Molecular hydrogen and its ions in dark interstellar clouds and star forming regions}",
       school = {University of Arizona},
         year = 2002,
        month = nov,
       adsurl = {https://ui.adsabs.harvard.edu/abs/2002PhDT........28K}
}

@ARTICLE{Lacy1991,
       author = {{Lacy}, J.~H. and {Carr}, J.~S. and {Evans}, Neal J., II and et al.},
        title = "{Discovery of Interstellar Methane: Observations of Gaseous and Solid CH 4 Absorption toward Young Stars in Molecular Clouds}",
      journal = {\apj},
         year = 1991,
        month = aug,
       volume = {376},
        pages = {556},
          doi = {10.1086/170304},
       adsurl = {https://ui.adsabs.harvard.edu/abs/1991ApJ...376..556L}
}

@ARTICLE{Lacy1994,
       author = {{Lacy}, J.~H. and {Knacke}, R. and {Geballe}, T.~R. and {Tokunaga}, A.~T.},
        title = "{Detection of Absorption by H 2 in Molecular Clouds: A Direct Measurement of the H 2:CO Ratio}",
      journal = {\apjl},
         year = 1994,
        month = jun,
       volume = {428},
        pages = {L69},
          doi = {10.1086/187395},
       adsurl = {https://ui.adsabs.harvard.edu/abs/1994ApJ...428L..69L}
}

@ARTICLE{Larsson2007,
       author = {{Larsson}, B. and {Liseau}, R. and {Pagani}, L. and {et } al.},
        title = "{Molecular oxygen in the {\ensuremath{\rho}} Ophiuchi cloud}",
      journal = {\aap},
         year = 2007,
        month = may,
       volume = {466},
       number = {3},
        pages = {999-1003},
          doi = {10.1051/0004-6361:20065500},
archivePrefix = {arXiv},
       eprint = {astro-ph/0702474},
 primaryClass = {astro-ph},
       adsurl = {https://ui.adsabs.harvard.edu/abs/2007A&A...466..999L}
}

@ARTICLE{Latter1993,
       author = {{Latter}, William B. and {Walker}, Christopher K. and {Maloney}, Philip R.},
        title = "{Detection of the Carbon Monoxide Ion (CO +) in the Interstellar Medium and a Planetary Nebula}",
      journal = {\apjl},
         year = 1993,
        month = dec,
       volume = {419},
        pages = {L97},
          doi = {10.1086/187146},
       adsurl = {https://ui.adsabs.harvard.edu/abs/1993ApJ...419L..97L}
}

@ARTICLE{Lis2010,
       author = {{Lis}, D.~C. and {Pearson}, J.~C. and {Neufeld}, D.~A. and et al.},
        title = "{Herschel/HIFI discovery of interstellar chloronium (H$_{2}$Cl$^{+}$)}",
      journal = {\aap},
         year = 2010,
        month = oct,
       volume = {521},
          eid = {L9},
        pages = {L9},
          doi = {10.1051/0004-6361/201014959},
archivePrefix = {arXiv},
       eprint = {1007.1461},
 primaryClass = {astro-ph.GA},
       adsurl = {https://ui.adsabs.harvard.edu/abs/2010A&A...521L...9L}
}

@ARTICLE{Liseau2012,
       author = {{Liseau}, R. and {Goldsmith}, P.~F. and {Larsson}, B. and et al.},
        title = "{Multi-line detection of O$_{2}$ toward {\ensuremath{\rho}} Ophuichi A}",
      journal = {\aap},
         year = 2012,
        month = may,
       volume = {541},
          eid = {A73},
        pages = {A73},
          doi = {10.1051/0004-6361/201118575},
archivePrefix = {arXiv},
       eprint = {1202.5637},
 primaryClass = {astro-ph.GA},
       adsurl = {https://ui.adsabs.harvard.edu/abs/2012A&A...541A..73L}
}

@ARTICLE{Liszt1978,
       author = {{Liszt}, H.~S. and {Turner}, B.~E.},
        title = "{Microwave detection of interstellar NO.}",
      journal = {\apjl},
         year = 1978,
        month = sep,
       volume = {224},
        pages = {L73-L76},
          doi = {10.1086/182762},
       adsurl = {https://ui.adsabs.harvard.edu/abs/1978ApJ...224L..73L}
}

@ARTICLE{Liszt1998,
       author = {{Liszt}, H.~S. and {Lucas}, R.},
        title = "{CO in absorption and emission toward compact extragalactic radio continuum sources}",
      journal = {\aap},
         year = 1998,
        month = nov,
       volume = {339},
        pages = {561-574},
       adsurl = {https://ui.adsabs.harvard.edu/abs/1998A&A...339..561L}
}

@ARTICLE{Magain1987,
       author = {{Magain}, P. and {Gillet}, D.},
        title = "{Detection of interstellar CH and CH+ towards SN 1987A.}",
      journal = {\aap},
         year = 1987,
        month = oct,
       volume = {184},
        pages = {L5-L6},
       adsurl = {https://ui.adsabs.harvard.edu/abs/1987A&A...184L...5M}
}

@ARTICLE{Mangum2015,
       author = {{Mangum}, Jeffrey G. and {Shirley}, Yancy L.},
        title = "{How to Calculate Molecular Column Density}",
      journal = {\pasp},
         year = 2015,
        month = mar,
       volume = {127},
       number = {949},
        pages = {266},
          doi = {10.1086/680323},
archivePrefix = {arXiv},
       eprint = {1501.01703},
 primaryClass = {astro-ph.IM},
       adsurl = {https://ui.adsabs.harvard.edu/abs/2015PASP..127..266M}
}

@ARTICLE{Mangum2007,
       author = {{Mangum}, J.~G. and {Emerson}, D.~T. and {Greisen}, E.~W.},
        title = "{The On The Fly imaging technique}",
      journal = {\aap},
         year = 2007,
        month = nov,
       volume = {474},
       number = {2},
        pages = {679-687},
          doi = {10.1051/0004-6361:20077811},
archivePrefix = {arXiv},
       eprint = {0709.0553},
 primaryClass = {astro-ph},
       adsurl = {https://ui.adsabs.harvard.edu/abs/2007A&A...474..679M}
}

@ARTICLE{Marcelino2009,
       author = {{Marcelino}, N{\'u}ria and {Cernicharo}, Jos{\'e} and {Tercero}, Bel{\'e}n and {Roueff}, Evelyne},
        title = "{Discovery of Fulminic Acid, HCNO, in Dark Clouds}",
      journal = {\apjl},
         year = 2009,
        month = jan,
       volume = {690},
       number = {1},
        pages = {L27-L30},
          doi = {10.1088/0004-637X/690/1/L27},
archivePrefix = {arXiv},
       eprint = {0811.2679},
 primaryClass = {astro-ph},
       adsurl = {https://ui.adsabs.harvard.edu/abs/2009ApJ...690L..27M}
}

@ARTICLE{Marcelino2018,
       author = {{Marcelino}, N. and {Ag{\'u}ndez}, M. and {Cernicharo}, J. and {Roueff}, E. and {Tafalla}, M.},
        title = "{Discovery of the elusive radical NCO and confirmation of H$_{2}$NCO$^{+}$ in space}",
      journal = {\aap},
         year = 2018,
        month = may,
       volume = {612},
          eid = {L10},
        pages = {L10},
          doi = {10.1051/0004-6361/201833074},
archivePrefix = {arXiv},
       eprint = {1804.05617},
 primaryClass = {astro-ph.GA},
       adsurl = {https://ui.adsabs.harvard.edu/abs/2018A&A...612L..10M}
}

@ARTICLE{Marcelino2023,
       author = {{Marcelino}, N. and {Puzzarini}, C. and {Ag{\'u}ndez}, M. and {Fuentetaja}, R. and {Tercero}, B. and {de Vicente}, P. and {Cernicharo}, J.},
        title = "{First detection of the HSO radical in space}",
      journal = {\aap},
         year = 2023,
        month = jun,
       volume = {674},
          eid = {L13},
        pages = {L13},
          doi = {10.1051/0004-6361/202346935},
       adsurl = {https://ui.adsabs.harvard.edu/abs/2023A&A...674L..13M}
}

@ARTICLE{Martin1961,
       author = {{Martin}, D.~W. and {McDaniel}, E.~W. and {Meeks}, M.~L.},
        title = "{On the Possible Occurence of H$_{3}$\^\{+\} in Interstellar Space.}",
      journal = {\apj},
         year = 1961,
        month = nov,
       volume = {134},
        pages = {1012-1013},
          doi = {10.1086/147232},
       adsurl = {https://ui.adsabs.harvard.edu/abs/1961ApJ...134.1012M}
}

@ARTICLE{Martin1979,
       author = {{Martin}, R.~N. and {Ho}, P.~T.~P.},
        title = "{Detection of extragalactic ammonia.}",
      journal = {\aap},
         year = 1979,
        month = apr,
       volume = {74},
       number = {1},
        pages = {L7-L9},
       adsurl = {https://ui.adsabs.harvard.edu/abs/1979A&A....74L...7M}
}

@ARTICLE{Martin2003,
       author = {{Mart{\'\i}n}, S. and {Mauersberger}, R. and {Mart{\'\i}n-Pintado}, J. and {Garc{\'\i}a-Burillo}, S. and {Henkel}, C.},
        title = "{First detections of extragalactic SO$_{2}$, NS and NO}",
      journal = {\aap},
         year = 2003,
        month = dec,
       volume = {411},
        pages = {L465-L468},
          doi = {10.1051/0004-6361:20031442},
archivePrefix = {arXiv},
       eprint = {astro-ph/0309663},
 primaryClass = {astro-ph},
       adsurl = {https://ui.adsabs.harvard.edu/abs/2003A&A...411L.465M}
}

@ARTICLE{Martin2006,
       author = {{Mart{\'\i}n}, S. and {Mauersberger}, R. and {Mart{\'\i}n-Pintado}, J. and {Henkel}, C. and {Garc{\'\i}a-Burillo}, S.},
        title = "{A 2 Millimeter Spectral Line Survey of the Starburst Galaxy NGC 253}",
      journal = {\apjs},
         year = 2006,
        month = jun,
       volume = {164},
       number = {2},
        pages = {450-476},
          doi = {10.1086/503297},
archivePrefix = {arXiv},
       eprint = {astro-ph/0602360},
 primaryClass = {astro-ph},
       adsurl = {https://ui.adsabs.harvard.edu/abs/2006ApJS..164..450M}
}

@ARTICLE{Matthews1984,
       author = {{Matthews}, H.~E. and {Irvine}, W.~M. and {Friberg}, P. and {Brown}, R.~D. and {Godfrey}, P.~D.},
        title = "{A new interstellar molecule: triearbon monoxide}",
      journal = {\nat},
         year = 1984,
        month = jul,
       volume = {310},
       number = {5973},
        pages = {125-126},
          doi = {10.1038/310125a0},
       adsurl = {https://ui.adsabs.harvard.edu/abs/1984Natur.310..125M}
}

@ARTICLE{Mauersberger1990,
       author = {{Mauersberger}, R. and {Henkel}, C. and {Sage}, L.~J.},
        title = "{Dense gas in nearby galaxies. III. HC3N as an extragalactic density probe.}",
      journal = {\aap},
         year = 1990,
        month = sep,
       volume = {236},
        pages = {63},
       adsurl = {https://ui.adsabs.harvard.edu/abs/1990A&A...236...63M}
}

@ARTICLE{Mauersberger1991,
       author = {{Mauersberger}, R. and {Henkel}, C.},
        title = "{Dense gas in nearby galaxies. IV. The detection of N2H+, SiO, H13CO+, H13CN and HN13C.}",
      journal = {\aap},
         year = 1991,
        month = may,
       volume = {245},
        pages = {457},
       adsurl = {https://ui.adsabs.harvard.edu/abs/1991A&A...245..457M}
}

@ARTICLE{McGuire2012,
       author = {{McGuire}, Brett A. and {Loomis}, Ryan A. and {Charness}, Cameron M. and {Corby}, Joanna F. and {Blake}, Geoffrey A. and {Hollis}, Jan M. and {Lovas}, Frank J. and {Jewell}, Philip R. and {Remijan}, Anthony J.},
        title = "{Interstellar Carbodiimide (HNCNH): A New Astronomical Detection from the GBT PRIMOS Survey via Maser Emission Features}",
      journal = {\apjl},
         year = 2012,
        month = oct,
       volume = {758},
       number = {2},
          eid = {L33},
        pages = {L33},
          doi = {10.1088/2041-8205/758/2/L33},
archivePrefix = {arXiv},
       eprint = {1209.1590},
 primaryClass = {astro-ph.GA},
       adsurl = {https://ui.adsabs.harvard.edu/abs/2012ApJ...758L..33M}
}

@ARTICLE{McKellar1940,
       author = {{McKellar}, A.},
        title = "{Evidence for the Molecular Origin of Some Hitherto Unidentified Interstellar Lines}",
      journal = {\pasp},
         year = 1940,
        month = jun,
       volume = {52},
       number = {307},
        pages = {187},
          doi = {10.1086/125159},
       adsurl = {https://ui.adsabs.harvard.edu/abs/1940PASP...52..187M}
}

@ARTICLE{Menten2011,
       author = {{Menten}, K.~M. and {Wyrowski}, F. and {Belloche}, A. and {G{\"u}sten}, R. and {Dedes}, L. and {M{\"u}ller}, H.~S.~P.},
        title = "{Submillimeter absorption from SH$^{+}$, a new widespread interstellar radical, $^{13}$CH$^{+}$ and HCl}",
      journal = {\aap},
         year = 2011,
        month = jan,
       volume = {525},
          eid = {A77},
        pages = {A77},
          doi = {10.1051/0004-6361/201014363},
archivePrefix = {arXiv},
       eprint = {1009.2825},
 primaryClass = {astro-ph.GA},
       adsurl = {https://ui.adsabs.harvard.edu/abs/2011A&A...525A..77M}
}

@ARTICLE{Meyer1991,
       author = {{Meyer}, David M. and {Roth}, Katherine C.},
        title = "{Discovery of Interstellar NH}",
      journal = {\apjl},
         year = 1991,
        month = aug,
       volume = {376},
        pages = {L49},
          doi = {10.1086/186100},
       adsurl = {https://ui.adsabs.harvard.edu/abs/1991ApJ...376L..49M}
}

@ARTICLE{Monje2011,
       author = {{Monje}, R.~R. and {Phillips}, T.~G. and {Peng}, R. and {Lis}, D.~C. and {Neufeld}, D.~A. and {Emprechtinger}, M.},
        title = "{Discovery of Hydrogen Fluoride in the Cloverleaf Quasar at z = 2.56}",
      journal = {\apjl},
         year = 2011,
        month = dec,
       volume = {742},
       number = {2},
          eid = {L21},
        pages = {L21},
          doi = {10.1088/2041-8205/742/2/L21},
archivePrefix = {arXiv},
       eprint = {1201.4882},
 primaryClass = {astro-ph.CO},
       adsurl = {https://ui.adsabs.harvard.edu/abs/2011ApJ...742L..21M}
}

@ARTICLE{Morris1975,
       author = {{Morris}, M. and {Gilmore}, W. and {Palmer}, P. and {Turner}, B.~E. and {Zuckerman}, B.},
        title = "{Detection of interstellar SiS and a study of the IRC +10216 molecular envelope.}",
      journal = {\apjl},
         year = 1975,
        month = jul,
       volume = {199},
        pages = {L47-L51},
          doi = {10.1086/181846},
       adsurl = {https://ui.adsabs.harvard.edu/abs/1975ApJ...199L..47M}
}

@ARTICLE{Muller2011,
       author = {{Muller}, S. and {Beelen}, A. and {Gu{\'e}lin}, M. and {Aalto}, S. and {Black}, J.~H. and {Combes}, F. and {Curran}, S.~J. and {Theule}, P. and {Longmore}, S.~N.},
        title = "{Molecules at z = 0.89. A 4-mm-rest-frame absorption-line survey toward PKS 1830-211}",
      journal = {\aap},
         year = 2011,
        month = nov,
       volume = {535},
          eid = {A103},
        pages = {A103},
          doi = {10.1051/0004-6361/201117096},
archivePrefix = {arXiv},
       eprint = {1104.3361},
 primaryClass = {astro-ph.CO},
       adsurl = {https://ui.adsabs.harvard.edu/abs/2011A&A...535A.103M}
}

@ARTICLE{Muller2013,
       author = {{Muller}, S. and {Beelen}, A. and {Black}, J.~H. and {Curran}, S.~J. and {Horellou}, C. and {Aalto}, S. and {Combes}, F. and {Gu{\'e}lin}, M. and {Henkel}, C.},
        title = "{A precise and accurate determination of the cosmic microwave background temperature at z = 0.89}",
      journal = {\aap},
         year = 2013,
        month = mar,
       volume = {551},
          eid = {A109},
        pages = {A109},
          doi = {10.1051/0004-6361/201220613},
archivePrefix = {arXiv},
       eprint = {1212.5456},
 primaryClass = {astro-ph.CO},
       adsurl = {https://ui.adsabs.harvard.edu/abs/2013A&A...551A.109M}
}

@ARTICLE{Muller2014,
       author = {{Muller}, S. and {Combes}, F. and {Gu{\'e}lin}, M. and et al.},
        title = "{An ALMA Early Science survey of molecular absorption lines toward PKS 1830-211. Analysis of the absorption profiles}",
      journal = {\aap},
         year = 2014,
        month = jun,
       volume = {566},
          eid = {A112},
        pages = {A112},
          doi = {10.1051/0004-6361/201423646},
archivePrefix = {arXiv},
       eprint = {1404.7667},
 primaryClass = {astro-ph.GA},
       adsurl = {https://ui.adsabs.harvard.edu/abs/2014A&A...566A.112M}
}

@ARTICLE{Muller2015,
       author = {{M{\"u}ller}, Holger S.~P. and {Muller}, S{\'e}bastien and {Schilke}, Peter and {Bergin}, Edwin A. and {Black}, John H. and {Gerin}, Maryvonne and {Lis}, Dariusz C. and {Neufeld}, David A. and {Suri}, S{\"u}meyye},
        title = "{Detection of extragalactic argonium, ArH$^{+}$, toward PKS 1830-211}",
      journal = {\aap},
         year = 2015,
        month = oct,
       volume = {582},
          eid = {L4},
        pages = {L4},
          doi = {10.1051/0004-6361/201527254},
archivePrefix = {arXiv},
       eprint = {1509.06917},
 primaryClass = {astro-ph.GA},
       adsurl = {https://ui.adsabs.harvard.edu/abs/2015A&A...582L...4M}
}

@ARTICLE{Muller2016,
       author = {{Muller}, S. and {Kawaguchi}, K. and {Black}, J.~H. and {Amano}, T.},
        title = "{Detection of extragalactic CF$^{+}$ toward PKS 1830-211. Chemical differentiation in the absorbing gas}",
      journal = {\aap},
         year = 2016,
        month = may,
       volume = {589},
          eid = {L5},
        pages = {L5},
          doi = {10.1051/0004-6361/201628494},
archivePrefix = {arXiv},
       eprint = {1604.00414},
 primaryClass = {astro-ph.GA},
       adsurl = {https://ui.adsabs.harvard.edu/abs/2016A&A...589L...5M}
}

@ARTICLE{Muller2017,
       author = {{Muller}, S. and {M{\"u}ller}, H.~S.~P. and {Black}, J.~H. andet al.},
        title = "{Detection of CH$^{+}$, SH$^{+}$, and their $^{13}$C- and $^{34}$S-isotopologues toward PKS 1830-211}",
      journal = {\aap},
         year = 2017,
        month = oct,
       volume = {606},
          eid = {A109},
        pages = {A109},
          doi = {10.1051/0004-6361/201731405},
archivePrefix = {arXiv},
       eprint = {1707.07446},
 primaryClass = {astro-ph.GA},
       adsurl = {https://ui.adsabs.harvard.edu/abs/2017A&A...606A.109M}
}

@ARTICLE{Neufeld1997,
       author = {{Neufeld}, David A. and {Zmuidzinas}, Jonas and {Schilke}, Peter and {Phillips}, Thomas G.},
        title = "{Discovery of Interstellar Hydrogen Fluoride 1}",
      journal = {\apjl},
         year = 1997,
        month = oct,
       volume = {488},
       number = {2},
        pages = {L141-L144},
          doi = {10.1086/310942},
archivePrefix = {arXiv},
       eprint = {astro-ph/9708013},
 primaryClass = {astro-ph},
       adsurl = {https://ui.adsabs.harvard.edu/abs/1997ApJ...488L.141N}
}

@ARTICLE{Neufeld2006,
       author = {{Neufeld}, D.~A. and {Schilke}, P. and {Menten}, K. and {et } al.},
        title = "{Discovery of interstellar CF$^{+}$}",
      journal = {\aap},
         year = 2006,
        month = aug,
       volume = {454},
       number = {2},
        pages = {L37-L40},
          doi = {10.1051/0004-6361:200600015},
archivePrefix = {arXiv},
       eprint = {astro-ph/0603201},
 primaryClass = {astro-ph},
       adsurl = {https://ui.adsabs.harvard.edu/abs/2006A&A...454L..37N}
}

@ARTICLE{Neufeld2012,
       author = {{Neufeld}, D.~A. and {Falgarone}, E. and {Gerin}, M. and {et } al.},
        title = "{Discovery of interstellar mercapto radicals (SH) with the GREAT instrument on SOFIA}",
      journal = {\aap},
         year = 2012,
        month = jun,
       volume = {542},
          eid = {L6},
        pages = {L6},
          doi = {10.1051/0004-6361/201218870},
archivePrefix = {arXiv},
       eprint = {1202.3142},
 primaryClass = {astro-ph.GA},
       adsurl = {https://ui.adsabs.harvard.edu/abs/2012A&A...542L...6N}
}

@ARTICLE{Nguyen1991,
       author = {{Nguyen-Q-Rieu} and {Henkel}, C. and {Jackson}, J.~M. and {Mauersberger}, R.},
        title = "{Detection of HNCO in external galaxies.}",
      journal = {\aap},
         year = 1991,
        month = jan,
       volume = {241},
        pages = {L33},
       adsurl = {https://ui.adsabs.harvard.edu/abs/1991A&A...241L..33N}
}

@ARTICLE{Ohishi1989,
       author = {{Ohishi}, Masatoshi and {Kaifu}, Norio and {Kawaguchi}, Kentarou and et al.},
        title = "{Detection of a New Circumstellar Carbon Chain Molecule C 4Si}",
      journal = {\apjl},
         year = 1989,
        month = oct,
       volume = {345},
        pages = {L83},
          doi = {10.1086/185558},
       adsurl = {https://ui.adsabs.harvard.edu/abs/1989ApJ...345L..83O}
}

@ARTICLE{Ohishi1991,
       author = {{Ohishi}, Masatoshi and {Suzuki}, Hiroko and {Ishikawa}, Shin-Ichi and et al.},
        title = "{Detection of a New Carbon-Chain Molecule, CCO}",
      journal = {\apjl},
         year = 1991,
        month = oct,
       volume = {380},
        pages = {L39},
          doi = {10.1086/186168},
       adsurl = {https://ui.adsabs.harvard.edu/abs/1991ApJ...380L..39O}
}

@ARTICLE{Ohishi1994,
       author = {{Ohishi}, Masatoshi and {McGonagle}, Douglas and {Irvine}, William M. and {Yamamoto}, Satoshi and {Saito}, Shuji},
        title = "{Detection of a New Interstellar Molecule, H 2CN}",
      journal = {\apjl},
         year = 1994,
        month = may,
       volume = {427},
        pages = {L51},
          doi = {10.1086/187362},
       adsurl = {https://ui.adsabs.harvard.edu/abs/1994ApJ...427L..51O}
}

@ARTICLE{Ohishi1996,
       author = {{Ohishi}, Masatoshi and {Ishikawa}, Shin-Ichi and {Amano}, Takayoshi and et al.},
        title = "{Detection of a New Interstellar Molecular Ion, H 2COH + (Protonated Formaldehyde)}",
      journal = {\apjl},
         year = 1996,
        month = nov,
       volume = {471},
        pages = {L61},
          doi = {10.1086/310325},
       adsurl = {https://ui.adsabs.harvard.edu/abs/1996ApJ...471L..61O}
}

@ARTICLE{Oka1980,
       author = {{Oka}, T.},
        title = "{Observation of the infrared spectrum of H$_{3}$ $^{ + }$}",
      journal = {\prl},
         year = 1980,
        month = aug,
       volume = {45},
       number = {7},
        pages = {531-534},
          doi = {10.1103/PhysRevLett.45.531},
       adsurl = {https://ui.adsabs.harvard.edu/abs/1980PhRvL..45..531O}
}

@ARTICLE{Ossenkopf2010,
       author = {{Ossenkopf}, V. and {M{\"u}ller}, H.~S.~P. and {Lis}, D.~C. and et al.},
        title = "{Detection of interstellar oxidaniumyl: Abundant H$_{2}$O$^{+}$ towards the star-forming regions DR21, Sgr B2, and NGC6334}",
      journal = {\aap},
         year = 2010,
        month = jul,
       volume = {518},
          eid = {L111},
        pages = {L111},
          doi = {10.1051/0004-6361/201014577},
archivePrefix = {arXiv},
       eprint = {1005.2521},
 primaryClass = {astro-ph.GA},
       adsurl = {https://ui.adsabs.harvard.edu/abs/2010A&A...518L.111O}
}

@BOOK{Osterbrock2006,
       author = {{Osterbrock}, Donald E. and {Ferland}, Gary J.},
        title = "{Astrophysics of gaseous nebulae and active galactic nuclei}",
         year = 2006,
       adsurl = {https://ui.adsabs.harvard.edu/abs/2006agna.book.....O}
}

@ARTICLE{Parise2012,
       author = {{Parise}, B. and {Bergman}, P. and {Du}, F.},
        title = "{Detection of the hydroperoxyl radical HO$_{2}$ toward {\ensuremath{\rho}} Ophiuchi A. Additional constraints on the water chemical network}",
      journal = {\aap},
         year = 2012,
        month = may,
       volume = {541},
          eid = {L11},
        pages = {L11},
          doi = {10.1051/0004-6361/201219379},
archivePrefix = {arXiv},
       eprint = {1205.0361},
 primaryClass = {astro-ph.GA},
       adsurl = {https://ui.adsabs.harvard.edu/abs/2012A&A...541L..11P}
}

@ARTICLE{Penzias1971a,
       author = {{Penzias}, A.~A. and {Jefferts}, K.~B. and {Wilson}, R.~W.},
        title = "{Interstellar 12C16O, 13C16O, and 12C18O}",
      journal = {\apj},
         year = 1971,
        month = may,
       volume = {165},
        pages = {229},
          doi = {10.1086/150893},
       adsurl = {https://ui.adsabs.harvard.edu/abs/1971ApJ...165..229P}
}

@ARTICLE{Penzias1971b,
       author = {{Penzias}, A.~A. and {Solomon}, P.~M. and {Wilson}, R.~W. and {Jefferts}, K.~B.},
        title = "{Interstellar Carbon Monosulfide}",
      journal = {\apjl},
         year = 1971,
        month = sep,
       volume = {168},
        pages = {L53},
          doi = {10.1086/180784},
       adsurl = {https://ui.adsabs.harvard.edu/abs/1971ApJ...168L..53P}
}

@ARTICLE{Pety2012,
       author = {{Pety}, J. and {Gratier}, P. and {Guzm{\'a}n}, V. and {Roueff}, E. and {Gerin}, M. and {Goicoechea}, J.~R. and {Bardeau}, S. and {Sievers}, A. and {Le Petit}, F. and {Le Bourlot}, J. and {Belloche}, A. and {Talbi}, D.},
        title = "{The IRAM-30 m line survey of the Horsehead PDR. II. First detection of the l-C$_{3}$H$^{+}$ hydrocarbon cation}",
      journal = {\aap},
         year = 2012,
        month = dec,
       volume = {548},
          eid = {A68},
        pages = {A68},
          doi = {10.1051/0004-6361/201220062},
archivePrefix = {arXiv},
       eprint = {1210.8178},
 primaryClass = {astro-ph.GA},
       adsurl = {https://ui.adsabs.harvard.edu/abs/2012A&A...548A..68P}
}

@ARTICLE{Pulliam2010,
       author = {{Pulliam}, R.~L. and {Savage}, C. and {Ag{\'u}ndez}, M. and et al.},
        title = "{Identification of KCN in IRC+10216: Evidence for Selective Cyanide Chemistry}",
      journal = {\apjl},
         year = 2010,
        month = dec,
       volume = {725},
       number = {2},
        pages = {L181-L185},
          doi = {10.1088/2041-8205/725/2/L181},
       adsurl = {https://ui.adsabs.harvard.edu/abs/2010ApJ...725L.181P}
}

@PHDTHESIS{Quenard2016,
       author = {{Qu{\'e}nard}, David},
        title = "{3D Modeling of Star Formation Regions: The contribution of the GASS GUI to radiative transfer codes}",
       school = {Universite de Toulouse Paul Sabatier, France},
         year = 2016,
        month = sep,
       adsurl = {https://ui.adsabs.harvard.edu/abs/2016PhDT........14Q}
}

@ARTICLE{Remijan2008,
       author = {{Remijan}, Anthony J. and {Hollis}, J.~M. and {Lovas}, F.~J. and et al.},
        title = "{Detection of Interstellar Cyanoformaldehyde (CNCHO)}",
      journal = {\apjl},
         year = 2008,
        month = mar,
       volume = {675},
       number = {2},
        pages = {L85},
          doi = {10.1086/533529},
       adsurl = {https://ui.adsabs.harvard.edu/abs/2008ApJ...675L..85R}
}

@ARTICLE{Rangwala2011,
       author = {{Rangwala}, Naseem and {Maloney}, Philip R. and {Glenn}, Jason and et al.},
        title = "{Observations of Arp 220 Using Herschel-SPIRE: An Unprecedented View of the Molecular Gas in an Extreme Star Formation Environment}",
      journal = {\apj},
         year = 2011,
        month = dec,
       volume = {743},
       number = {1},
          eid = {94},
        pages = {94},
          doi = {10.1088/0004-637X/743/1/94},
archivePrefix = {arXiv},
       eprint = {1106.5054},
 primaryClass = {astro-ph.CO},
       adsurl = {https://ui.adsabs.harvard.edu/abs/2011ApJ...743...94R}
}

@ARTICLE{ReyMontejo2024,
       author = {{Rey-Montejo}, Marta and {Jim{\'e}nez-Serra}, Izaskun and {Mart{\'\i}n-Pintado}, Jes{\'u}s and {Rivilla}, V{\'\i}ctor M. and {Meg{\'\i}as}, Andr{\'e}s and {San Andr{\'e}s}, David and {Sanz-Novo}, Miguel and {Colzi}, Laura and {Zeng}, Shaoshan and {L{\'o}pez-Gallifa}, {\'A}lvaro and {Mart{\'\i}nez-Henares}, Antonio and {Mart{\'\i}n}, Sergio and {Tercero}, Bel{\'e}n and {de Vicente}, Pablo and {Requena-Torres}, Miguel},
        title = "{Discovery of MgS and NaS in the Interstellar Medium and Tentative Detection of CaO}",
      journal = {\apj},
         year = 2024,
        month = nov,
       volume = {975},
       number = {2},
          eid = {174},
        pages = {174},
          doi = {10.3847/1538-4357/ad736e},
archivePrefix = {arXiv},
       eprint = {2407.07693},
 primaryClass = {astro-ph.GA},
       adsurl = {https://ui.adsabs.harvard.edu/abs/2024ApJ...975..174R}
}

@ARTICLE{Rickard1975,
       author = {{Rickard}, L.~J. and {Palmer}, P. and {Morris}, M. and {Zuckerman}, B. and {Turner}, B.~E.},
        title = "{Detection of extragalactic carbon monoxide at millimeter wavelengths.}",
      journal = {\apjl},
         year = 1975,
        month = jul,
       volume = {199},
        pages = {L75-L78},
          doi = {10.1086/181852},
       adsurl = {https://ui.adsabs.harvard.edu/abs/1975ApJ...199L..75R}
}

@ARTICLE{Rickard1977,
       author = {{Rickard}, L.~J. and {Palmer}, P. and {Turner}, B.~E. and {Morris}, M. and {Zuckerman}, B.},
        title = "{Observations of extragalactic molecules. II. HCN and CS.}",
      journal = {\apj},
         year = 1977,
        month = jun,
       volume = {214},
        pages = {390-393},
          doi = {10.1086/155261},
       adsurl = {https://ui.adsabs.harvard.edu/abs/1977ApJ...214..390R}
}

@ARTICLE{Ridgway1976,
       author = {{Ridgway}, S.~T. and {Hall}, D.~N.~B. and {Wojslaw}, R.~S. and {Kleinmann}, S.~G. and {Weinberger}, D.~A.},
        title = "{Circumstellar acetylene in the infrared spectrum of IRC +10216.}",
      journal = {\nat},
         year = 1976,
        month = nov,
       volume = {264},
        pages = {345-346},
          doi = {10.1038/264345a0},
       adsurl = {https://ui.adsabs.harvard.edu/abs/1976Natur.264..345R}
}

@ARTICLE{Rivilla2019,
       author = {{Rivilla}, V.~M. and {Beltr{\'a}n}, M.~T. and {Vasyunin}, A. and et al.},
        title = "{First ALMA maps of HCO, an important precursor of complex organic molecules, towards IRAS 16293-2422}",
      journal = {\mnras},
         year = 2019,
        month = feb,
       volume = {483},
       number = {1},
        pages = {806-823},
          doi = {10.1093/mnras/sty3078},
archivePrefix = {arXiv},
       eprint = {1811.01650},
 primaryClass = {astro-ph.GA},
       adsurl = {https://ui.adsabs.harvard.edu/abs/2019MNRAS.483..806R}
}

@ARTICLE{Rivilla2020,
       author = {{Rivilla}, V{\'\i}ctor M. and {Mart{\'\i}n-Pintado}, Jes{\'u}s and et al.},
        title = "{Prebiotic Precursors of the Primordial RNA World in Space: Detection of NH$_{2}$OH}",
      journal = {\apjl},
         year = 2020,
        month = aug,
       volume = {899},
       number = {2},
          eid = {L28},
        pages = {L28},
          doi = {10.3847/2041-8213/abac55},
archivePrefix = {arXiv},
       eprint = {2008.00228},
 primaryClass = {astro-ph.GA},
       adsurl = {https://ui.adsabs.harvard.edu/abs/2020ApJ...899L..28R}
}

@ARTICLE{Rivilla2021,
       author = {{Rivilla}, V.~M. and {Jim{\'e}nez-Serra}, I. and {Garc{\'\i}a de la Concepci{\'o}n}, J. and {Mart{\'\i}n-Pintado}, J. and {Colzi}, L. and {Rodr{\'\i}guez-Almeida}, L.~F. and {Tercero}, B. and {Rico-Villas}, F. and {Zeng}, S. and {Mart{\'\i}n}, S. and {Requena-Torres}, M.~A. and {de Vicente}, P.},
        title = "{Detection of the cyanomidyl radical (HNCN): a new interstellar species with the NCN backbone}",
      journal = {\mnras},
         year = 2021,
        month = sep,
       volume = {506},
       number = {1},
        pages = {L79-L84},
          doi = {10.1093/mnrasl/slab074},
archivePrefix = {arXiv},
       eprint = {2106.09652},
 primaryClass = {astro-ph.GA},
       adsurl = {https://ui.adsabs.harvard.edu/abs/2021MNRAS.506L..79R}
}

@ARTICLE{Rivilla2022,
       author = {{Rivilla}, V{\'\i}ctor M. and {Garc{\'\i}a De La Concepci{\'o}n}, Juan and {Jim{\'e}nez-Serra}, Izaskun and et al.},
        title = "{Ionize Hard: Interstellar PO+ Detection}",
      journal = {Frontiers in Astronomy and Space Sciences},
         year = 2022,
        month = apr,
       volume = {9},
          eid = {829288},
        pages = {829288},
          doi = {10.3389/fspas.2022.829288},
archivePrefix = {arXiv},
       eprint = {2202.13928},
 primaryClass = {astro-ph.GA},
       adsurl = {https://ui.adsabs.harvard.edu/abs/2022FrASS...9.9288R}
}

@ARTICLE{Rodriguez2021,
       author = {{Rodr{\'\i}guez-Almeida}, Lucas F. and {Jim{\'e}nez-Serra}, Izaskun and {Rivilla}, V{\'\i}ctor M. and et al.},
        title = "{Thiols in the Interstellar Medium: First Detection of HC(O)SH and Confirmation of C$_{2}$H$_{5}$SH}",
      journal = {\apjl},
         year = 2021,
        month = may,
       volume = {912},
       number = {1},
          eid = {L11},
        pages = {L11},
          doi = {10.3847/2041-8213/abf7cb},
archivePrefix = {arXiv},
       eprint = {2104.08036},
 primaryClass = {astro-ph.GA},
       adsurl = {https://ui.adsabs.harvard.edu/abs/2021ApJ...912L..11R}
}

@ARTICLE{Rydbeck1973,
       author = {{Rydbeck}, O.~E.~H. and {Elld{\'e}r}, J. and {Irvine}, W.~M.},
        title = "{Radio Detection of Interstellar CH}",
      journal = {\nat},
         year = 1973,
        month = dec,
       volume = {246},
       number = {5434},
        pages = {466-468},
          doi = {10.1038/246466a0},
       adsurl = {https://ui.adsabs.harvard.edu/abs/1973Natur.246..466R}
}

@ARTICLE{Sage1995,
       author = {{Sage}, Leslie J. and {Ziurys}, L.~M.},
        title = "{Toward Extragalactic Chemistry: Detections of N 2H + and SiO in Nearby Galaxies}",
      journal = {\apj},
         year = 1995,
        month = jul,
       volume = {447},
        pages = {625},
          doi = {10.1086/175904},
       adsurl = {https://ui.adsabs.harvard.edu/abs/1995ApJ...447..625S}
}

@ARTICLE{Saito1972,
       author = {{Saito}, Shuji},
        title = "{Laboratory Observations of the 1\_\{01\}<-0\_\{00\} Transitions for the HCO and DCO Free Radicals by Microwave Spectroscopy}",
      journal = {\apjl},
         year = 1972,
        month = dec,
       volume = {178},
        pages = {L95},
          doi = {10.1086/181092},
       adsurl = {https://ui.adsabs.harvard.edu/abs/1972ApJ...178L..95S}
}

@ARTICLE{Saito1987,
       author = {{Saito}, Shuji and {Kawaguchi}, Kentarou and {Yamamoto}, Satoshi and et al.},
        title = "{Laboratory Detection and Astronomical Identification of a New Free Radical, CCS( 3 Sigma -)}",
      journal = {\apjl},
         year = 1987,
        month = jun,
       volume = {317},
        pages = {L115},
          doi = {10.1086/184923},
       adsurl = {https://ui.adsabs.harvard.edu/abs/1987ApJ...317L.115S}
}

@ARTICLE{SanzNovo2024a,
       author = {{Sanz-Novo}, Miguel and {Rivilla}, V{\'\i}ctor M. and {Jim{\'e}nez-Serra}, Izaskun and {Mart{\'\i}n-Pintado}, Jes{\'u}s and {Colzi}, Laura and {Zeng}, Shaoshan and {Meg{\'\i}as}, Andr{\'e}s and {L{\'o}pez-Gallifa}, {\'A}lvaro and {Mart{\'\i}nez-Henares}, Antonio and {Massalkhi}, Sarah and {Tercero}, Bel{\'e}n and {de Vicente}, Pablo and {San Andr{\'e}s}, David and {Mart{\'\i}n}, Sergio and {Requena-Torres}, Miguel A.},
        title = "{Interstellar Detection of O-protonated Carbonyl Sulfide, HOCS$^{+}$}",
      journal = {\apj},
         year = 2024,
        month = apr,
       volume = {965},
       number = {2},
          eid = {149},
        pages = {149},
          doi = {10.3847/1538-4357/ad2c01},
archivePrefix = {arXiv},
       eprint = {2402.15405},
 primaryClass = {astro-ph.GA},
       adsurl = {https://ui.adsabs.harvard.edu/abs/2024ApJ...965..149S}
}

@ARTICLE{SanzNovo2024b,
       author = {{Sanz-Novo}, Miguel and {Rivilla}, V{\'\i}ctor M. and {M{\"u}ller}, Holger S.~P. and {Jim{\'e}nez-Serra}, Izaskun and {Mart{\'\i}n-Pintado}, Jes{\'u}s and {Colzi}, Laura and {Zeng}, Shaoshan and {Meg{\'\i}as}, Andr{\'e}s and {L{\'o}pez-Gallifa}, {\'A}lvaro and {Mart{\'\i}nez-Henares}, Antonio and {Tercero}, Bel{\'e}n and {de Vicente}, Pablo and {San Andr{\'e}s}, David and {Mart{\'\i}n}, Sergio and {Requena-Torres}, Miguel A.},
        title = "{Discovery of Thionylimide, HNSO, in Space: The first N-, S-, and O-bearing Interstellar Molecule}",
      journal = {\apjl},
         year = 2024,
        month = apr,
       volume = {965},
       number = {2},
          eid = {L26},
        pages = {L26},
          doi = {10.3847/2041-8213/ad3945},
archivePrefix = {arXiv},
       eprint = {2404.01044},
 primaryClass = {astro-ph.GA},
       adsurl = {https://ui.adsabs.harvard.edu/abs/2024ApJ...965L..26S}
}

@ARTICLE{Seaquist1986,
       author = {{Seaquist}, E.~R. and {Bell}, M.~B.},
        title = "{Detection of the Hydrocarbon Ring Molecule C 3H 2 in the Radio Galaxy Centaurus A (= NGC 5128)}",
      journal = {\apjl},
         year = 1986,
        month = apr,
       volume = {303},
        pages = {L67},
          doi = {10.1086/184654},
       adsurl = {https://ui.adsabs.harvard.edu/abs/1986ApJ...303L..67S}
}

@ARTICLE{Shirley2015,
       author = {{Shirley}, Yancy L.},
        title = "{The Critical Density and the Effective Excitation Density of Commonly Observed Molecular Dense Gas Tracers}",
      journal = {\pasp},
         year = 2015,
        month = mar,
       volume = {127},
       number = {949},
        pages = {299},
          doi = {10.1086/680342},
archivePrefix = {arXiv},
       eprint = {1501.01629},
 primaryClass = {astro-ph.IM},
       adsurl = {https://ui.adsabs.harvard.edu/abs/2015PASP..127..299S}
}

@ARTICLE{Sinclair1973,
       author = {{Sinclair}, M.~W. and {Fourikis}, N. and {Ribes}, J.~C. and {Robinson}, B.~J. and {Brown}, R.~D. and {Godfrey}, P.~D.},
        title = "{Detection of interstellar thioformaldehyde}",
      journal = {Australian Journal of Physics},
         year = 1973,
        month = feb,
       volume = {26},
        pages = {85},
          doi = {10.1071/PH730085},
       adsurl = {https://ui.adsabs.harvard.edu/abs/1973AuJPh..26...85S}
}

@ARTICLE{Snyder1969,
       author = {{Snyder}, Lewis E. and {Buhl}, David and {Zuckerman}, B. and {Palmer}, Patrick},
        title = "{Microwave Detection of Interstellar Formaldehyde}",
      journal = {\prl},
         year = 1969,
        month = mar,
       volume = {22},
       number = {13},
        pages = {679-681},
          doi = {10.1103/PhysRevLett.22.679},
       adsurl = {https://ui.adsabs.harvard.edu/abs/1969PhRvL..22..679S}
}

@ARTICLE{Snyder1971,
       author = {{Snyder}, Lewis E. and {Buhl}, David},
        title = "{Observations of Radio Emission from Interstellar Hydrogen Cyanide}",
      journal = {\apjl},
         year = 1971,
        month = jan,
       volume = {163},
        pages = {L47},
          doi = {10.1086/180664},
       adsurl = {https://ui.adsabs.harvard.edu/abs/1971ApJ...163L..47S}
}

@ARTICLE{Snyder1972,
       author = {{Snyder}, L.~E. and {Buhl}, D.},
        title = "{Detection of several new interstellar molecules.}",
      journal = {Annals of the New York Academy of Sciences},
         year = 1972,
        month = may,
       volume = {194},
       number = {1},
        pages = {17-24},
          doi = {10.1111/j.1749-6632.1972.tb12687.x},
       adsurl = {https://ui.adsabs.harvard.edu/abs/1972NYASA.194...17S}
}

@ARTICLE{Snyder1975,
       author = {{Snyder}, L.~E. and {Hollis}, J.~M. and {Ulich}, B.~L. and et al.},
        title = "{Radio detection of interstellar sulfur dioxide.}",
      journal = {\apjl},
         year = 1975,
        month = jun,
       volume = {198},
        pages = {L81-L84},
          doi = {10.1086/181817},
       adsurl = {https://ui.adsabs.harvard.edu/abs/1975ApJ...198L..81S}
}

@BOOK{Sobolev1960,
       author = {{Sobolev}, V.~V.},
        title = "{Moving envelopes of stars}",
         year = 1960,
       adsurl = {https://ui.adsabs.harvard.edu/abs/1960mes..book.....S}
}

@ARTICLE{Souza1977,
       author = {{Souza}, S.~P. and {Lutz}, B.~L.},
        title = "{Detection of C$_{2}$ in the interstellar spectrum of Cygnus OB2 Number 12 (IV Cygni Number 12).}",
      journal = {\apjl},
         year = 1977,
        month = aug,
       volume = {216},
        pages = {L49-L51},
          doi = {10.1086/182507},
       adsurl = {https://ui.adsabs.harvard.edu/abs/1977ApJ...216L..49S}
}

@BOOK{Spitzer1978,
       author = {{Spitzer}, Lyman},
        title = "{Physical processes in the interstellar medium}",
         year = 1978,
          doi = {10.1002/9783527617722},
       adsurl = {https://ui.adsabs.harvard.edu/abs/1978ppim.book.....S}
}

@ARTICLE{Stark1979,
       author = {{Stark}, A.~A. and {Wolff}, R.~S.},
        title = "{Some observations of extragalactic HCO$^{+}$ and HCN.}",
      journal = {\apj},
         year = 1979,
        month = apr,
       volume = {229},
        pages = {118-120},
          doi = {10.1086/156935},
       adsurl = {https://ui.adsabs.harvard.edu/abs/1979ApJ...229..118S}
}

@ARTICLE{Sutton1985,
       author = {{Sutton}, E.~C. and {Blake}, G.~A. and {Masson}, C.~R. and {Phillips}, T.~G.},
        title = "{Molecular line survey of Orion A from 215 to 247 GHz.}",
      journal = {\apjs},
         year = 1985,
        month = jul,
       volume = {58},
        pages = {341-378},
          doi = {10.1086/191045},
       adsurl = {https://ui.adsabs.harvard.edu/abs/1985ApJS...58..341S}
}

@ARTICLE{Swings1937,
       author = {{Swings}, P. and {Rosenfeld}, L.},
        title = "{Considerations Regarding Interstellar Molecules}",
      journal = {\apj},
         year = 1937,
        month = nov,
       volume = {86},
        pages = {483-486},
          doi = {10.1086/143880},
       adsurl = {https://ui.adsabs.harvard.edu/abs/1937ApJ....86..483S}
}

@ARTICLE{Tenenbaum2007,
       author = {{Tenenbaum}, E.~D. and {Woolf}, N.~J. and {Ziurys}, L.~M.},
        title = "{Identification of Phosphorus Monoxide (X$^{2}${\ensuremath{\Pi}}$_{r}$) in VY Canis Majoris: Detection of the First PO Bond in Space}",
      journal = {\apjl},
         year = 2007,
        month = sep,
       volume = {666},
       number = {1},
        pages = {L29-L32},
          doi = {10.1086/521361},
       adsurl = {https://ui.adsabs.harvard.edu/abs/2007ApJ...666L..29T}
}

@ARTICLE{Tenenbaum2008,
       author = {{Tenenbaum}, E.~D. and {Ziurys}, L.~M.},
        title = "{A Search for Phosphine in Circumstellar Envelopes: PH$_{3}$ in IRC +10216 and CRL 2688?}",
      journal = {\apjl},
         year = 2008,
        month = jun,
       volume = {680},
       number = {2},
        pages = {L121},
          doi = {10.1086/589973},
       adsurl = {https://ui.adsabs.harvard.edu/abs/2008ApJ...680L.121T}
}

@ARTICLE{Tenenbaum2009,
       author = {{Tenenbaum}, E.~D. and {Ziurys}, L.~M.},
        title = "{Millimeter Detection of AlO (X $^{2}${\ensuremath{\Sigma}}$^{+}$): Metal Oxide Chemistry in the Envelope of VY Canis Majoris}",
      journal = {\apjl},
         year = 2009,
        month = mar,
       volume = {694},
       number = {1},
        pages = {L59-L63},
          doi = {10.1088/0004-637X/694/1/L59},
       adsurl = {https://ui.adsabs.harvard.edu/abs/2009ApJ...694L..59T}
}

@ARTICLE{Tenenbaum2010,
       author = {{Tenenbaum}, E.~D. and {Ziurys}, L.~M.},
        title = "{Exotic Metal Molecules in Oxygen-rich Envelopes: Detection of AlOH (X$^{1}${\ensuremath{\Sigma}}$^{+}$) in VY Canis Majoris}",
      journal = {\apjl},
         year = 2010,
        month = mar,
       volume = {712},
       number = {1},
        pages = {L93-L97},
          doi = {10.1088/2041-8205/712/1/L93},
       adsurl = {https://ui.adsabs.harvard.edu/abs/2010ApJ...712L..93T}
}

@ARTICLE{Tercero2020,
       author = {{Tercero}, B. and {Cernicharo}, J. and {Cuadrado}, S. and {de Vicente}, P. and {Gu{\'e}lin}, M.},
        title = "{New molecular species at redshift z = 0.89}",
      journal = {\aap},
         year = 2020,
        month = apr,
       volume = {636},
          eid = {L7},
        pages = {L7},
          doi = {10.1051/0004-6361/202037837},
archivePrefix = {arXiv},
       eprint = {2004.02486},
 primaryClass = {astro-ph.GA},
       adsurl = {https://ui.adsabs.harvard.edu/abs/2020A&A...636L...7T}
}

@ARTICLE{Thaddeus1972,
       author = {{Thaddeus}, P. and {Kutner}, M.~L. and {Penzias}, A.~A. and {Wilson}, R.~W. and {Jefferts}, K.~B.},
        title = "{Interstellar Hydrogen Sulfide}",
      journal = {\apjl},
         year = 1972,
        month = sep,
       volume = {176},
        pages = {L73},
          doi = {10.1086/181023},
       adsurl = {https://ui.adsabs.harvard.edu/abs/1972ApJ...176L..73T}
}

@ARTICLE{Thaddeus1981,
       author = {{Thaddeus}, P. and {Guelin}, M. and {Linke}, R.~A.},
        title = "{Three new 'nonterrestrial' molecules}",
      journal = {\apjl},
         year = 1981,
        month = may,
       volume = {246},
        pages = {L41-L45},
          doi = {10.1086/183549},
       adsurl = {https://ui.adsabs.harvard.edu/abs/1981ApJ...246L..41T}
}

@ARTICLE{Thaddeus1984,
       author = {{Thaddeus}, P. and {Cummins}, S.~E. and {Linke}, R.~A.},
        title = "{Identification of the SiCC radical toward IRC +10216 : the first molecular ring in an astronomical source.}",
      journal = {\apjl},
         year = 1984,
        month = aug,
       volume = {283},
        pages = {L45-L48},
          doi = {10.1086/184330},
       adsurl = {https://ui.adsabs.harvard.edu/abs/1984ApJ...283L..45T}
}

@ARTICLE{Thaddeus1985a,
       author = {{Thaddeus}, P. and {Gottlieb}, C.~A. and {Hjalmarson}, A. and {Johansson}, L.~E.~B. and {Irvine}, W.~M. and {Friberg}, P. and {Linke}, R.~A.},
        title = "{Astronomical identification of the C3 H radical.}",
      journal = {\apjl},
         year = 1985,
        month = jul,
       volume = {294},
        pages = {L49-L53},
          doi = {10.1086/184507},
       adsurl = {https://ui.adsabs.harvard.edu/abs/1985ApJ...294L..49T}
}

@ARTICLE{Thaddeus1985,
       author = {{Thaddeus}, P. and {Vrtilek}, J.~M. and {Gottlieb}, C.~A.},
        title = "{Laboratory and astronomical identification of cyclopropenylidene, C3H2.}",
      journal = {\apjl},
         year = 1985,
        month = dec,
       volume = {299},
        pages = {L63-L66},
          doi = {10.1086/184581},
       adsurl = {https://ui.adsabs.harvard.edu/abs/1985ApJ...299L..63T}
}

@ARTICLE{Thaddeus2008,
       author = {{Thaddeus}, P. and {Gottlieb}, C.~A. and {Gupta}, H. and {Br{\"u}nken}, S. and {McCarthy}, M.~C. and {Ag{\'u}ndez}, M. and {Gu{\'e}lin}, M. and {Cernicharo}, J.},
        title = "{Laboratory and Astronomical Detection of the Negative Molecular Ion C$_{3}$N$^{-}$}",
      journal = {\apj},
         year = 2008,
        month = apr,
       volume = {677},
       number = {2},
        pages = {1132-1139},
          doi = {10.1086/528947},
       adsurl = {https://ui.adsabs.harvard.edu/abs/2008ApJ...677.1132T}
}

@ARTICLE{Thompson1978,
       author = {{Thompson}, R.~I. and {Lebofsky}, M.~J. and {Rieke}, G.~H.},
        title = "{The 2 - 2.5 micron spectrum of NGC 1068: a detection of extragalactic molecular hydrogen.}",
      journal = {\apjl},
         year = 1978,
        month = jun,
       volume = {222},
        pages = {L49-L53},
          doi = {10.1086/182690},
       adsurl = {https://ui.adsabs.harvard.edu/abs/1978ApJ...222L..49T}
}

@BOOK{Thompson2017,
       author = {{Thompson}, A. Richard and {Moran}, James M. and {Swenson}, George W., Jr.},
        title = "{Interferometry and Synthesis in Radio Astronomy, 3rd Edition}",
         year = 2017,
          doi = {10.1007/978-3-319-44431-4},
       adsurl = {https://ui.adsabs.harvard.edu/abs/2017isra.book.....T}
}

@BOOK{Townes1955,
       author = {{Townes}, C.~H. and {Schawlow}, A.~L.},
        title = "{Microwave Spectroscopy}",
         year = 1955,
       adsurl = {https://ui.adsabs.harvard.edu/abs/1955misp.book.....T}
}

@ARTICLE{Tucker1974,
       author = {{Tucker}, K.~D. and {Kutner}, M.~L. and {Thaddeus}, P.},
        title = "{The Ethynyl Radical C$_{2}$H-A New Interstellar Molecule}",
      journal = {\apjl},
         year = 1974,
        month = nov,
       volume = {193},
        pages = {L115},
          doi = {10.1086/181646},
       adsurl = {https://ui.adsabs.harvard.edu/abs/1974ApJ...193L.115T}
}

@ARTICLE{Turner1974,
       author = {{Turner}, B.~E.},
        title = "{U93.174: a New Interstellar Line with Quadrupole Hyperfine Splitting}",
      journal = {\apjl},
         year = 1974,
        month = oct,
       volume = {193},
        pages = {L83},
          doi = {10.1086/181638},
       adsurl = {https://ui.adsabs.harvard.edu/abs/1974ApJ...193L..83T}
}

@ARTICLE{Turner1975,
       author = {{Turner}, B.~E. and {Liszt}, H.~S. and {Kaifu}, N. and {Kisliakov}, A.~G.},
        title = "{Microwave detection of interstellar cyanamide.}",
      journal = {\apjl},
         year = 1975,
        month = nov,
       volume = {201},
        pages = {L149-L152},
          doi = {10.1086/181963},
       adsurl = {https://ui.adsabs.harvard.edu/abs/1975ApJ...201L.149T}
}

@ARTICLE{Turner1977,
       author = {{Turner}, B.~E.},
        title = "{Microwave detection of interstellar ketene.}",
      journal = {\apjl},
         year = 1977,
        month = apr,
       volume = {213},
        pages = {L75-L79},
          doi = {10.1086/182413},
       adsurl = {https://ui.adsabs.harvard.edu/abs/1977ApJ...213L..75T}
}

@ARTICLE{Turner1987,
       author = {{Turner}, B.~E. and {Bally}, John},
        title = "{Detection of Interstellar PN: The First Identified Phosphorus Compound in the Interstellar Medium}",
      journal = {\apjl},
         year = 1987,
        month = oct,
       volume = {321},
        pages = {L75},
          doi = {10.1086/185009},
       adsurl = {https://ui.adsabs.harvard.edu/abs/1987ApJ...321L..75T}
}

@ARTICLE{Ulich1976,
       author = {{Ulich}, B.~L. and {Haas}, R.~W.},
        title = "{Absolute calibration of millimeter-wavelength spectral lines.}",
      journal = {\apjs},
         year = 1976,
        month = mar,
       volume = {30},
        pages = {247-258},
          doi = {10.1086/190361},
       adsurl = {https://ui.adsabs.harvard.edu/abs/1976ApJS...30..247U}
}

@ARTICLE{Ulich1977,
       author = {{Ulich}, B.~L. and {Hollis}, J.~M. and {Snyder}, L.~E.},
        title = "{Radio detection of nitroxyl (HNO): the first interstellar NO bond.}",
      journal = {\apjl},
         year = 1977,
        month = oct,
       volume = {217},
        pages = {L105-L108},
          doi = {10.1086/182549},
       adsurl = {https://ui.adsabs.harvard.edu/abs/1977ApJ...217L.105U}
}

@ARTICLE{Turner1971,
       author = {{Turner}, B.~E.},
        title = "{Detection of Interstellar Cyanoacetylene}",
      journal = {\apjl},
         year = 1971,
        month = jan,
       volume = {163},
        pages = {L35},
          doi = {10.1086/180662},
       adsurl = {https://ui.adsabs.harvard.edu/abs/1971ApJ...163L..35T}
}

@ARTICLE{Turner1992a,
       author = {{Turner}, B.~E.},
        title = "{Detection of SiN in IRC +10216}",
      journal = {\apjl},
         year = 1992,
        month = mar,
       volume = {388},
        pages = {L35},
          doi = {10.1086/186324},
       adsurl = {https://ui.adsabs.harvard.edu/abs/1992ApJ...388L..35T}
}

@ARTICLE{Turner1992b,
       author = {{Turner}, B.~E.},
        title = "{Detection of Interstellar SO +: A Diagnostic of Dissociative Shock Chemistry}",
      journal = {\apjl},
         year = 1992,
        month = sep,
       volume = {396},
        pages = {L107},
          doi = {10.1086/186528},
       adsurl = {https://ui.adsabs.harvard.edu/abs/1992ApJ...396L.107T}
}

@ARTICLE{Turner1994,
       author = {{Turner}, B.~E. and {Steimle}, T.~C. and {Meerts}, Leo},
        title = "{Detection of Sodium Cyanide (NaCN) in IRC 10216}",
      journal = {\apjl},
         year = 1994,
        month = may,
       volume = {426},
        pages = {L97},
          doi = {10.1086/174043},
       adsurl = {https://ui.adsabs.harvard.edu/abs/1994ApJ...426L..97T}
}

@ARTICLE{Usero2004,
       author = {{Usero}, A. and {Garc{\'\i}a-Burillo}, S. and {Fuente}, A. and {Mart{\'\i}n-Pintado}, J. and {Rodr{\'\i}guez-Fern{\'a}ndez}, N.~J.},
        title = "{Molecular gas chemistry in AGN. I. The IRAM 30 m survey of NGC 1068}",
      journal = {\aap},
         year = 2004,
        month = jun,
       volume = {419},
        pages = {897-912},
          doi = {10.1051/0004-6361:20035774},
archivePrefix = {arXiv},
       eprint = {astro-ph/0402556},
 primaryClass = {astro-ph},
       adsurl = {https://ui.adsabs.harvard.edu/abs/2004A&A...419..897U}
}

@ARTICLE{Vancittert1934,
       author = {{van Cittert}, P.~H.},
        title = "{Die Wahrscheinliche Schwingungsverteilung in Einer von Einer Lichtquelle Direkt Oder Mittels Einer Linse Beleuchteten Ebene}",
      journal = {Physica},
         year = 1934,
        month = jan,
       volume = {1},
       number = {1},
        pages = {201-210},
          doi = {10.1016/S0031-8914(34)90026-4},
       adsurl = {https://ui.adsabs.harvard.edu/abs/1934Phy.....1..201V}
}

@ARTICLE{Vandertak2008,
       author = {{van der Tak}, F.~F.~S. and {Aalto}, S. and {Meijerink}, R.},
        title = "{Detection of extragalactic H\_3O\^+}",
      journal = {\aap},
         year = 2008,
        month = jan,
       volume = {477},
       number = {1},
        pages = {L5-L8},
          doi = {10.1051/0004-6361:20078824},
archivePrefix = {arXiv},
       eprint = {0711.2109},
 primaryClass = {astro-ph},
       adsurl = {https://ui.adsabs.harvard.edu/abs/2008A&A...477L...5V}
}

@ARTICLE{vanderWerf2010,
       author = {{van der Werf}, P.~P. and {Isaak}, K.~G. and {Meijerink}, R. and et al.},
        title = "{Black hole accretion and star formation as drivers of gas excitation and chemistry in Markarian 231}",
      journal = {\aap},
         year = 2010,
        month = jul,
       volume = {518},
          eid = {L42},
        pages = {L42},
          doi = {10.1051/0004-6361/201014682},
archivePrefix = {arXiv},
       eprint = {1005.2877},
 primaryClass = {astro-ph.GA},
       adsurl = {https://ui.adsabs.harvard.edu/abs/2010A&A...518L..42V}
}

@ARTICLE{vanDishoeck1988,
       author = {{van Dishoeck}, Ewine F. and {Black}, John H.},
        title = "{The Photodissociation and Chemistry of Interstellar CO}",
      journal = {\apj},
         year = 1988,
        month = nov,
       volume = {334},
        pages = {771},
          doi = {10.1086/166877},
       adsurl = {https://ui.adsabs.harvard.edu/abs/1988ApJ...334..771V}
}

@ARTICLE{vanDishoeck1993,
       author = {{van Dishoeck}, Ewine F. and {Jansen}, David J. and {Schilke}, Peter and {Phillips}, T.~G.},
        title = "{Detection of the Interstellar NH 2 Radical}",
      journal = {\apjl},
         year = 1993,
        month = oct,
       volume = {416},
        pages = {L83},
          doi = {10.1086/187076},
       adsurl = {https://ui.adsabs.harvard.edu/abs/1993ApJ...416L..83V}
}

@ARTICLE{vanDishoeck1996,
       author = {{van Dishoeck}, E.~F. and {Helmich}, F.~P. and {de Graauw}, T. and et al.},
        title = "{A search for interstellar gas-phase CO\_2\_. Gas: solid state abundance ratios.}",
      journal = {\aap},
         year = 1996,
        month = nov,
       volume = {315},
        pages = {L349-L352},
       adsurl = {https://ui.adsabs.harvard.edu/abs/1996A&A...315L.349V}
}

@ARTICLE{Vanloon1999,
       author = {{van Loon}, Jacco Th. and {Zijlstra}, Albert A. and {Groenewegen}, M.~A.~T.},
        title = "{Luminous carbon stars in the Magellanic Clouds}",
      journal = {\aap},
         year = 1999,
        month = jun,
       volume = {346},
        pages = {805-810},
archivePrefix = {arXiv},
       eprint = {astro-ph/9902284},
 primaryClass = {astro-ph},
       adsurl = {https://ui.adsabs.harvard.edu/abs/1999A&A...346..805V}
}

@INPROCEEDINGS{Vastel2015,
       author = {{Vastel}, C. and {Bottinelli}, S. and {Caux}, E. and {Glorian}, J.-M. and {Boiziot}, M.},
        title = "{CASSIS: a tool to visualize and analyse instrumental and synthetic spectra.}",
    booktitle = {SF2A-2015: Proceedings of the Annual meeting of the French Society of Astronomy and Astrophysics},
         year = 2015,
       editor = {{Martins}, F. and {Boissier}, S. and {Buat}, V. and {Cambr{\'e}sy}, L. and {Petit}, P.},
        month = dec,
        pages = {313-316},
       adsurl = {https://ui.adsabs.harvard.edu/abs/2015sf2a.conf..313V}
}

@ARTICLE{Vastel2018,
       author = {{Vastel}, Charlotte and {Qu{\'e}nard}, D. and {Le Gal}, R. and et al.},
        title = "{Sulphur chemistry in the L1544 pre-stellar core}",
      journal = {\mnras},
         year = 2018,
        month = aug,
       volume = {478},
       number = {4},
        pages = {5514-5532},
          doi = {10.1093/mnras/sty1336},
archivePrefix = {arXiv},
       eprint = {1806.01102},
 primaryClass = {astro-ph.GA},
       adsurl = {https://ui.adsabs.harvard.edu/abs/2018MNRAS.478.5514V}
}

@ARTICLE{Vastel2019,
       author = {{Vastel}, C. and {Loison}, J.~C. and {Wakelam}, V. and {Lefloch}, B.},
        title = "{Isocyanogen formation in the cold interstellar medium}",
      journal = {\aap},
         year = 2019,
        month = may,
       volume = {625},
          eid = {A91},
        pages = {A91},
          doi = {10.1051/0004-6361/201935010},
archivePrefix = {arXiv},
       eprint = {1904.07570},
 primaryClass = {astro-ph.GA},
       adsurl = {https://ui.adsabs.harvard.edu/abs/2019A&A...625A..91V}
}

@ARTICLE{Wallerstein1971,
       author = {{Wallerstein}, George},
        title = "{Spectroscopic Observations of VY Canis Majoris during 1969-1971}",
      journal = {\apj},
         year = 1971,
        month = oct,
       volume = {169},
        pages = {195},
          doi = {10.1086/151131},
       adsurl = {https://ui.adsabs.harvard.edu/abs/1971ApJ...169..195W}
}

@ARTICLE{Weaver1965,
       author = {{Weaver}, Harold and {Williams}, David R.~W. and {Dieter}, N.~H. and {Lum}, W.~T.},
        title = "{Observations of a Strong Unidentified Microwave Line and of Emission from the OH Molecule}",
      journal = {\nat},
         year = 1965,
        month = oct,
       volume = {208},
       number = {5005},
        pages = {29-31},
          doi = {10.1038/208029a0},
       adsurl = {https://ui.adsabs.harvard.edu/abs/1965Natur.208...29W}
}

@ARTICLE{Weinreb1963,
       author = {{Weinreb}, S. and {Barrett}, A.~H. and {Meeks}, M.~L. and {Henry}, J.~C.},
        title = "{Radio Observations of OH in the Interstellar Medium}",
      journal = {\nat},
         year = 1963,
        month = nov,
       volume = {200},
       number = {4909},
        pages = {829-831},
          doi = {10.1038/200829a0},
       adsurl = {https://ui.adsabs.harvard.edu/abs/1963Natur.200..829W}
}

@ARTICLE{Weliachew1971,
       author = {{Weliachew}, L.},
        title = "{Detection of Interstellar OH in Two External Galaxies}",
      journal = {\apjl},
         year = 1971,
        month = jul,
       volume = {167},
        pages = {L47},
          doi = {10.1086/180757},
       adsurl = {https://ui.adsabs.harvard.edu/abs/1971ApJ...167L..47W}
}

@ARTICLE{Whiteoak1980,
       author = {{Whiteoak}, J.~B. and {Gardner}, F.~F. and {Hoglund}, B.},
        title = "{The detection of CH in external galaxies}",
      journal = {\mnras},
         year = 1980,
        month = jan,
       volume = {190},
        pages = {17P-22P},
          doi = {10.1093/mnras/190.1.17P},
       adsurl = {https://ui.adsabs.harvard.edu/abs/1980MNRAS.190P..17W}
}

@ARTICLE{Wilson1970,
       author = {{Wilson}, R.~W. and {Jefferts}, K.~B. and {Penzias}, A.~A.},
        title = "{Carbon Monoxide in the Orion Nebula}",
      journal = {\apjl},
         year = 1970,
        month = jul,
       volume = {161},
        pages = {L43},
          doi = {10.1086/180567},
       adsurl = {https://ui.adsabs.harvard.edu/abs/1970ApJ...161L..43W}
}

@ARTICLE{Wilson1971,
       author = {{Wilson}, R.~W. and {Penzias}, A.~A. and {Jefferts}, K.~B. and {Kutner}, M. and {Thaddeus}, P.},
        title = "{Discovery of Interstellar Silicon Monoxide}",
      journal = {\apjl},
         year = 1971,
        month = aug,
       volume = {167},
        pages = {L97},
          doi = {10.1086/180769},
       adsurl = {https://ui.adsabs.harvard.edu/abs/1971ApJ...167L..97W}
}

@ARTICLE{Weiss2010,
       author = {{Wei{\ss}}, A. and {Requena-Torres}, M.~A. and {G{\"u}sten}, R. and et al.},
        title = "{HIFI spectroscopy of low-level water transitions in M 82}",
      journal = {\aap},
         year = 2010,
        month = oct,
       volume = {521},
          eid = {L1},
        pages = {L1},
          doi = {10.1051/0004-6361/201015078},
archivePrefix = {arXiv},
       eprint = {1007.1167},
 primaryClass = {astro-ph.CO},
       adsurl = {https://ui.adsabs.harvard.edu/abs/2010A&A...521L...1W}
}

@ARTICLE{Welty2013,
       author = {{Welty}, Daniel E. and {Howk}, J. Christopher and {Lehner}, Nicolas and {Black}, John H.},
        title = "{Detection of interstellar C$_{2}$ and C$_{3}$ in the Small Magellanic Cloud}",
      journal = {\mnras},
         year = 2013,
        month = jan,
       volume = {428},
       number = {2},
        pages = {1107-1115},
          doi = {10.1093/mnras/sts093},
archivePrefix = {arXiv},
       eprint = {1209.6420},
 primaryClass = {astro-ph.GA},
       adsurl = {https://ui.adsabs.harvard.edu/abs/2013MNRAS.428.1107W}
}

@BOOK{Williams2014,
       author = {{Williams}, David A. and {Viti}, Serena},
        title = "{Observational Molecular Astronomy}",
         year = 2014,
       adsurl = {https://ui.adsabs.harvard.edu/abs/2014oma..book.....W}
}

@ARTICLE{Wilson1995,
       author = {{Wilson}, Christine D.},
        title = "{The Metallicity Dependence of the CO-to-H 2 Conversion Factor from Observations of Local Group Galaxies}",
      journal = {\apjl},
         year = 1995,
        month = aug,
       volume = {448},
        pages = {L97},
          doi = {10.1086/309615},
archivePrefix = {arXiv},
       eprint = {astro-ph/9506103},
 primaryClass = {astro-ph},
       adsurl = {https://ui.adsabs.harvard.edu/abs/1995ApJ...448L..97W}
}

@BOOK{Wilson2013,
       author = {{Wilson}, Thomas L. and {Rohlfs}, Kristen and {H{\"u}ttemeister}, Susanne},
        title = "{Tools of Radio Astronomy}",
         year = 2013,
          doi = {10.1007/978-3-642-39950-3},
       adsurl = {https://ui.adsabs.harvard.edu/abs/2013tra..book.....W}
}

@ARTICLE{Winnewisser1975,
       author = {{Winnewisser}, G. and {Churchwell}, E.},
        title = "{Detection of formic acid in Sagittarius B2 by its {}2$_{11}$-{}2$_{12}$ transition.}",
      journal = {\apjl},
         year = 1975,
        month = aug,
       volume = {200},
        pages = {L33-L36},
          doi = {10.1086/181890},
       adsurl = {https://ui.adsabs.harvard.edu/abs/1975ApJ...200L..33W}
}

@ARTICLE{Woods1975,
       author = {{Woods}, R. Claude and {Dixon}, Thomas A. and {Saykally}, Richard J. and {Szanto}, Peter G.},
        title = "{Laboratory microwave spectrum of HCO$^{ + }$}",
      journal = {\prl},
         year = 1975,
        month = nov,
       volume = {35},
       number = {19},
        pages = {1269-1272},
          doi = {10.1103/PhysRevLett.35.1269},
       adsurl = {https://ui.adsabs.harvard.edu/abs/1975PhRvL..35.1269W}
}

@ARTICLE{Woods1983,
       author = {{Woods}, R.~C. and {Gudeman}, C.~S. and {Dickman}, R.~L. and et al.},
        title = "{The / abundance ratio in molecular clouds.}",
      journal = {\apj},
         year = 1983,
        month = jul,
       volume = {270},
        pages = {583-588},
          doi = {10.1086/161150},
       adsurl = {https://ui.adsabs.harvard.edu/abs/1983ApJ...270..583W}
}

@ARTICLE{Wootten1986,
       author = {{Wootten}, A. and {Boulanger}, F. and {Bogey}, M. and {Combes}, F. and {Encrenaz}, P.~J. and {Gerin}, M. and {Ziurys}, L.},
        title = "{A search for interstellar H3O+.}",
      journal = {\aap},
         year = 1986,
        month = sep,
       volume = {166},
        pages = {L15-L18},
       adsurl = {https://ui.adsabs.harvard.edu/abs/1986A&A...166L..15W}
}

@ARTICLE{Wyrowski2010,
       author = {{Wyrowski}, F. and {Menten}, K.~M. and {G{\"u}sten}, R. and {Belloche}, A.},
        title = "{First interstellar detection of OH$^{+}$}",
      journal = {\aap},
         year = 2010,
        month = jul,
       volume = {518},
          eid = {A26},
        pages = {A26},
          doi = {10.1051/0004-6361/201014364},
archivePrefix = {arXiv},
       eprint = {1004.2627},
 primaryClass = {astro-ph.GA},
       adsurl = {https://ui.adsabs.harvard.edu/abs/2010A&A...518A..26W}
}

@ARTICLE{Yamamoto1987,
       author = {{Yamamoto}, Satoshi and {Saito}, Shuji and {Ohishi}, Masatoshi and {Suzuki}, Hiroko and {Ishikawa}, Shin-Ichi and {Kaifu}, Norio and {Murakami}, Akinori},
        title = "{Laboratory and Astronomical Detection of the Cyclic H 3H Radical}",
      journal = {\apjl},
         year = 1987,
        month = nov,
       volume = {322},
        pages = {L55},
          doi = {10.1086/185036},
       adsurl = {https://ui.adsabs.harvard.edu/abs/1987ApJ...322L..55Y}
}

@ARTICLE{Yamamoto1987a,
       author = {{Yamamoto}, Satoshi and {Saito}, Shuji and {Kawaguchi}, Kentarou and {Kaifu}, Norio and {Suzuki}, Hiroko and {Ohishi}, Masatoshi},
        title = "{Laboratory Detection of a New Carbon-Chain Molecule C 3S and Its Astronomical Identification}",
      journal = {\apjl},
         year = 1987,
        month = jun,
       volume = {317},
        pages = {L119},
          doi = {10.1086/184924},
       adsurl = {https://ui.adsabs.harvard.edu/abs/1987ApJ...317L.119Y}
}

@ARTICLE{Zack2011,
       author = {{Zack}, L.~N. and {Halfen}, D.~T. and {Ziurys}, L.~M.},
        title = "{Detection of FeCN (x $^{4}${\ensuremath{\Delta}} $_{i}$ ) in IRC+10216: A New Interstellar Molecule}",
      journal = {\apjl},
         year = 2011,
        month = jun,
       volume = {733},
       number = {2},
          eid = {L36},
        pages = {L36},
          doi = {10.1088/2041-8205/733/2/L36},
       adsurl = {https://ui.adsabs.harvard.edu/abs/2011ApJ...733L..36Z}
}

@ARTICLE{Zernike1938,
       author = {{Zernike}, F.},
        title = "{The concept of degree of coherence and its application to optical problems}",
      journal = {Physica},
         year = 1938,
        month = aug,
       volume = {5},
       number = {8},
        pages = {785-795},
          doi = {10.1016/S0031-8914(38)80203-2},
       adsurl = {https://ui.adsabs.harvard.edu/abs/1938Phy.....5..785Z}
}

@ARTICLE{Ziurys1986,
       author = {{Ziurys}, L.~M. and {Turner}, B.~E.},
        title = "{HCNH +: A New Interstellar Molecular Ion}",
      journal = {\apjl},
         year = 1986,
        month = mar,
       volume = {302},
        pages = {L31},
          doi = {10.1086/184631},
       adsurl = {https://ui.adsabs.harvard.edu/abs/1986ApJ...302L..31Z}
}

@ARTICLE{Ziurys1987,
       author = {{Ziurys}, L.~M.},
        title = "{Detection of Interstellar PN: The First Phosphorus-bearing Species Observed in Molecular Clouds}",
      journal = {\apjl},
         year = 1987,
        month = oct,
       volume = {321},
        pages = {L81},
          doi = {10.1086/185010},
       adsurl = {https://ui.adsabs.harvard.edu/abs/1987ApJ...321L..81Z}
}

@ARTICLE{Ziurys1994,
       author = {{Ziurys}, L.~M. and {Apponi}, A.~J. and {Hollis}, J.~M. and {Snyder}, L.~E.},
        title = "{Detection of Interstellar N 2O: A New Molecule Containing an N-O Bond}",
      journal = {\apjl},
         year = 1994,
        month = dec,
       volume = {436},
        pages = {L181},
          doi = {10.1086/187662},
       adsurl = {https://ui.adsabs.harvard.edu/abs/1994ApJ...436L.181Z}
}

@ARTICLE{Ziurys1994b,
       author = {{Ziurys}, L.~M. and {Apponi}, A.~J. and {Phillips}, T.~G.},
        title = "{Exotic Fluoride Molecules in IRC +10216: Confirmation of AlF and Searches for MgF and CaF}",
      journal = {\apj},
         year = 1994,
        month = oct,
       volume = {433},
        pages = {729},
          doi = {10.1086/174682},
       adsurl = {https://ui.adsabs.harvard.edu/abs/1994ApJ...433..729Z}
}

@ARTICLE{Ziurys1995,
       author = {{Ziurys}, L.~M. and {Apponi}, A.~J. and {Guelin}, M. and {Cernicharo}, J.},
        title = "{Detection of MgCN in IRC +10216: A New Metal-bearing Free Radical}",
      journal = {\apjl},
         year = 1995,
        month = may,
       volume = {445},
        pages = {L47},
          doi = {10.1086/187886},
       adsurl = {https://ui.adsabs.harvard.edu/abs/1995ApJ...445L..47Z}
}

@ARTICLE{Ziurys2002,
       author = {{Ziurys}, L.~M. and {Savage}, C. and {Highberger}, J.~L. and et al.},
        title = "{More Metal Cyanide Species: Detection of AlNC (X $^{1}${\ensuremath{\Sigma}}$^{+}$) toward IRC +10216}",
      journal = {\apjl},
         year = 2002,
        month = jan,
       volume = {564},
       number = {1},
        pages = {L45-L48},
          doi = {10.1086/338775},
       adsurl = {https://ui.adsabs.harvard.edu/abs/2002ApJ...564L..45Z}
}

@ARTICLE{Zuckerman1971,
       author = {{Zuckerman}, B. and {Ball}, John A. and {Gottlieb}, Carl A.},
        title = "{Microwave Detection of Interstellar Formic Acid}",
      journal = {\apjl},
         year = 1971,
        month = jan,
       volume = {163},
        pages = {L41},
          doi = {10.1086/180663},
       adsurl = {https://ui.adsabs.harvard.edu/abs/1971ApJ...163L..41Z}
}

@ARTICLE{Zuckerman1972,
       author = {{Zuckerman}, B. and {Morris}, M. and {Palmer}, Patrick and {Turner}, B.~E.},
        title = "{Observations of cs, HCN, U89.2, and U90.7 in NGC 2264}",
      journal = {\apjl},
         year = 1972,
        month = may,
       volume = {173},
        pages = {L125},
          doi = {10.1086/180931},
       adsurl = {https://ui.adsabs.harvard.edu/abs/1972ApJ...173L.125Z}
}





\end{document}